\documentclass[usenatbib,times,usegraphicx,useAMS]{mn2e}
\usepackage{amssymb,url,times,bm}
\usepackage{xcolor}

{\left\lbrace\begin{array}{@{}l@{}}}%
{\end{array}\right.}

\defcitealias{adams_04}{A04}

\title[External photoevaporation of protoplanetary discs]{External photoevaporation of protoplanetary discs in sparse stellar groups: the impact of dust growth}

\author[Facchini, Clarke and Bisbas]{Stefano Facchini$^{1,2}$\thanks{facchini@mpe.mpg.de}, Cathie J. Clarke$^1$ and Thomas G. Bisbas$^{2,3,4}$\\
$^1$ Institute of Astronomy, Madingley Road, Cambridge CB3 OHA, UK\\
$^2$ Max-Planck-Institut f\"{u}r Extraterrestrische Physik, Giessenbachstrasse 1, 85748 Garching, Germany\\
$^3$ University College London, Gower Place, London, WC1E 6BT, UK\\
$^4$ Department of Astronomy, University of Florida, Gainesville, FL 32611, USA\\
}
\date{Submission date}
\pagerange{\pageref{firstpage}--\pageref{lastpage}} \pubyear{2015}

\begin{document}
\label{firstpage}
\bibliographystyle{mn2e_long}
\maketitle

\begin{abstract}

We estimate the mass loss rates of photoevaporative winds launched from the outer edge of protoplanetary discs impinged by an ambient radiation field. We focus on mild/moderate environments (the number of stars in the group/cluster is $N\gtrsim50$), and explore disc sizes ranging between $20$ and $250$\,AU. We evaluate the steady-state structures of the photoevaporative winds by coupling temperature estimates obtained with a PDR code with 1D radial hydrodynamical equations. We also consider the impact of dust dragging and grain growth on the final mass loss rates. We find that these winds are much more significant than have been appreciated hitherto when grain growth is included in the modelling: in particular, mass loss rates $\gtrsim10^{-8}M_\odot/$yr are predicted even for modest background field strengths ($\gtrsim30\,G_0$) in the case of discs that extend to $R > 150$\,AU. Grain growth significantly affects the final mass loss rates by reducing the average cross section at FUV wavelengths, and thus allowing a much more vigorous flow. The radial profiles of observable quantities (in particular surface density, temperature and velocity patterns) indicate that these winds have characteristic features that are now potentially observable with ALMA. In particular, such discs should have extended gaseous emission that is dust depleted in the outer regions, characterised by a non-Keplerian rotation curve, and with a radially increasing temperature gradient.

\end{abstract}



\section{Introduction}
\label{sec:introduction}

There is observational evidence that the environment of star forming regions can significantly affect the evolution of protostellar and protoplanetary discs. Until recent years the only evidence of such interplay has come from single objects showing clear signatures of an on-going interaction between their disc and the ambient environment \citep[e.g.][]{1993ApJ...410..696O,2000AJ....119.2919B,2005A&A...441..195V}. With the advent of sub-millimetre surveys, we can now start accessing statistical samples of relevant properties of protoplanetary discs (mostly their mass and outer radius) and infer whether they do show imprints of being affected by the environment. Such studies have shown that this is indeed the case in the most extreme star forming region within $500$\,pc from us, the Orion Nebula Cluster (ONC). In this cluster, \citet{2012A&A...546L...1D} have suggested that typical disc sizes decrease as a function of the stellar surface density of the environment in which they are embedded. Even more interestingly, \citet{2014ApJ...784...82M} \citep[see also][]{2010ApJ...725..430M} have shown that the dust component of discs tends to be less massive in the immediate vicinity of the dominant O star in the ONC, $\theta^1$C. \citet{2015ApJ...802...77M} have not observed such a trend in another O-stars bearing cluster, NGC 2024. For milder star forming regions, we still lack observational evidence of how important the environment can be in shaping the structure and the evolution of discs. Upcoming spatially resolved surveys of discs might be able to clarify this. Earlier studies have analysed whether disc frequency depends on distance from the most massive stars in OB stars-bearing clusters. Most of them agree that in the close proximity to the luminous O stars, disc fraction declines by a factor of $\sim2$. This trend has been observed for example in NGC 2244 \citep{2007ApJ...660.1532B}, in NGC 6611 \citep{2007A&A...462..245G,2009A&A...496..453G}, in NGC 6357 \citep{2012A&A...539A.119F} and in Cyg OB2 (Guarcello et al. in prep.). However, these results are still controversial: for example, \citet{2015ApJ...811...10R} have recently argued that such decreasing trend in disc fraction could be enhanced by sample incompleteness in some of these studies. In less populous clusters, in which the number of sources near O stars is smaller, a more robust result is that no evidence of spatial gradient is observed \citep[e.g.][]{2011ApJ...733..113R}.

There are two main environmental mechanisms that can affect protoplanetary discs: star - disc interactions, and photoevaporation. In dense clusters, during the lifetime of a disc \citep[$\sim3-10$\,Myr, e.g.][]{2001ApJ...553L.153H,2009AIPC.1158....3M,2010A&A...510A..72F}, gravitational encounters with other stars may perturb the disc, truncating it to a defined outer radius, and steepening the surface density profile \citep[see][and references therein]{1997MNRAS.287..148H,2014A&A...565A.130B,2014MNRAS.441.2094R,2015A&A...577A.115V}. Very few objects that might be undergoing such encounters have been detected \citep[][]{2014ApJ...792...68S,2006A&A...452..897C,2015MNRAS.449.1996D}, since the timescale of observability is very short.

An arguably more important mechanism is the photoevaporation caused by the energetic radiation permeating the young associations. When stars form in groups, EUV (Extreme Ultraviolet) and FUV (Far Ultraviolet) radiation of the most massive stars heats the outer regions of protoplanetary discs, and can drive a gaseous flow from the disc surface \citep[][hereafter \citetalias{adams_04}]{hollenbach_94,adams_04}.  Such a scenario has been very successful in explaining the so-called `proplyds' \citep{johnstone_98,1999ApJ...515..669S,2000ApJ...539..258R}, i.e. dark silhouette discs or cocoon-like structures observed in star forming regions. Such objects have been observed for example in the ONC \citep{1993ApJ...410..696O,2000AJ....119.2919B}, in Cyg OB2 \citep{2012ApJ...746L..21W}, in Carina \citep{2003ApJ...587L.105S}, and in other star forming regions showing massive OB associations (i.e. luminous sources of high energy radiation). When the UV flux impinging onto the discs is less severe than in these star forming regions, the effect of such mass loss on the evolution of protoplanetary discs, and the impact on their planet formation potential, is however rarely considered.

To be more quantitative, external photoevaporation can be summarised in three different regimes. When the EUV component is dominant, due to the proximity of a massive O star (e.g. $\theta^1$C in the ONC), the EUV flux reaches the surface of the disc, heating it up to $\sim10^4$ K. A completely ionised flow is formed, driving gas outwards and photoevaporating the disc \citep[][]{hollenbach_94}. When the EUV flux is less severe, the FUV field can generate a neutral wind that is optically thick to the EUV radiation, since the FUV component has a larger penetrating depth than the EUV one. When this is the case, two different regimes may occur, depending on the disc size $R_{\rm d}$. We define the gravitational radius $R_{\rm g}$ of the disc as the radius at which the thermal energy is equal to the gravitational binding energy (equivalently, as the radius at which the sound speed is equal to the escape velocity):

\begin{equation}
R_{\rm g} = \frac{GM_*\mu m_{\rm H}}{k_{\rm B}T}\approx 140\,{\rm AU}\left(\frac{T}{1000\,{\rm K}}\right)^{-1} \left( \frac{M_*}{M_\odot} \right),
\end{equation}
where $T$ is the temperature, $M_*$ is the mass of the central star, and $\mu$ the mean molecular weight (in this case we used atomic gas with $\mu=1.3$). When $R_{\rm d}>R_{\rm g}$, the wind is launched supersonically \citep[e.g.][]{johnstone_98}. Such discs are defined to be supercritical. However, moderate FUV fields are not able to heat up the outer regions of discs to high enough temperatures such that $R_{\rm g}<R_{\rm d}$. \citetalias{adams_04} have shown that when this is the case, the flow structure can be described by a non isothermal Parker wind, with the addition of a centrifugal term. This model uses the reasonable assumption that the mass loss rate from the outer rim of the disc dominates the mass loss rate from the surface of the disc \citepalias[see the Appendix in][]{adams_04}, where the gas is more embedded in the gravitational potential well of the central star. Since the outer regions of discs contain the bulk of the mass and since the rate of disc evolution is determined by the outer disc radius, such external winds have important implications for disc lifetimes and surface density profiles. Indeed, as noted by \citet{2007MNRAS.376.1350C} and confirmed by \citet{2013ApJ...774....9A}, such winds accelerate disc clearing not on account of the mass lost in the wind (which is a fraction of that accreted onto the star over the disc lifetime) but instead because they modify the disc's outer boundary, preventing disc spreading and keeping the viscous timescale relatively short.

\citet{2003ARA&A..41...57L}, \citet{2003AJ....126.1916P} and others have shown that the probability density function for cluster membership number $N$ for clusters in the Solar Neighbourhood scales with $1/N^2$. Therefore the cumulative probability for a star to be born in a cluster of size $\leq N$ is inversely proportional to $N$ \citep[e.g.][]{2010ARA&A..48...47A}. Within 2\,kpc of the Sun, the median value of this distribution (also taking into account isolated stars) is $\sim300$ \citep{2006ApJ...641..504A,2007prpl.conf..361A}. The majority of protoplanetary discs are then very likely to evolve embedded in relatively `mild' (low $N$) environments. Clusters and groups with $N\leq500$ have typical FUV fields $G_{\rm FUV}\leq3000\,G_0$ \citep[e.g.][]{2008ApJ...675.1361F}, where $G_0$ is the local FUV interstellar field \citep[$1.6\times10^{-3}$\,erg\,s$^{-1}$\,cm$^{-2}$,][]{1968BAN....19..421H}. \citet{2008ApJ...675.1361F} have also shown that the FUV flux depends strongly on the most massive star in the cluster, and on the intra-cluster column density of absorbing material. Stars in small groups $(N\sim50-100)$ are likely to be impinged by an environmental FUV field of $\sim30-300\,G_0$, i.e. slightly larger than the local field. The majority of protoplanetary discs are then very likely to lie in the subcritical regime of external photoevaporation, at least in the Solar Neighbourhood.

Note that at such low values the external radiation becomes comparable to the radiation from the central star even at large radii \citep[e.g.][]{2003ApJ...591L.159B}. In particular, \citet{2014ApJ...784..127F} looked at a sample of CTTSs and estimated the FUV field close to the central object by deriving the FUV continuum and hot gas lines profiles from high spectral resolution HST observations at FUV wavelengths. They obtained FUV fields of $\sim10^7\,G_0$ at $\sim1\,$AU, which are due to both stellar and accretion contributions. By considering geometrical dilution only, FUV fields of $300$ and $30\,G_0$ are thus expected at $180$ and $580$\,AU, respectively. However, the bulk of the disc is shielded by its inner regions, especially for low flaring angles, and most of the radiation is absorbed $\sim2$ scale-heights above the midplane, as shown in thermo-chemical models \citep[for example,][compute an FUV intensity of $\sim100\,G_0$ at $\sim100\,$AU and $z/R\sim0.2$ for HD 100546]{2012A&A...541A..91B}. Besides, the midplane of the outer disc can be directly impinged by the ambient radiation, and gas can be removed more easily. Thus in the outer regions of discs external photoevaporation could play a dominant (or at least comparable) role with respect to the photoevaporation driven by the FUV radiation of the central star \citep{2009ApJ...690.1539G,2015ApJ...804...29G}. Evidently this is an issue that will only be finally settled
using 2D radiation hydrodynamical simulations including both internal
and external FUV sources.

The subcritical regime of external photoevaporation is the most difficult to probe observationally. However, there are some observed signatures that might be indicating that such a mechanism is occurring even in quite isolated systems (i.e. very low external UV fluxes). A potential example of this is the disc around IM Lup. \citet{2009A&A...501..269P} have shown that outside the mm-bright disc \citep[$R_{\rm out,dust}=400$\,AU;][]{2008A&A...489..633P}, the gas structure as probed by CO emission lines indicates a steep decrease in the surface density profile, which resembles the profiles obtained by \citetalias{adams_04} in their flow models. \citet{2015ApJ...810..112O} have also detected a DCO$^+$ double ringed structure, which could be tracing a radially increasing temperature gradient in the outer regions of the disc in agreement with the predictions from external photoevaporative models (though the authors interpret the data in terms of non-thermal desorption of CO ices). Note that similar features are also predicted by photoevaporation models where the winds are driven by the FUV field emitted by the central star and by the accreting material \citep{2009ApJ...690.1539G,2015ApJ...804...29G}. From simple calculations of external photoevaporation models the gas flow in the outer regions of this disc would be highly dust depleted, since the drag force is very weak for the large grains probed by submm observations, and such an effect is compatible with the dust depleted outer regions of the disc. There are other systems where this second effect (a radially decreasing dust-to-gas ratio, as probed by line/continuum size discrepancies) is observed \citep[e.g.][]{2007A&A...467..163P,2012ApJ...744..162A,2013ApJ...774...16R,2013A&A...557A.133D}. Note however that \citet{birnstiel_14} have shown that such discrepancies can also be explained by the size-dependent radial drift of dust particles, without appealing to an outer disc wind. We will suggest alternative observational diagnostics that can discriminate between the two scenarios.

In this paper, we firstly aim to compute the mass loss rates of discs affected by external photoevaporation in the subcritical regime, by extending the parameter space investigated by \citetalias{adams_04} to larger (but still subcritical) discs, and milder ambient fields. Our method of solution broadly
follows that of \citetalias{adams_04} but  with some important differences
relating to the effect of non-isothermality in
correctly locating the critical point of the flow.
We also iterate towards a self-consistent solution which takes
into account the fact that only small grains are entrained
in the flow. Note that the effect of partial entrainment
of dust grains upon the thermal structure of the flow has
been recently applied to the internal photoevaporation scenario by \citet{2015ApJ...804...29G}. We finally present typical radial profiles of the main hydrodynamical quantities, and propose potential observational signatures of ongoing photoevaporation in this subcritical regime.

\begin{figure}
\center
\includegraphics[width=\columnwidth]{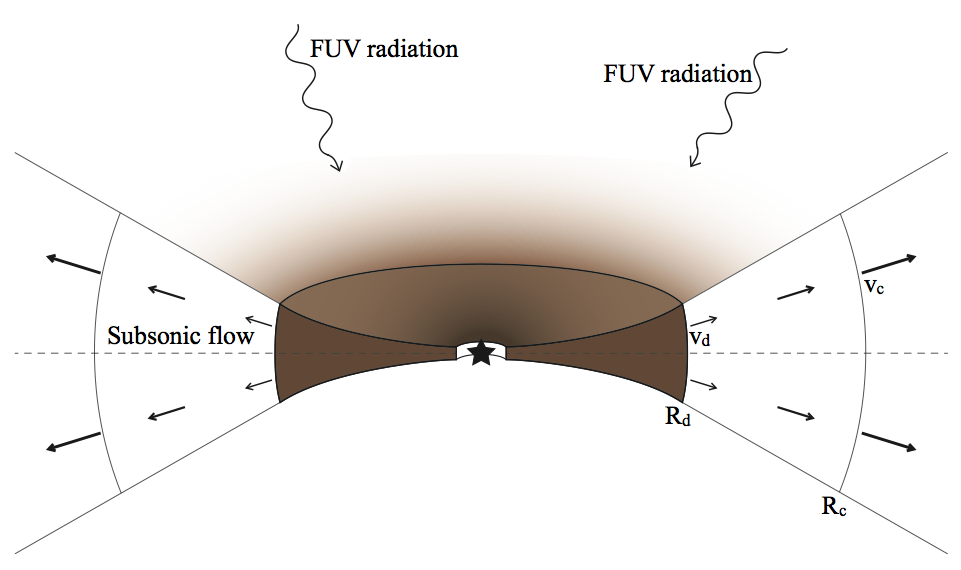}
\caption{Schematic illustrating a photoevaporating disc impinged by an ambient FUV radiation, in the subcritical regime ($R_{\rm d}<R_{\rm g}$). The radiation can be directional or more isotropic, depending on the source of irradiation. From the outer edge of the disc, a subsonic radial flow develops, until it reaches the critical radius $R_{\rm c}$, where the flow velocity is close to the sound speed. From this point outwards the velocity of the wind is approximately uniform. In this subcritical regime, the mass loss from the disc's outer rim is more significant than the mass loss from the disc surface.}
\label{fig:cartoon}
\end{figure}

The paper is structured as follows. In Section \ref{sec:adams} we briefly summarise the photoevaporative model by \citetalias{adams_04}, and explain how our description differs from theirs. In Sections \ref{sec:pdr} - \ref{sec:phys_sol} we detail the main ingredients of our model; respectively the thermal properties, gas hydrodynamics, dust hydrodynamics, and the iteration procedure to obtain the final solutions. In Section \ref{sec:results} we present our results, which are then discussed in Section \ref{sec:discussion}. In Section \ref{sec:concl} we summarise our conclusions.

\section{Comparison with previous work}
\label{sec:adams}

We consider the subcritical regime ($R_{\rm d} < R_{\rm g}$) previously investigated
by \citetalias{adams_04} and assume that the disc is irradiated by an isotropic FUV ambient radiation. \citetalias{adams_04} have shown that the ratio between the mass loss rate from the disc surface $\dot{M}_{\rm sur}$ and the mass loss from the disc edge $\dot{M}$ scales as $\dot{M}_{\rm sur}/\dot{M}\sim (R_{\rm d}/R_{\rm g})^{1/2}$. Since they focus on the subcritical regime, where $R_{\rm d}<R_{\rm g}$, the mass loss rate is dominated by the radial flow emerging from the outer rim of the disc, and not by the material flowing from its surface. Thus the subsonic wind can be described with a 1D radial model (see Fig. \ref{fig:cartoon} for a schematic of the model). \citetalias{adams_04} computed the temperature as a function of local gas density $n$ and FUV (from the ambient radiation) optical depth $\tau$ by using the photo-dissociation region (PDR) code by \citet{1999ApJ...527..795K}. They then coupled this temperature dependence with the steady
state momentum/continuity equations in order to iteratively find a self-consistent steady state solution of the gaseous flow. The model we present in this paper is very similar to the one proposed by \citetalias{adams_04}, but contains some key differences, that will be described in more detail between Sections \ref{sec:pdr} - \ref{sec:dust}. In particular:

\begin{enumerate}
\item We take deviations from isothermality into account in locating
the critical point of the flow, in contrast to \citetalias{adams_04} who
impose the condition that the flow is transonic at the location
corresponding to the sonic point for {\it isothermal} gas. This
self-consistent location of the critical point results in
a different
location and local flow velocity at this point compared with the
isothermal solution. Our results demonstrate that this difference is the most important one, since it modifies mass loss rates significantly. Having located the critical point of the flow
we are able to integrate inwards to the disc edge (in contrast
to \citetalias{adams_04} who adopt an iterative scheme in integrating outwards from
the disc edge).
\item The self-consistent location of the critical point allows us to
obtain solutions over a larger range of parameter space than
\citetalias{adams_04}. Specifically we are able to find solutions for lower values
of the interstellar field $G_{\rm FUV}$ (i.e. down to $30\,G_0$) and for a wider range of outer disc
radii (out to $R_{\rm d} = 250$\,AU).
\item Having determined a flow solution at fixed dust to gas ratio
we then take account of the fact that only the small grains are
entrained in the flow and re-compute the flow solution implementing the reduced dust to gas ratio in the flow.
\item Our code computing the temperature structure is different from \citetalias{adams_04} (see Section \ref{sec:pdr} below).
\end{enumerate}

\begin{figure*}
\begin{center}
\includegraphics[width=1.02\columnwidth]{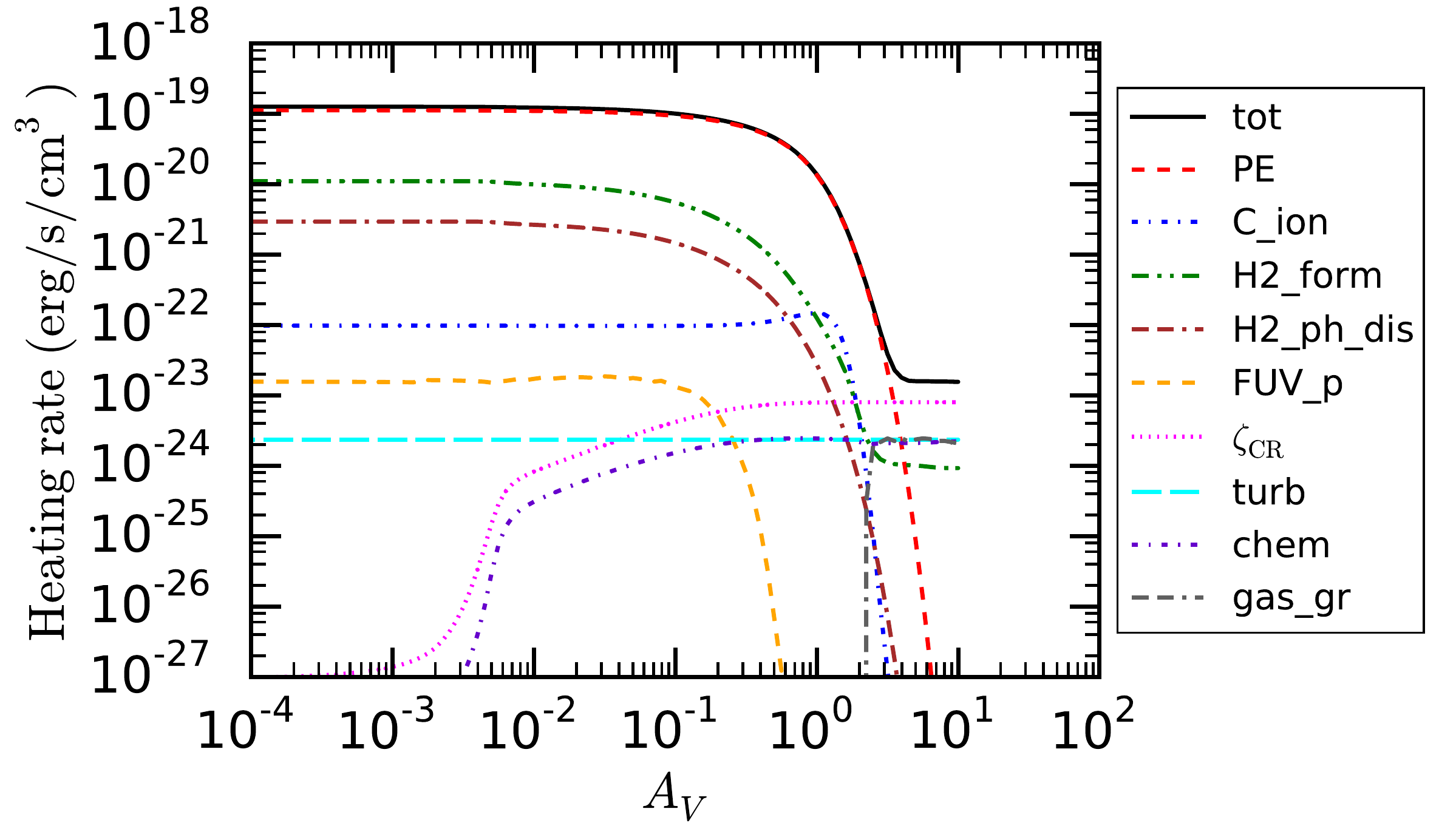}
\includegraphics[width=0.975\columnwidth]{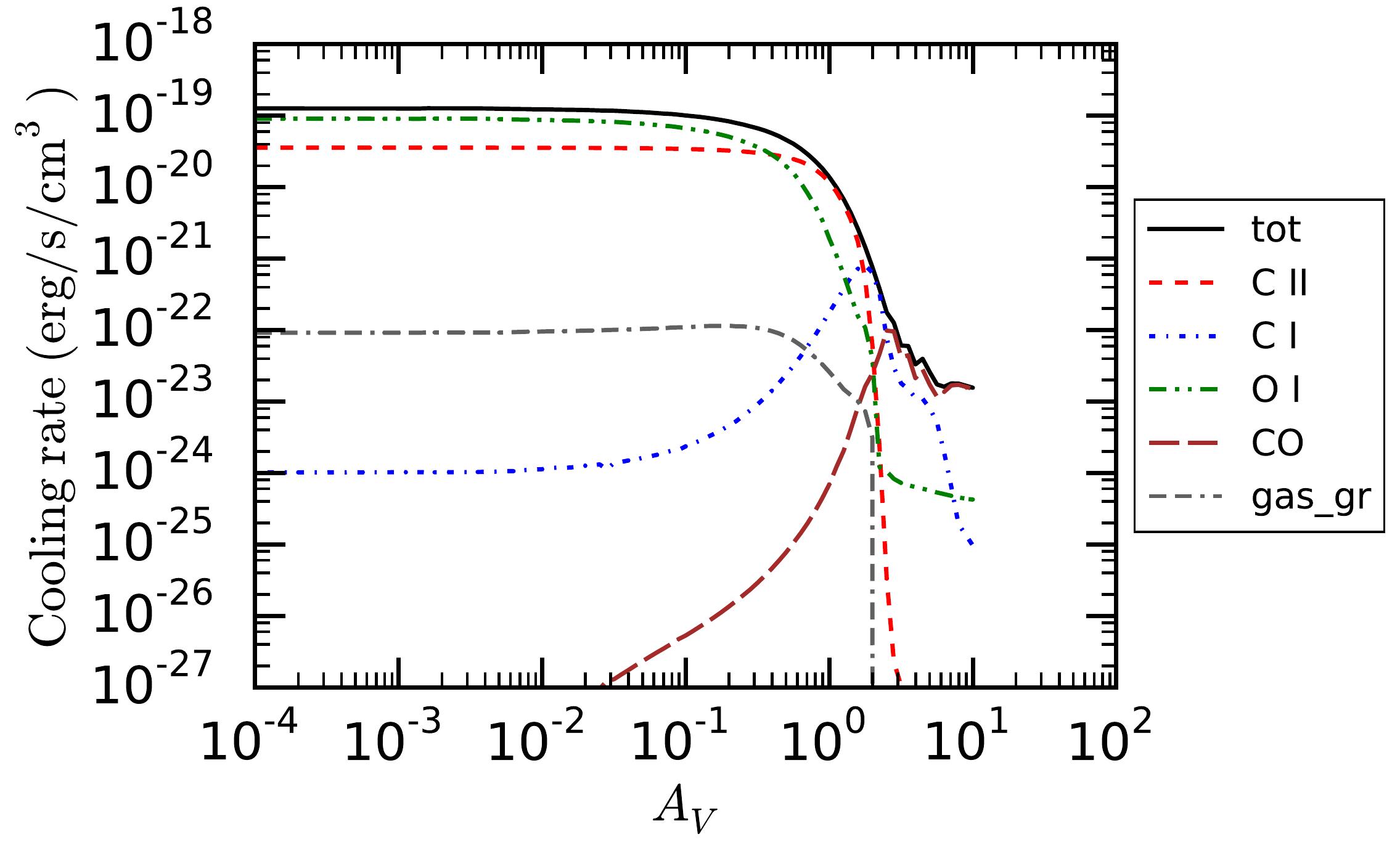}

\end{center}
\caption{Heating and cooling rates per unit volume for a slab of material of density $n=10^4$\,cm$^{-3}$ impinged by a FUV field of $G_{\rm FUV}=300\,G_0$, after it has reached thermo-chemical equilibrium computed with the reduced chemical network. The dust-to-gas ratio is $10^{-2}$ and the PAH-to-dust ratio is $2.6\times10^{-2}$. The heating functions (labelled in the legend) are respectively: total heating rate (tot); photoelectric heating (PE); C ionisation (C\_ion); H$_2$ formation (H2\_form); H$_2$ photodissociation (H2\_ph\_dis); FUV pumping (FUV\_p); cosmic rays ($\zeta_{\rm CR}$); turbulent heating (turb); chemical heating (chem); gas-grain collisions (gas\_gr). More details can be found in \citet{bisbas_12}.
}
\label{fig:heat_cool}
\end{figure*}

\section{Temperature estimates - PDR}
\label{sec:pdr}

FUV radiation ($6\le{\rm h}\nu<13.6\,{\rm eV}$) plays a crucial role in determining the thermal structure and corresponding chemistry in so-called photodissociation regions (PDRs) where gas
undergoes a transition between ionised and  molecular state. If a disc is interacting with an ionising radiation field impinging externally, then between the ionisation front (if present) and its dense gas, the region is optically thin to FUV radiation but optically thick to  EUV radiation (${\rm h}\nu\ge13.6\,{\rm eV}$). This causes the gas to be in a state of partial
dissociation where the temperature is set by the FUV radiation field.
Under these conditions, determination of the gas temperature
requires modelling of various heating and cooling processes
in terms of a complex and non-linear set of differential
equations describing a network of chemical reactions. In the last two decades, many groups have managed to develop numerical codes that are able to solve such problems \citep[see][for a paper detailing inter-comparison between such codes]{rollig_07}. Below we describe the 3D-PDR code \citep{bisbas_12} that we use
to obtain the gas temperature as a function of column density, grain
opacity, local gas density and FUV field.

\subsection{3D-PDR}

The {\sc 3d-pdr} code \citep{bisbas_12} is a three-dimensional time-dependent code treating photodissociation regions of arbitrary density distribution. Using an iteration scheme it solves the chemistry at every cell consisting of the gaseous structure until thermal and chemical equilibrium has been reached. {\sc 3d-pdr} uses a set of various heating and cooling functions fully described in \citet{bisbas_12}, and shown in Fig. \ref{fig:heat_cool} for a specific case. The code has been used in various one-dimensional \citep{2014MNRAS.443..111B,2015ApJ...803...37B} and three-dimensional \citep{2013ApJ...770...49O,2014MNRAS.440L..81O,2015ApJ...799..235G} applications. The 1D version of the code \textsc{ucl-pdr} has been benchmarked with many other PDR codes \citep{rollig_07}.

In this paper, we have adopted the code modifications described in \citet{2014MNRAS.443..111B}. We use a reduced version of the UMIST 2012 network \citep{2013A&A...550A..36M} which contains 33 species (including e$^-$) and 330 reactions. Table \ref{tab:species} shows the initial abundances used by the {\sc 3d-pdr} code at the beginning of the calculations \citep[taken from][]{2009ARA&A..47..481A}. Using the full chemical network of 215 species causes difference in the temperature by up to $\sim10-15$\%. Since other unknown parameters (such as polyciclic aromatic hydrocarbon abundances, see below) lead to even larger uncertainties, we opted for the reduced network to save computational time. We consider one-dimensional uniform density profiles with $10^2<n<10^8\,{\rm cm}^{-3}$ interacting with radiation fields of $30$, $300$ and $3000\,G_0$. The spatial extent of each density profile is chosen so that the visual extinction, $A_{V}$, is in the range $10^{-7}\leq A_{V}\leq10$. Density is sampled every $0.1$ dex, and $A_V$ every $0.05$ dex. The results are then interpolated with cubic-spline functions to a much finer grid. The cosmic ray ionisation rate, $\zeta_{\rm CR}$, is taken to be $\zeta_{\rm CR}=5\times10^{-17}\,{\rm s}^{-1}$. The treatment of PAHs is discussed in the next Section. We evolve the chemistry in each simulation for $10$\,Myr. We have checked that the temperature balance is reached on shorter timescales, $\sim10$\,kyr. {\it A posteriori}, we have verified that the flow timescale is always longer than $10$\,kyr, thus temperature balance within the flow is a reasonable assumption.

\begin{table}
\caption{Abundances of species used in the present work, from \citet{2009ARA&A..47..481A}.}
\centering
\label{tab:species}
\begin{tabular}{l l l l}
\hline
H     & $4\times10^{-1}$   & Mg$^+$ & $3.981\times10^{-5}$ \\
H$_2$ & $3\times10^{-1}$   & C$^+$  & $2.69\times10^{-4}$  \\
He    & $8.5\times10^{-2}$ & O      & $4.898\times10^{-4}$ \\
\hline
\end{tabular}
\end{table}

\subsection{Dust and PAHs}
\label{subsection:pdr_dust}

Dust plays a key role in the determination of the gas temperature. For example, for the density range addressed here, the main heating mechanism of the gas is usually photoelectric heating from the atomic layers of PAHs \citep[e.g.][]{2001ApJS..134..263W,2012ApJ...747...81C}, but it can also affect the temperature through other mechanisms such as H$_2$ formation and gas-grain collisions. Secondly, dust can have a significant impact on the chemistry, which eventually sets the abundances of the main coolants, e.g. by controlling the amount of ices. Note that all these mechanisms depend on the total surface area of dust grains. Finally, dust sets the attenuation factor of the FUV radiation via the penetrating depth $\tau$, where we set $\tau=1.8A_V=N_{\rm H}\sigma_{\rm FUV}$ and $N_{\rm H}$ is the hydrogen column density. A detailed description of this last effect will be described in Section \ref{sec:dust}.

\begin{figure*}
\begin{center}
\includegraphics[width=\columnwidth]{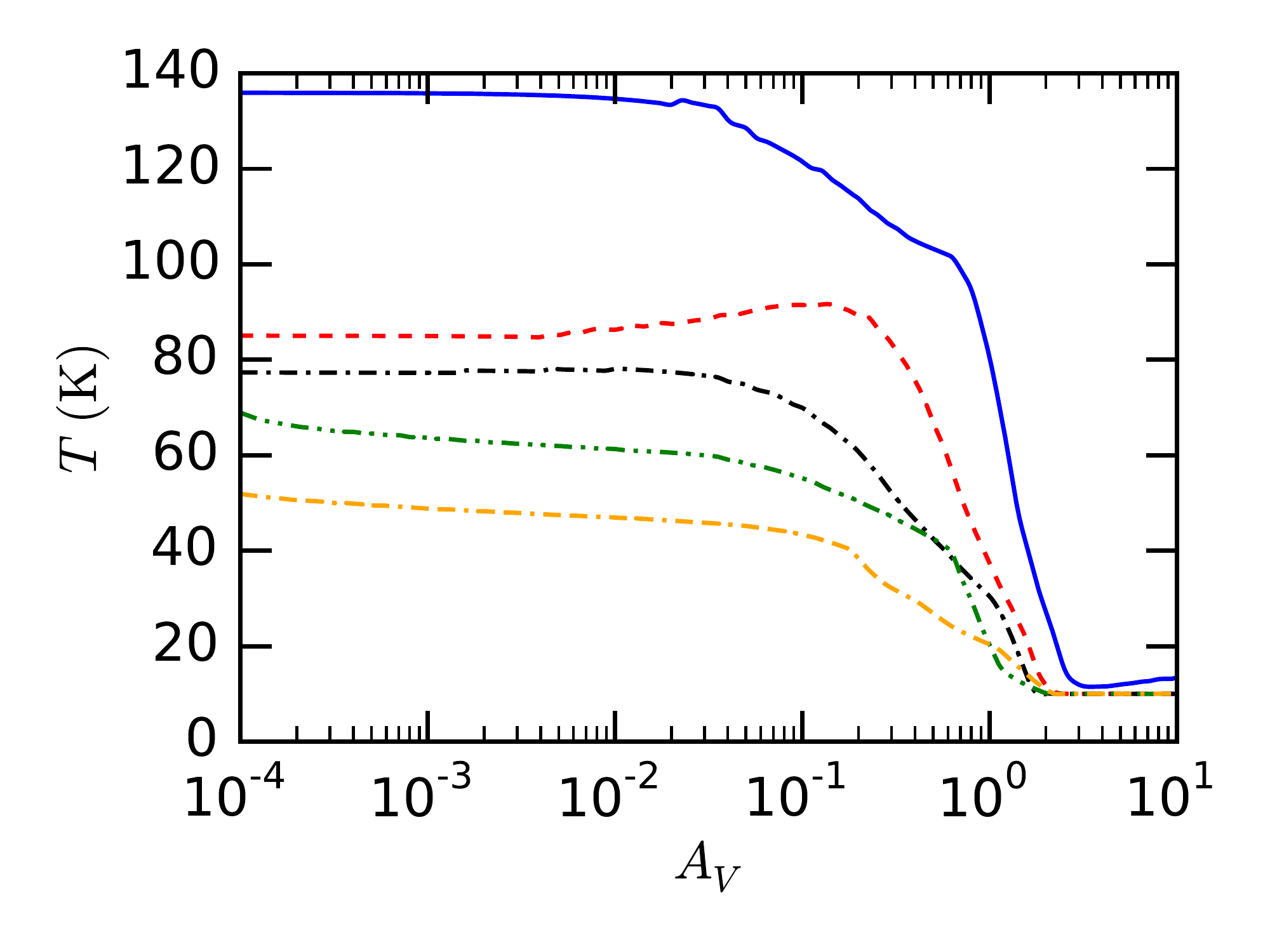}
\includegraphics[width=\columnwidth]{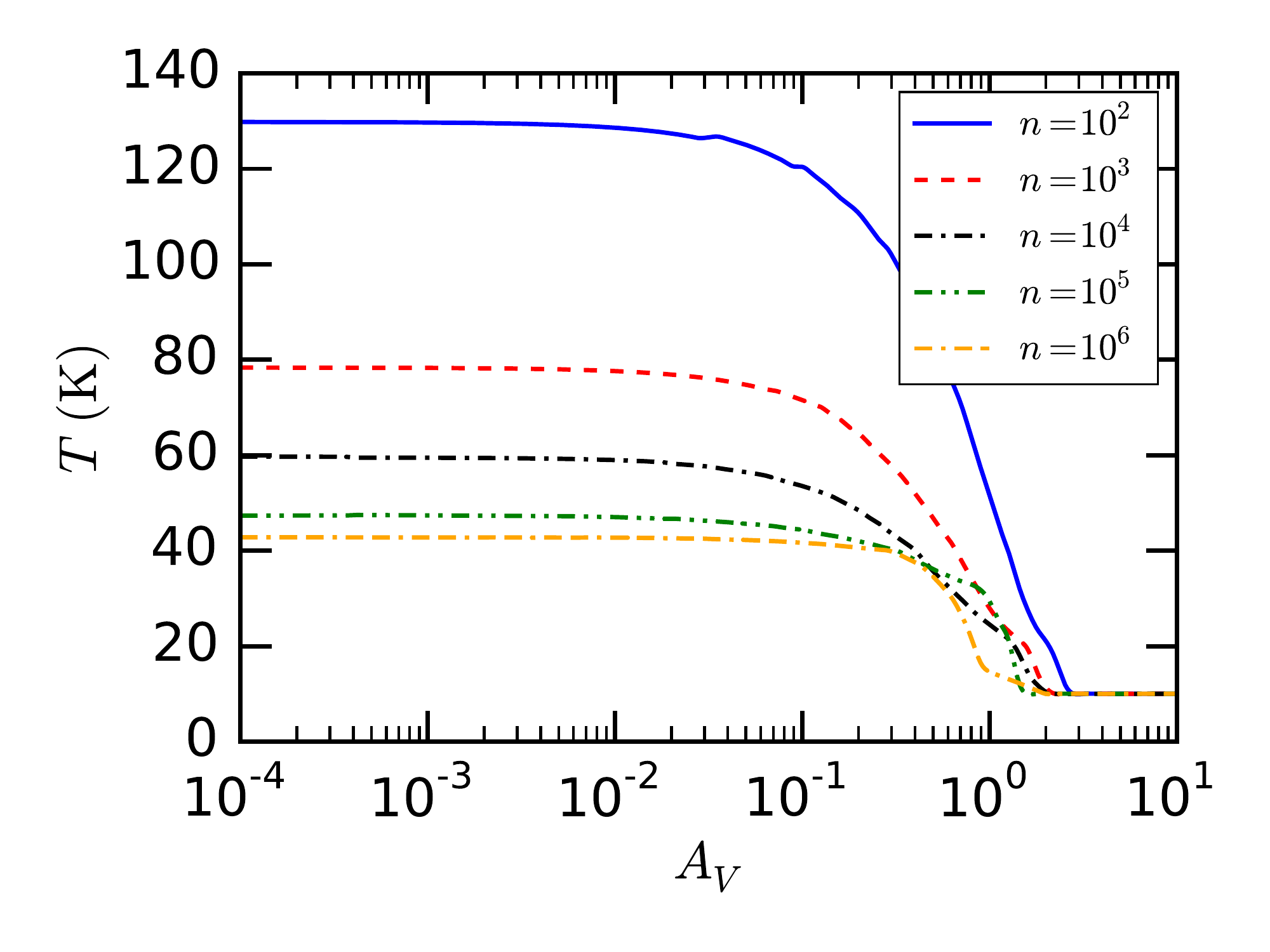}\\
\includegraphics[width=\columnwidth]{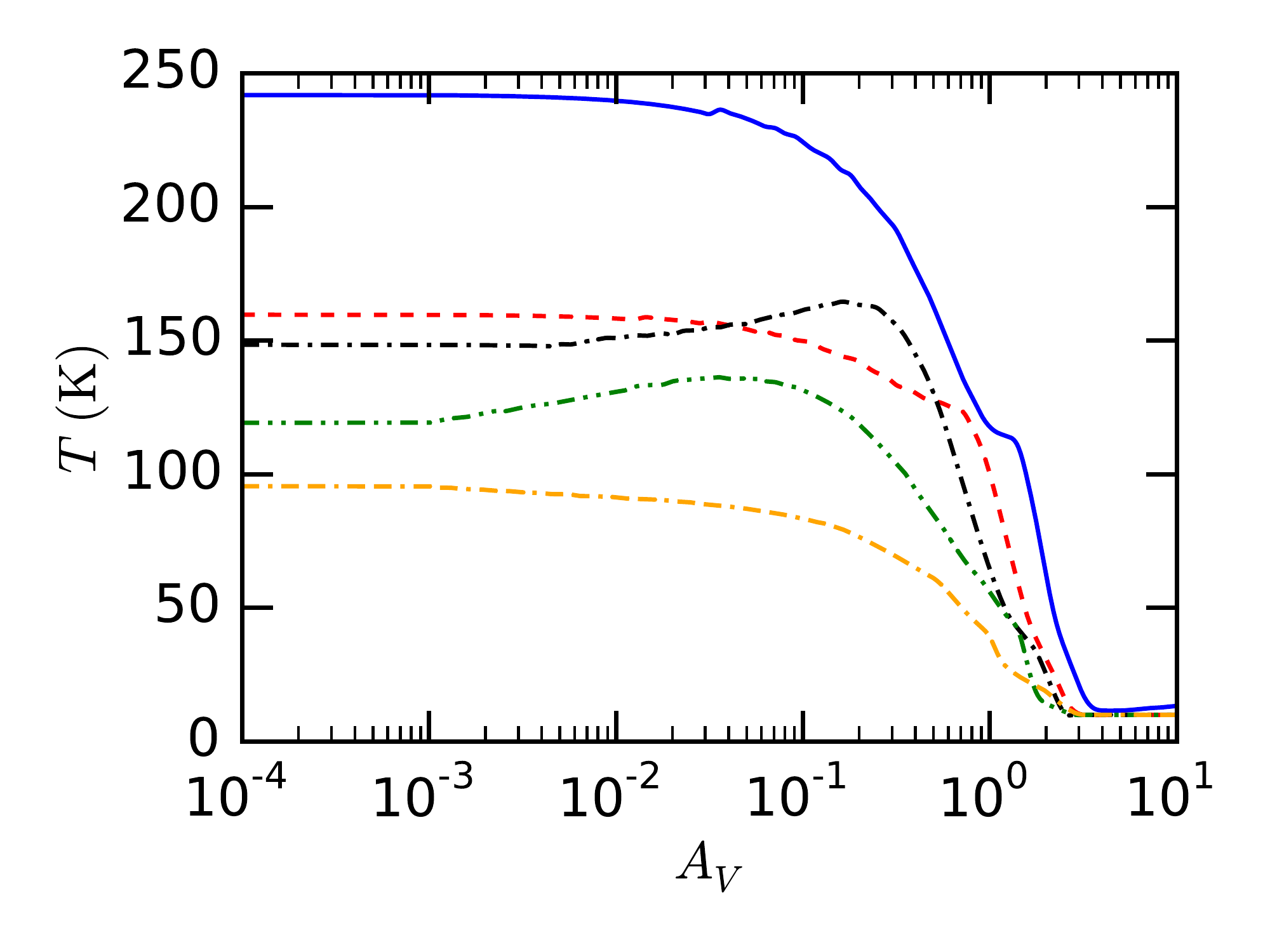}
\includegraphics[width=\columnwidth]{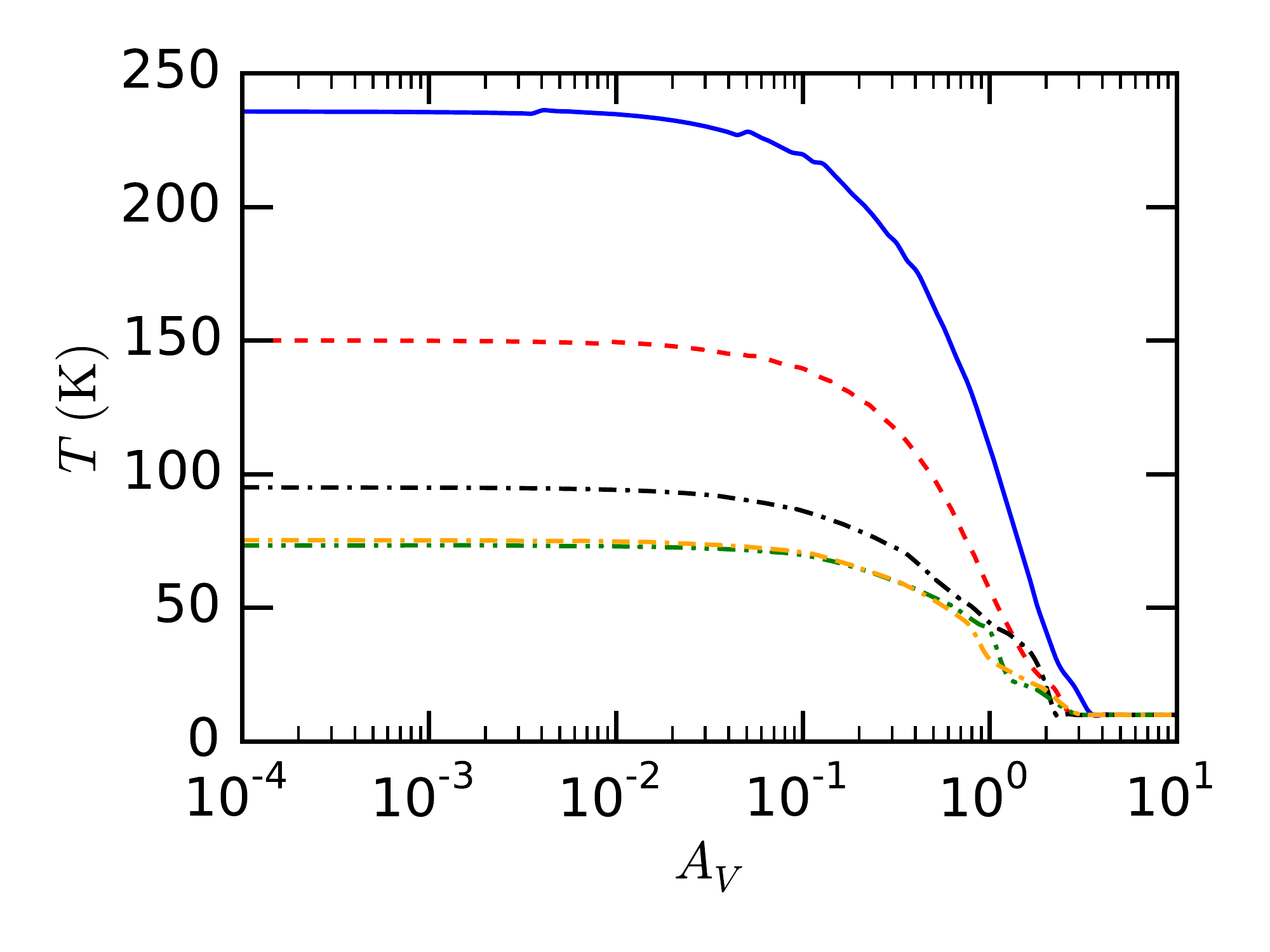}\\
\includegraphics[width=\columnwidth]{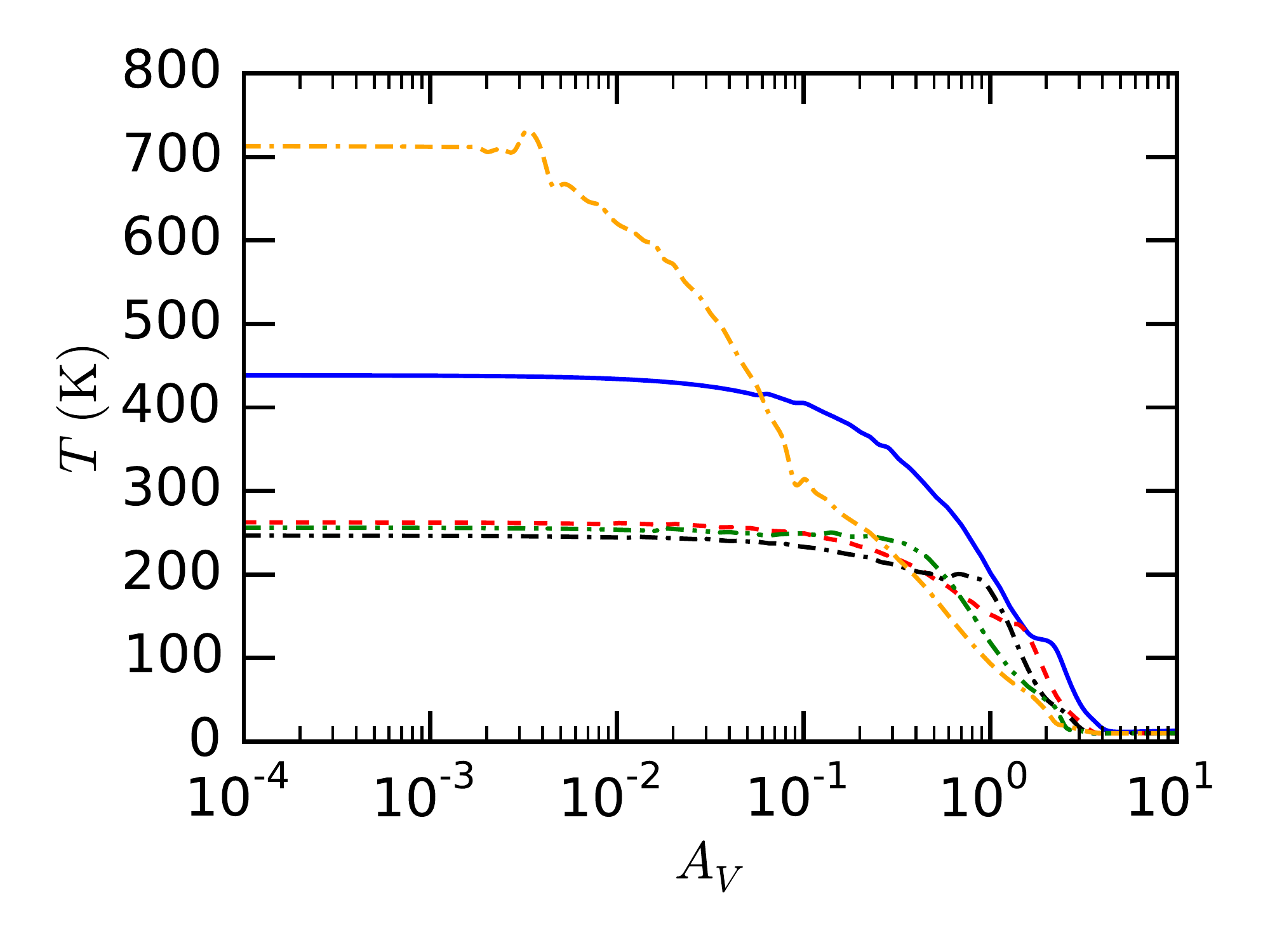}
\includegraphics[width=\columnwidth]{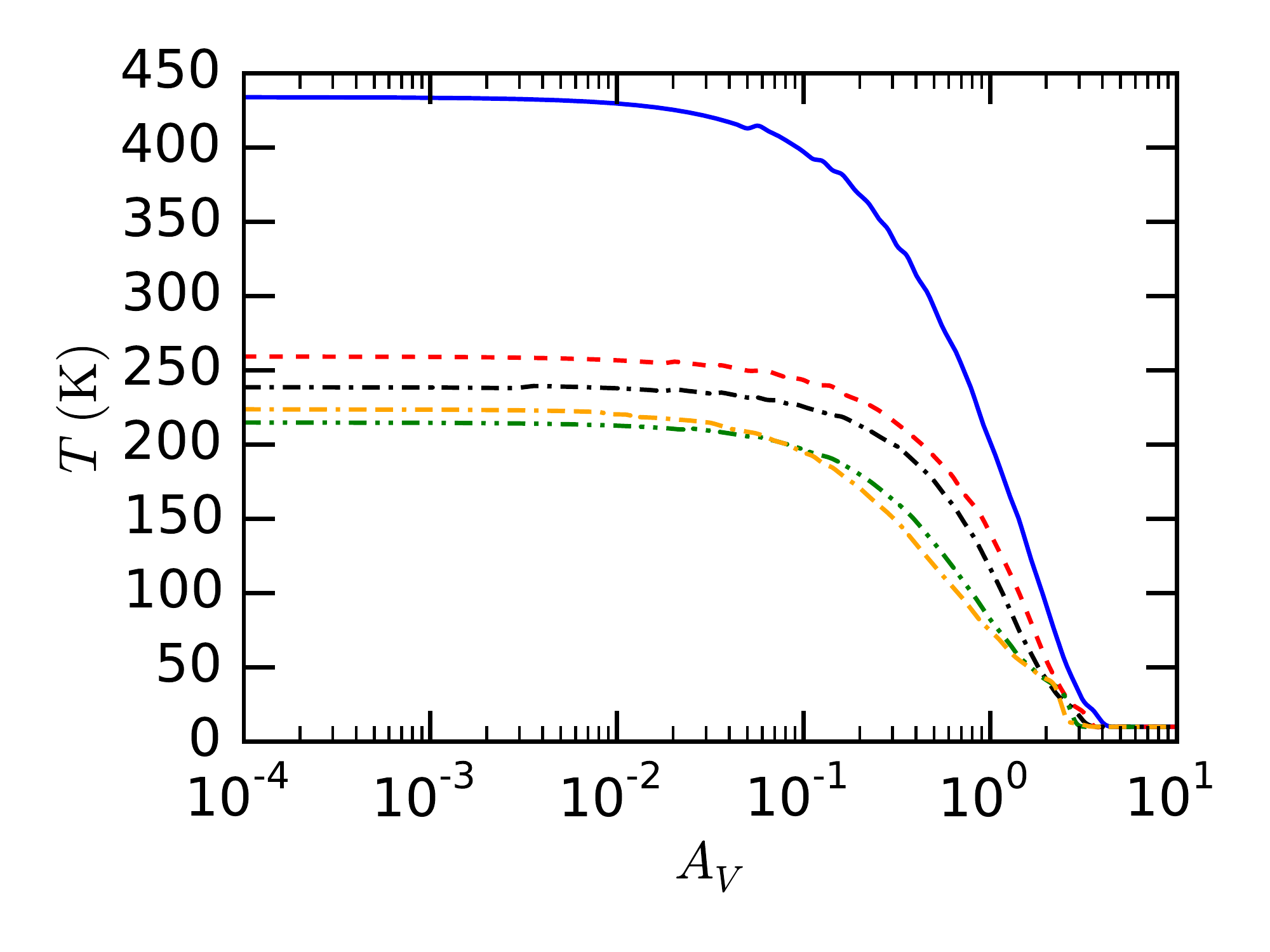}
\end{center}
\caption{Gas temperature as a function of visual extinction $A_V$ for three different values of the ambient FUV field: $30$, $300$ and $3000\,G_0$, from top to bottom. The left panels are associated to a grain size distribution with maximum grain size $s_{\rm max}=3.5\,\mu$m, the right panels to a distribution with $s_{\rm max}=1\,$mm. Line colours indicate different gas numerical densities $n$, ranging between $10^2-10^6$\,cm$^{-3}$ (the legend is reported in the right top panel).}
\label{fig:pdr}
\end{figure*}

Typical interstellar dust can be modelled with a simple distribution $d\tilde{n}/ds\propto s^{-q}$, where $\tilde{n}$ is the numerical density of dust particles, and $s$ is the grain size. It is well known that in the interstellar medium $q$ assumes a typical value $q\sim3.5$ \citep[historically labelled as MRN distribution, from][]{1977ApJ...217..425M}. More recently, interferometric observations have suggested a shallower distribution in the optically thin regions of protoplanetary discs at submm - mm wavelengths, where $q \sim 2.5 - 3$ \citep[e.g.][]{2010A&A...521A..66R,2010A&A...512A..15R,2012A&A...540A...6R}, and dust evolution models have suggested the same qualitative result \citep[e.g.][]{2011A&A...525A..11B}. Moreover, the maximum grain size does increase substantially in the midplane of protoplanetary discs \citep[see the recent review by][and references therein]{2014prpl.conf..339T}, to sizes $\gtrsim1$\,mm. Grain growth can alter the heating and cooling processes mentioned above.

The only observation indirectly constraining the grain size distribution within the photoevaporative winds is by \citet{1999ApJ...515..669S}, who estimated the cross section in supercritical winds at FUV wavelengths from the location of the ionisation front around proplyds in Orion. Their best estimate is $\sigma_{\rm FUV}=8\times10^{-22}$\,cm$^{2}$, where the error bars are assumed to be very large, since the used sample is very small ($\sim10$ objects) and the estimate is model dependent. \citet{1999ApJ...515..669S} and later \citetalias{adams_04} noticed that such cross section is $\sim0.3$ the typical ISM one, thus indicating moderate grain growth of the dust entrained in the wind.

We compute the cross section at FUV wavelengths ($\lambda=0.1\,\mu$m) for an ISM-like dust distribution \citep[$q=3.5$ and $s_{\rm max}=0.25\,\mu$m, see e.g.][]{2011piim.book.....D} using the code described in \citet{2002MNRAS.334..589W} (see their section 4.1). We use the optical constants from \citet{1997A&A...323..566L}, for iceless silicate grains with a porosity of $0.3$. The absorption coefficients of the grains $Q_{\rm abs}(\lambda,s)$ are computed using Mie theory \citep{1983asls.book.....B}, Rayleigh-Gans theory or geometric optics in the appropriate limits \citep[][]{1993ApJ...402..441L}. Assuming a dust-to-gas ratio to $10^{-2}$ we obtain a cross section $\sigma_{\rm FUV}\approx2.6\times10^{-21}$\,cm$^2$, as expected. We then determine the maximum grain size of a distribution with $q=3.5$ that has an associated cross section $0.3$ times smaller than the ISM-like one. The best fit gives $s_{\rm max}=3.5\,\mu$m. This result confirms that the estimates by \citet{1999ApJ...515..669S} are probing moderate grain growth.

In the PDR code (and throughout the whole paper), we consider two MRN distributions ($q=3.5$) with maximum grain sizes $s_{\rm max}=3.5\,\mu$m and $s_{\rm max}=1$\,mm. We fix the dust-to-gas ratio to $10^{-2}$. The first distribution is chosen as to give the same cross section used by \citetalias{adams_04} in their models, in order to have a direct comparison with their results. The second distribution represents a disc where substantial grain growth has occurred, and the population of small grains, which contribute the most to the dust surface area when $q>3$, is reduced by a factor $\sim10^2$.

Polycyclic Aromatic Hydrocarbons (PAHs) are the dominant heating source, since photons are very efficient in removing electrons from such mono-layered molecules. However, the amount of PAHs in protoplanetary discs is not well constrained by observations. Infra-Red observations from Spitzer and from the ground have revealed strong PAH emission features in some Herbig Ae/Be stars, whereas very few T Tauri stars have shown them at a detectable level \citep[][and references therein]{2006A&A...459..545G}. This difference is probably due to the more intense UV field emitted by the Ae/Be stars \citep[e.g.][]{2007A&A...466..229V} exciting the PAHs. For the same reason, PAH emission is spatially concentrated in the inner region of the discs, where the UV flux from the central star is higher \citep{2007A&A...476..279G,2014A&A...563A..78M}. PAH abundance in discs is still being debated. Observations \citep[e.g.][]{2006A&A...459..545G,2010ApJ...714..778O} tend to indicate that PAH-to-dust mass ratio in discs is $\sim10\%$ of the PAH-to-dust mass ratio in the galactic ISM, which is $\approx4\times10^{-2}$ \citep{2007ApJ...657..810D,2008ARA&A..46..289T}, but the estimate is poorly constrained. In thermo-chemical disc models, different values of PAH-to-dust mass ratio have been used in the literature, e.g. $0.5\times10^{-2}$ \citep{2012A&A...541A..91B}, or $1.3\times10^{-2}$ \citep{2015arXiv151103431W}. Because of the level of the uncertainties in this value, we simply fix the PAH-to-gas mass ratio to $2.6\times10^{-4}$, as in \citet{2003ApJ...587..278W}, since we use their prescription to compute the photoelectric heating.  For the dust-to-gas ratio of $10^{-2}$ used in this paper, this value yields a PAH-to-dust mass ratio of $2.6\times10^{-2}$.

Another debated topic is whether grain growth affects the amount of PAHs, i.e. whether they should scale according to the amount of small grains. This is still largely unconstrained. As an example, one of the strongest PAH emission features is observed in Oph IRS 48 \citep{2007A&A...476..279G}, where grain growth has certainly occurred \citep[e.g.][]{2013Sci...340.1199V}. Even though some recent models \citep[e.g.][]{2015ApJ...804...29G} assume a reduced PAH abundance when small grains are depleted by effective grain growth, in this paper we prefer to keep the PAH abundance fixed, since there is no clear observational indication that PAHs would follow the small grains.

In order to compute the photoelectric heating due to PAHs, we adopt the treatment by  \citet{1994ApJ...427..822B}, with the additional modifications suggested by \citet{2003ApJ...587..278W} (with the $\phi_{\rm PAH}$ factor defined in their equation 21 equal to $0.4$).

\subsection{Results}
\label{subsec:pdr_results}

Fig. \ref{fig:pdr} shows the results of the \textsc{3d-pdr} code for FUV fields of $30$, $300$ and $3000\,G_0$ for the two grain size distributions. We present the gas temperature as a function of visual extinction $A_V$, for logarithmically sampled values of gas density $n$. As expected, the gas temperature increases with intensity of the external radiation. At a given $G_{\rm FUV}$, for the parameters we have chosen, temperature is usually a decreasing function of visual extinction $A_V$. For $s_{\rm max}=3.5\,\mu$m temperature does also increase with $A_V$ in the marginally optically thick regime, when $n\sim10^{3}-10^4\,$cm$^{-3}$. Similarly, temperature is not a monotonically decreasing function of density $n$. This non-monotonic behaviour is a well known result in PDR codes \citep{1999ApJ...527..795K}.

These results depends on the metallicity and on the dust-to-gas ratio, since they both regulate the heating and cooling functions. In particular, reducing the amount of small grains has the net effect of slightly decreasing the temperatures, as we can see comparing the temperatures obtained with the two different dust distributions. The smaller dust total surface area reduces some heating mechanisms, such as photoelectric heating, H$_2$ formation rate and H$_2$ photodissociation. However, the total heating is still dominated by the photoelectric effect on PAHs. Cooling is not affected significantly, since the main coolants are C\,\textsc{ii} and O\,\textsc{i} in the optically thin regime (see Fig. \ref{fig:heat_cool}). The net effect is that temperature is not affected significantly by the total amount of small grains, provided that the PAH abundance is kept fixed. The only exception is apparent in the high density and high FUV flux case ($n=10^6$ and $G_{\rm FUV}=3000\,G_0$). In this regime, heating from FUV pumping of H$_2$ molecules becomes comparable to the photoelectric heating \citep[][\citetalias{adams_04}]{1999RvMP...71..173H}. The abundance of H$_2$ molecules depends on the H$_2$ formation rate, which depends on the total dust surface area. By reducing the amount of small grains, H$_2$ formation rate drops, and so does the FUV pumping, resulting in a lower temperature.

For the cases considered here and displayed in Fig. \ref{fig:pdr}, at fixed density the temperature always shows a plateau at low optical depths, and then decreases steeply with visual extinction, until it approaches unity and the gas becomes optically thick to the external radiation. These two characteristics will shape the final profiles of the subsonic winds. Finally, the temperatures are lower than the ones computed by \citetalias{adams_04} (e.g. by a factor $\sim2$ when $G_{\rm FUV}=300\,G_0$), in the regions of parameter space that overlap, even in the hottest case when $s_{\rm max}=3.5\,\mu$m. For the same dust distribution, we obtain temperatures similar to the ones by \citet{1999ApJ...527..795K} (and therefore by \citetalias{adams_04}, who use the same PDR code) by using the prescription by \citet{1994ApJ...427..822B} \citep[without the correction by][]{2003ApJ...587..278W} for the PAH photoelectric heating. In particular, we reproduce the same trend observed by \citet{1999ApJ...527..795K} in their fig.~1.

\section{Hydrodynamics}
\label{sec:hydro}

In the last section we have shown how to derive the temperature of the gas in the PDR region as a function of local density $n$ and visual extinction $A_V$. We now want to couple these results to the hydrodynamical equations describing the flow departing from the edge of a protoplanetary disc.

\subsection{Equations for the wind structure}
\label{subsec:hydroeq}

We develop a 1D description of the problem in a quasi spherical geometry, where we self-consistently solve for the radial steady-state structure of the wind launched from the disc outer edge. The equations shown in this section are based on the similar model by \citetalias{adams_04}.

The time independent version of the continuity equation is:

\begin{equation}
\label{eq:cont}
\dot{M}=4\pi R^2 \mathcal{F}  \mu m_{\rm H}nv,
\end{equation}
where $m_{\rm H}$ is the hydrogen atom mass and $v$ is the velocity of the gas. $\mathcal{F}$ is the fraction of the solid angle subtended by the disc outer edge (see equation 14 by \citetalias{adams_04}). The flow might be not perfectly spherical, since the wind might not be pressure confined at the boundaries of the wedge defined by the solid angle $4\pi\mathcal{F}$. We could model the flow with a hyper-spherical continuity equation, where $\dot{M}\propto R^\alpha$ and $\alpha>2$ ($\alpha=2$ defines the spherical case, see equation \ref{eq:cont}). For simplicity, we consider the spherical case only in this paper.

Similarly, the time independent momentum equation is: 
\begin{equation}
v\frac{dv}{dR}+\frac{1}{\rho}\frac{dP}{dR}+\frac{GM_*}{R^2}-\frac{j^2}{R^3}=0,
\end{equation}
where $j^2=GM_*R_{\rm d}$. We are considering a flow with uniform specific angular momentum, given by the Keplerian angular momentum at the outer radius of the disc. We assume an ideal gas law for the pressure $P=nk_{\rm B}T$, where T is a function of density and optical depth (as shown in Section \ref{sec:pdr}).

Following \citetalias{adams_04}, we can express the above equations in a dimensionless form, where our parametrisation choice is slightly different. We define $\xi\equiv R/R_{\rm d}$, $f\equiv T/T_{\rm c}$, $g\equiv n/n_{\rm c}$ and $u\equiv v/c_{\rm s,c}$, where $c_{\rm s}$ is the sound speed of the gas. With the subscript $c$ we indicate quantities evaluated at a critical radius $R_{\rm c}>R_{\rm d}$, which is defined in Section \ref{subsec:crit_rad}. Note that with these dimensionless units the local sound speed $u_{\rm s}^2$ is coincident with the temperature $f$. We also define the parameters $\beta$:
\begin{equation}
\label{eq:beta}
\beta=\frac{GM_*\mu m_{\rm H}}{R_{\rm d}k_{\rm B} T_{\rm c} }=\frac{GM_*}{R_{\rm d}c_{\rm s,c}^2 },
\end{equation}
and the optical depth $\tau\equiv N_{\rm H}\sigma_{\rm FUV}$, where 
\begin{equation}
N_{\rm H} = \int^\infty_R{n(R')dR'}
\end{equation}
is the column density between $R$ and infinity. We thus obtain:
\begin{equation}
\label{eq:tau}
\frac{d\tau}{d\xi}=-\sigma_{\rm FUV} R_{\rm d} n_{\rm c} g = -\tau_{\rm d}g,
\end{equation}
where $\tau_{\rm d}=\sigma_{\rm FUV} R_{\rm d} n_{\rm c}$. Combining the continuity and the momentum equation, we obtain a single differential equation for the dimensionless velocity $u$:
\begin{equation}
\label{eq:dlnu}
\frac{d\ln{u}}{d\xi}(u^2-f-g\frac{\partial f}{\partial g}) = \frac{2}{\xi}(f+g\frac{\partial f}{\partial g}) - \beta\frac{\xi-1}{\xi^3}+\tau_{\rm d}g\frac{\partial f}{\partial \tau}.
\end{equation}

This last equation reduces to the standard Parker wind equation \citep[][]{1958ApJ...128..664P}, in the limit of isothermality and in the absence of the centrifugal term. By solving this equation, we can obtain the velocity structure of the flow, from which we can obtain the density structure of the gas via the continuity equation.

\subsection{The critical radius}

\label{subsec:crit_rad}
The structure of equation \ref{eq:dlnu} shows a natural definition of a critical point: when the r.h.s. of the equation equals $0$. When this happens, the l.h.s. of the same equation has to equal $0$ as well. This can happen when $u$ has a null gradient, or when the second multiplicand is $0$. We require this second case to be the one, since we are looking for the analogue of the transonic solution in the typical Parker wind problem \citep{clarke_book_07}. In the isothermal case, this same procedure defines the sonic radius $R_{\rm s}$, where $u^2=f$. We expect the critical radius to be of the same order of magnitude as the sonic radius. By requiring that the l.h.s. of the equation is $0$, but $du/d\xi\neq0$, we implicitly define a critical velocity:

\begin{equation}
\label{eq:u_crit}
u_{\rm c}^2= f + g\frac{\partial f}{\partial g},
\end{equation}
which is related to the sound speed $u_{\rm s}$ via
an additional term taking into account possible
departures from isothermality. The equation for the dimensionless critical radius $\xi_{\rm c}=R_{\rm c}/R_{\rm d}$ is then:

\begin{equation}
\label{eq:crit_rad}
g \tau_d \frac{\partial f}{\partial \tau} \xi_c^3 +  2 u_{\rm c}^2 \xi_{\rm c}^2 - \beta \xi_{\rm c} + \beta = 0,
\end{equation}
where all the quantities are evaluated at the critical radius. This problem has to be solved implicitly, since all the quantities depend on where the critical radius is located, and on the initial conditions of the problem. Locating the critical radius self-consistently is a key
ingredient of our approach in order to be able to integrate the
flow structure inwards from this point to the disc outer edge. Note that this is a key difference between our model and the one proposed by \citetalias{adams_04}. In Section \ref{sec:results} we will show that taking into account this non-isothermal term significantly affects the final mass loss rates.

For every set of parameters $M_*$, $R_{\rm d}$ and $G_{\rm FUV}$ there is a family of solutions
each of which has a different value of $T_{\rm c}$, mass loss rate and
pressure at $R_{\rm d}$. We require that our flow is in pressure equilibrium
with the disc at $R_{\rm d}$; therefore in principle we could use
the pressure at the disc edge as a further independent variable
which would define the mass loss rate and flow structure for a given
disc in a given environment \citepalias[cf.][]{adams_04}. This however
requires iterative solutions for various guessed flow velocities
at $R_{\rm d}$ and is not a preferred method in the case where the
critical point conditions are themselves a function of the
flow solution. Since we start each flow solution from the
self-consistently determined critical point of the flow, it is
operationally convenient to use $T_{\rm c}$ as the independent variable.
Each solution can then be mapped onto the corresponding value of
the pressure at the disc outer edge.

In order to evaluate the critical radius for a given value of $T_{\rm c}$ we
proceed as follows. We first guess the value of the critical radius
(i.e. as parametrised by $\xi_{\rm c}$); for an initial guess we take
the re-scaled sonic radius $\xi_{\rm s}$, using the isothermal limit
of equation \ref{eq:crit_rad}:

\begin{equation}
\label{eq:sonic_rad}
\xi_{\rm s}= \frac{\beta}{4}\Big[ 1+ \Big( 1-\frac{8}{\beta} \Big)^{1/2} \Big],
\end{equation}
and noting that $\beta$ is fixed for given $T_{\rm c}$ by equation \ref{eq:beta}. 

Once we have an initial guess for the critical radius, for every iteration on the value of $\xi_{\rm c}$ we calculate
the pair of values for the number density and optical depth
at this point for which the PDR models predict that $T=T_{\rm c}$
locally. This step requires an assumption about the density
structure of the flow {\it outside} the critical radius.
Here we follow \citet{johnstone_98} and \citetalias{adams_04} in assuming that
the velocity is approximately constant in this region so that
from the continuity equation we then obtain:

\begin{equation}
n(R)=n_{\rm c} \Big( \frac{R_{\rm c}}{R} \Big)^2,\ \ \ {\rm for}\ R>R_{\rm c},
\end{equation}
and consequently

\begin{equation}
\label{eq:tau_c}
\tau_{\rm c} = \sigma_{\rm FUV} \int_{R_{\rm c}}^{\infty} {n_{\rm c} \Big( \frac{R_{\rm c}}{R'} \Big)^2 dR'} = \sigma_{\rm FUV} R_{\rm c} n_{\rm c}.
\end{equation}

\begin{figure}
\center
\includegraphics[width=\columnwidth]{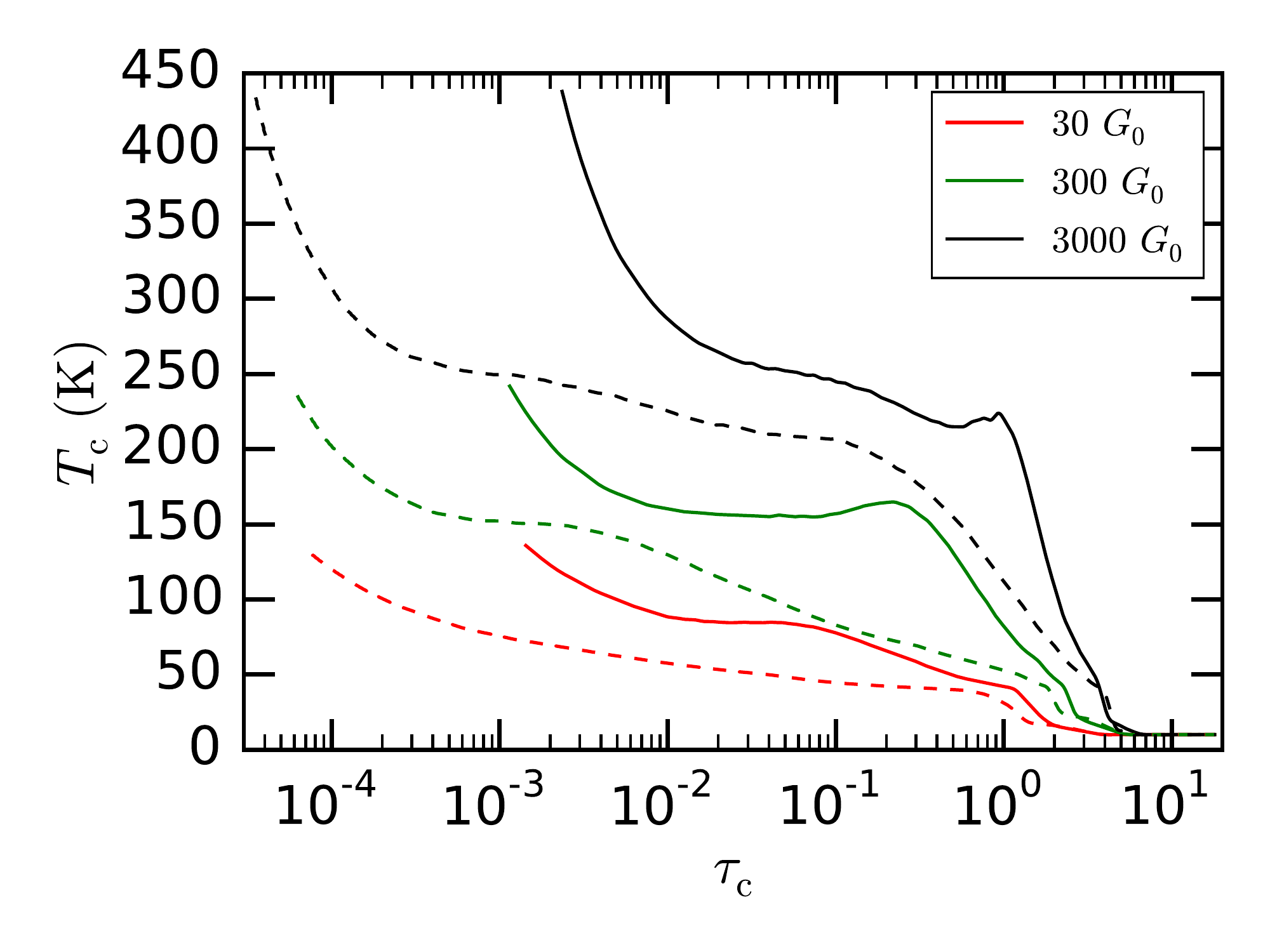}
\caption{The critical temperature, i.e. the input parameter to obtain a solution to the thermo - hydrodynamical equations, as a function of the optical depth evaluated at the critical radius $R_{\rm c}$, for $G_{\rm FUV}=30$, $300$ and $3000\,G_0$ for solution with $R_{\rm d}=120$. This implicit relation defines the boundary conditions at $R=R_{\rm c}$. Solid lines refer to $s_{\rm max}=3.5\,\mu$m, dashed lines to $s_{\rm max}=1\,$mm. Every curve is for one specific solution only, since it depends on the value of $\sigma_{\rm FUV}$, which varies for different flow solutions (see Section \ref{subsec:dust_gs}).
}
\label{fig:temp_init_cond}
\end{figure}

Figure \ref{fig:temp_init_cond} gives an example of $T_{\rm c}$ as a function of $\tau_{\rm c}$
for an initial guess $R_{\rm c}\sim500$\,AU and for a range of
FUV fluxes. For low FUV fluxes, the are no solutions in the low optical depth range ($\tau<10^{-3}$). This constraint is set by the lower limit we have implicitly imposed on the density, when we have not dealt with anything less dense than $10^2$\,cm$^{-3}$ in Section \ref{sec:pdr}.

We are now able to determine $\tau_{\rm c}$ and $n_{\rm c}$ by solving the equation:

\begin{equation}
\label{eq:t_crit}
T_{\rm c}-T(\tau_{\rm c})=0.
\end{equation}

We can evaluate the partial derivatives of $T$ with respect to
density and optical depth (at fixed $R_{\rm c}$) and use these in equations \ref{eq:u_crit} and \ref{eq:crit_rad}. We then solve for the next value of
$R_{\rm c}$ by solving equation \ref{eq:crit_rad} using a bisector method. We then iterate the whole process, until we reach convergence on the value of $\xi_{\rm c}$. By doing so, we have obtained the critical radius, and we have the initial conditions for $T$ (given by the input parameter $T_{\rm c}$), $v$, $n$ and $\tau$. The critical radius is always larger than the sonic radius (as defined by equation \ref{eq:sonic_rad}), but at the most by a factor of a few tenths of a dex (see an example in Fig. \ref{fig:crit_rad}). On the other hand,  the critical velocity is usually smaller than the sound speed at the critical radius, by $\sim20\%$ at the most (see an example in Fig. \ref{fig:v_crit}). Finally, we can estimate the mass-loss rate $\dot{M}$ from equation \ref{eq:cont}. We are ready to start integrating equation \ref{eq:dlnu} inwards, from the critical radius to the disc outer edge, in order to deduce the velocity profile of the subsonic flow between $R_{\rm d}$ and $R_{\rm c}$.

\subsubsection{Extending the solutions}

\begin{figure}
\center
\includegraphics[width=\columnwidth]{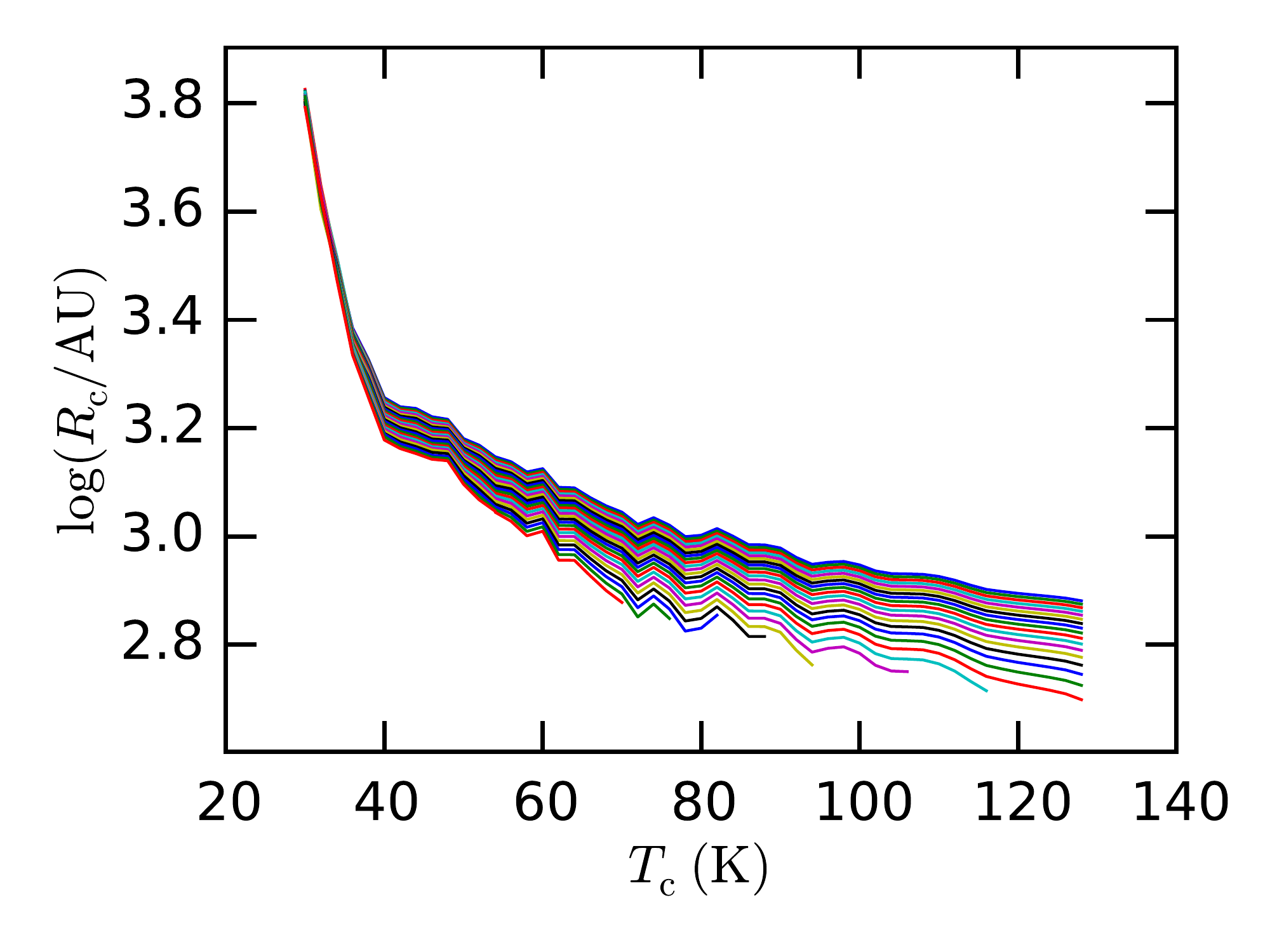}\\
\includegraphics[width=\columnwidth]{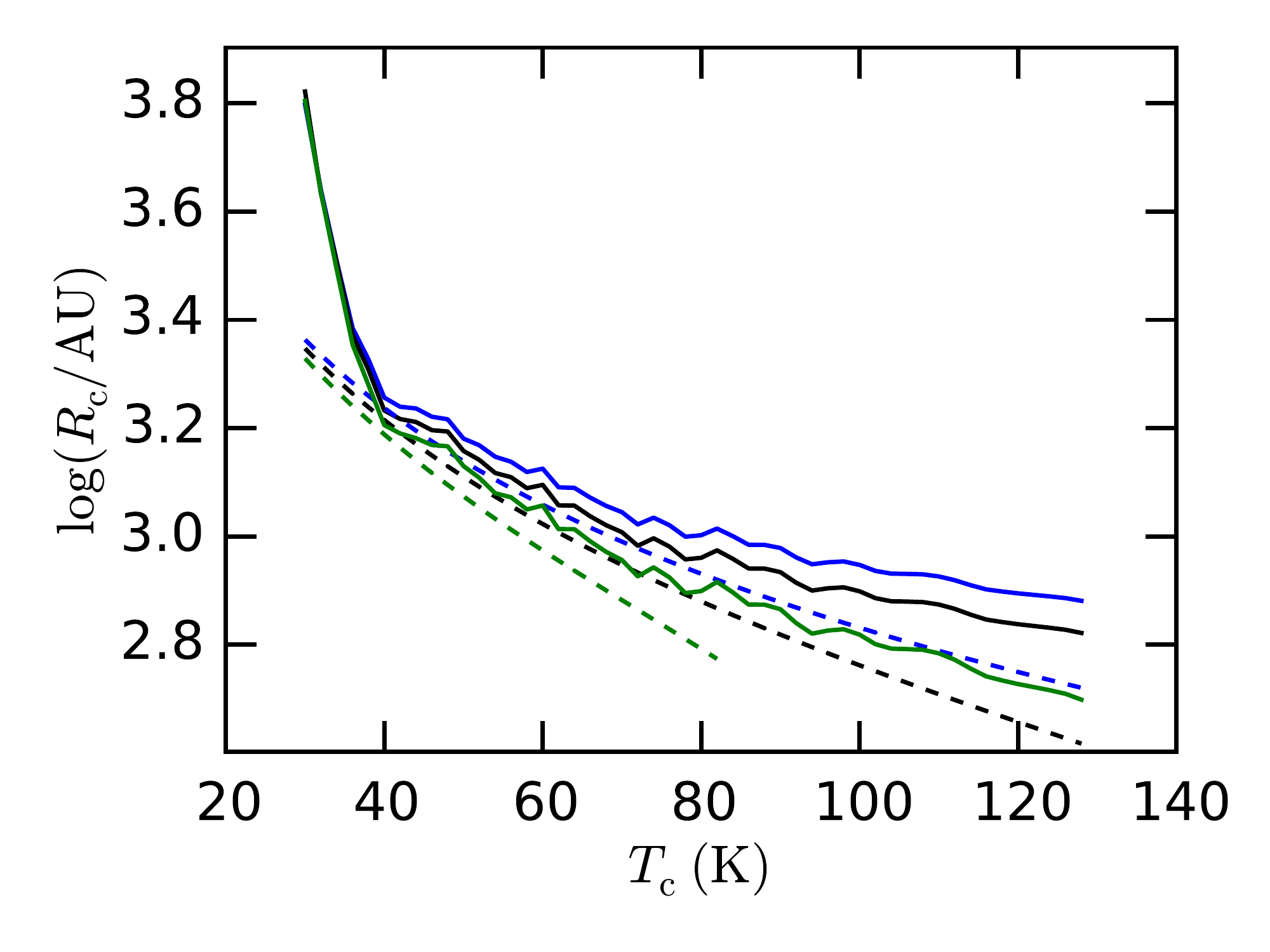}
\caption{Top panel: critical radius dependence on critical temperature for $G_{\rm FUV}=30\,G_0$ and $s_{\rm max}=1\,$mm, for a set of disc radii ranging from $20$ AU (top line) to $250$ AU (bottom line), sampled every $10$ AU. Every line illustrates a different disc radius $R_{\rm d}$. Bottom panel: solid lines are the same as in the top panel, for $R_{\rm d}=20$, $100$ and $180$\,AU (blue, black and green lines, respectively). Dashed lines indicate the sonic radius, as defined by equation \ref{eq:sonic_rad}.}
\label{fig:crit_rad}
\end{figure}

\begin{figure}
\center
\includegraphics[width=\columnwidth]{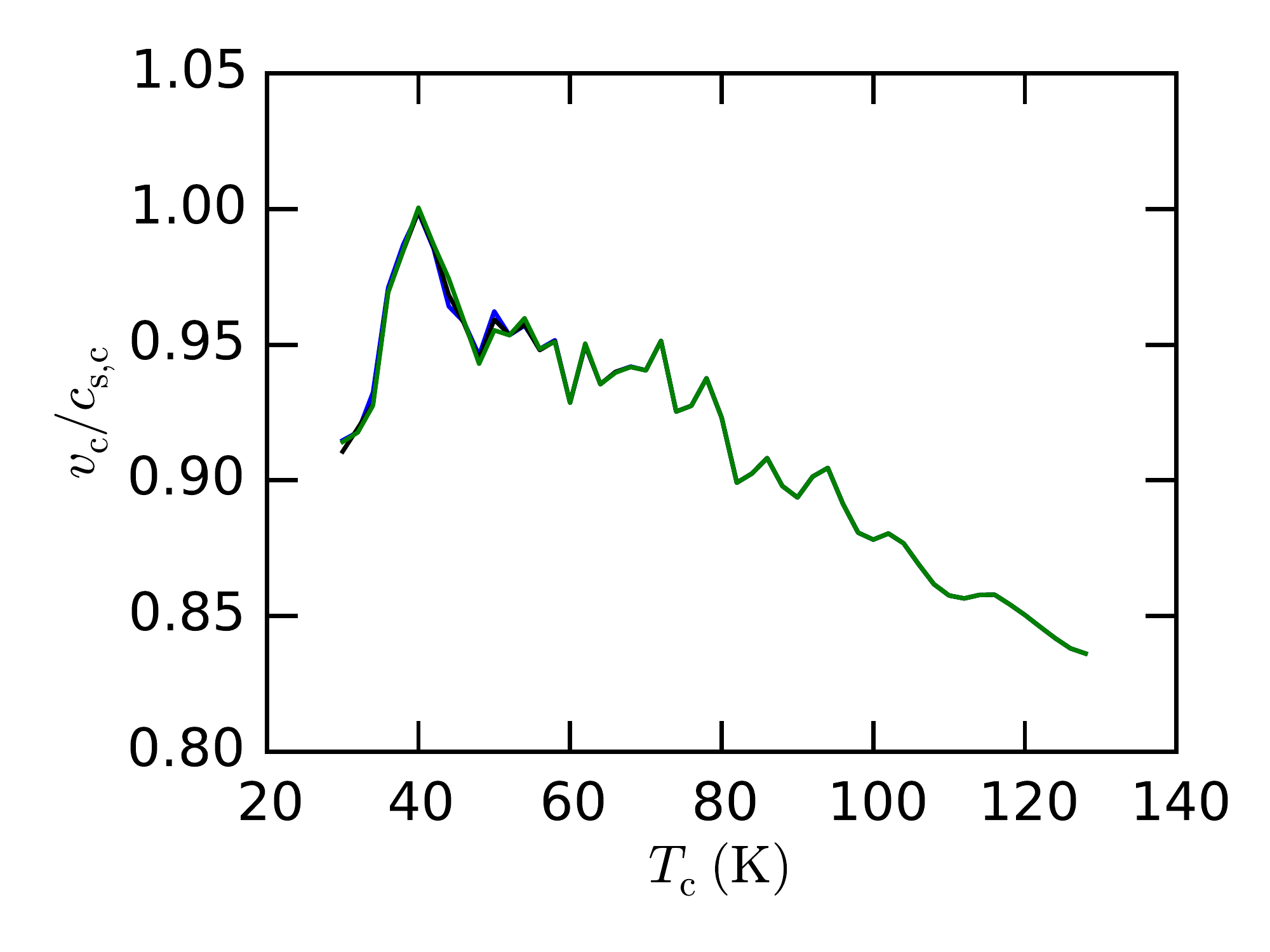}
\caption{Ratio of critical velocity to the sound speed at $R_{\rm c}$ for solutions with $G_{\rm FUV}=30\,G_0$ and $s_{\rm max}=1\,$mm, and $R_{\rm d}=20$, $100$ and $180$\,AU (blue, black and green lines, respectively).}
\label{fig:v_crit}
\end{figure}

In the definition of the sonic radius (see equation \ref{eq:sonic_rad}), by construction we are requiring that $\beta>8$. Thus we are implicitly setting an upper limit to the $T_{\rm c}$ we can set at a given $R_{\rm d}$:

\begin{equation}
T_{\rm c} < \frac{GM_*\mu m_{\rm H}}{8k_{\rm B}R_{\rm d}} \simeq 873 \Big( \frac{M_*}{M_\odot} \Big) \Big( \frac{20\,{\rm AU}}{R_{\rm d}} \Big)\,{\rm K}.
\end{equation}
This strongly limits the explorable region of parameter space, when $R_{\rm d}$ is large and the FUV flux (and therefore temperature) is high. In order to enlarge the parameter space, when $\beta<8$, as a first guess for $R_{\rm c}$  we choose the critical radius of the solution with the same $T_{\rm c}$ and the closest value of $R_{\rm d}$ showing a iteratively converged solution for the critical radius. In Fig. \ref{fig:crit_rad} we show $R_{\rm c}$ versus $T_{\rm c}$ when $G_{\rm FUV}=30\,G_0$, for a range of disc radii $R_{\rm d}$ between $20$ and $250$ AU, sampled every $10$ AU. The critical radius does not show a strong dependence on $R_{\rm d}$, at fixed $T_{\rm c}$. In this way we are able to obtain more solutions, even though in some regions of parameter space we do not manage to obtain one.

\subsection{Method of solution}
\label{subsec:num}

Once we have located the critical radius and determined the boundary condition for flow density, optical depth and velocity for a given $T_{\rm c}$, a disc radius and a FUV flux $G_{\rm FUV}$, we can integrate equation \ref{eq:dlnu} from $R_{\rm c}$ to $R_{\rm d}$ (in rescaled units, from $\xi_{\rm c}$ to $1$). We use a simple Euler code. At every step we compute $u_{i+1}$ from equation \ref{eq:dlnu}. Then, we calculate dimensionless density $g_{i+1}$ from the dimensionless form of equation \ref{eq:cont}, and $\tau_{i+1}$ from equation \ref{eq:tau}. At given $g_{i+1}$ and $\tau_{i+1}$ we can compute the new temperature partial derivatives $\partial f/\partial g_{i+1}$ and $\partial f/\partial \tau_{i+1}$ from the PDR output, and therefore obtain the next value of the temperature $f_{i+1}$. With this set of equations we can self-consistently solve for the flow steady-state. We use an initially logarithmically spaced grid, since the absolute value of the gradient of most of the quantities increases as the solution approaches $R_{\rm d}$. This happens mostly in the transition between the optically thin and the optically thick regime, when the temperature drops steeply (as shown in Fig. \ref{fig:pdr}). In order to better resolve this region we also apply an adaptive mesh algorithm, i.e. we require that $\delta\xi$ is mall enough to ensure that the relative change in velocity between two steps is less than $1\%$ ($|u_{i+1}-u_i|/u_i < 0.01$).

However, we need to apply slight modifications to the algorithm at the two boundaries of the integration, in the proximity of both the critical radius $R_{\rm c}$ and the disc outer radius $R_{\rm d}$.

\subsubsection{The critical point}

In Section \ref{subsec:crit_rad} we have defined the critical radius as the radius at which the r.h.s. of equation \ref{eq:dlnu} vanishes. Moreover, at this same radius the second multiplicand of the l.h.s. of the same equation is $0$. Since we are starting integrating the same equation from $R_{\rm c}$, we have a null value both at the numerator and at the denominator. This leads to the possibility of having multiple \emph{transcritical} solutions (the analogue version of the transonic solutions in the standard isothermal case). More specifically, we will have a solution that will be supersonic between $R_{\rm d}$ and $R_{\rm c}$, and one solution that will be subsonic. We are looking for the second one. Therefore we need to enforce the solution to relax onto the subsonic branch. We do so by following a procedure that is similar to the one used by \citet{murray-clay_09} for an analogous problem.

We expand equation \ref{eq:dlnu} around the critical point to first order in both radius and velocity: $\xi = \xi_{\rm c} + \delta\xi$, and $u=u_{\rm c} + \delta u$. We obtain the following relation:

\begin{equation}
\label{eq:du_erf}
\frac{\delta u}{\delta \xi}\Big|_{\xi_{\rm c}} = \frac{-B + \sqrt{B^2-4AD}}{2A},
\end{equation}
where:

\begin{equation}
A=2u_{\rm c}+\frac{2g}{u_{\rm c}}\frac{\partial f}{\partial g} + \frac{g^2}{u_{\rm c}}\frac{\partial^2 f}{\partial g^2};
\end{equation}

\begin{equation}
B=2 \tau_d g \frac{\partial f}{\partial \tau} + \frac{8g}{\xi}\frac{\partial f}{\partial g} + \frac{4g^2}{\xi}\frac{\partial^2 f}{\partial g^2} + 2\tau_{\rm d}g^2\frac{\partial^2 f}{\partial\tau\partial g};
\end{equation}

\begin{eqnarray}
\nonumber D=u_{\rm c} \times
&  \displaystyle \Big(\frac{2u_{\rm c}^2}{\xi^2}  -\beta\frac{2\xi-3}{\xi^4} + \frac{8g}{\xi^2} \frac{\partial f}{\partial g} + \frac{4\tau_{\rm d}g}{\xi} \frac{\partial f}{\partial \tau}\\ 
& \displaystyle +\frac{4g^2}{\xi^2} \frac{\partial^2 f}{\partial g^2} + \frac{4\tau_{\rm d}g^2}{\xi} \frac{\partial^2 f}{\partial \tau \partial g} + \tau_{\rm d}^2 g^2 \frac{\partial^2 f}{\partial \tau^2}\Big),
\end{eqnarray}
where all the quantities are evaluated at $\xi=\xi_{\rm c}$. In equation \ref{eq:du_erf} we chose the positive root, in order to pick the subsonic solution at $\xi < \xi_{\rm c}$.

Similarly to \citet{murray-clay_09}, we calculate the velocity by using:
\begin{equation}
\frac{du}{d\xi}=F_{\rm exact}\frac{du}{d\xi}\Big|_{\rm exact} + (1-F_{\rm exact})\frac{\delta u}{\delta \xi}\Big|_{\xi_{\rm c}},
\end{equation}
where $du/d\xi|_{\rm exact}$ is evaluated from equation \ref{eq:dlnu}, and

\begin{equation}
F_{\rm exact}=-{\rm erf}{\Big[ h\Big( 1-\frac{f}{u^2}-\frac{g}{u^2}\frac{\partial f}{\partial g} \Big) \Big]},
\end{equation}
where erf is the error function. The parameter $h=20$ gives the smoothing length of the transition between $du/d\xi|_{\rm exact}$ and $\delta u/\delta \xi|_{\xi_{\rm c}}$ when computing $u(\xi)$. We use this modified version of equation \ref{eq:dlnu} until $F_{\rm exact}=1$ to the level of machine precision.

\subsubsection{Temperature corrections near $R_{\rm d}$}
Some solutions become completely optically thick before reaching the disc outer radius (i.e. $R>R_{\rm d}$). In the PDR calculations, we have set a minimum temperature equal to $10$\,K. However, the temperature of the flow could be higher than that, due to the impinging radiation from the central star. When we compute the temperature of the flow, we therefore include heating from the central star, by using the following simple prescription:

\begin{equation}
T=\max{(T_{\rm PDR},T_{\rm rad})},
\end{equation}
where $T_{\rm PDR}$ is the temperature evaluated via the PDR code (i.e. the one used so far in the paper), and $T_{\rm rad}$ is given by:

\begin{equation}
\label{eq:t_disc}
T_{\rm rad} = 100\,{\rm K}\, \Big( \frac{R}{1\,{\rm AU}} \Big)^{-1/2},
\end{equation}
i.e. a temperature profile found to fit the spectral energy distributions of passive discs \citep[e.g.][]{2005ApJ...631.1134A,2007ApJ...671.1800A}.

\section{Dust component}
\label{sec:dust}

In Section \ref{sec:hydro}, we have reported the equations and the procedure to obtain a solution for the gas quantities between the critical radius and the disc's outer edge. However, this solution depends on the dust properties within the flow. In particular, as mentioned in Section \ref{subsection:pdr_dust}, the attenuation factor strongly depends on the grain size distribution and on the maximum grain size $s_{\rm entr}$ entrained in the wind. Such dependence can be summarised in how $\sigma_{\rm FUV}$ is related to these two properties of the dust material. In this section, we also describe the hydrodynamic equations for the dust particles, which will be used to determine the maximum grain size entrained in the flow, and thus the cross section.

\subsection{Cross section}
\label{subsec:dust_gs}

It is well known that protoplanetary discs witness substantial grain growth in the discs' midplane (see Section \ref{subsection:pdr_dust}). Such grain growth can be schematised in two effects: producing a shallower distribution, i.e. a lower $q$, and leading to a larger maximum grain size $s_{\rm max}$. As introduced in Section \ref{subsection:pdr_dust}, in this paper we focus on two distributions with the same power law index $q=3.5$ but different maximum grain sizes ($s_{\rm max}=3.5\,\mu$m and $1\,$mm). The same distributions have been used to compute the dependence of the temperature on $n$ and $A_V$. As mentioned above, we consider a dust-to-gas ratio of $0.01$ for both the distributions.

We compute the cross section at FUV wavelengths ($\lambda=0.1\,\mu$m) using the code of \citet{2002MNRAS.334..589W} as briefly described in Section \ref{subsection:pdr_dust}. In the equations reported below, we describe a general treatment where $q$ can assume different values. We take into account that there is a maximum grain size entrained in the photoevaporative wind, and we compute the cross sections for the truncated distribution:

\begin{equation}
\sigma_{\rm FUV}=\mu m_{\rm H} \frac{q_{\rm s}}{m_{\rm entr} \delta_{\rm gd} },
\end{equation}
where
\begin{equation}
q_{\rm s} = \int_{s_{\rm min}}^{s_{\rm entr}} {\pi s^2 Q_{\rm abs}(\lambda,s) s^{-q}ds},
\end{equation}
\begin{equation}
m_{\rm entr} = \frac{4}{3}\frac{\pi \bar{\rho}}{4-q} (s_{\rm entr}^{(4-q)}-s_{\rm min}^{(4-q)}),
\end{equation}
and
\begin{equation}
\delta_{\rm gd} = 100 \left( \frac{s_{\rm max}^{(4-q)}-s_{\rm min}^{(4-q)}}{s_{\rm entr}^{(4-q)}-s_{\rm min}^{(4-q)}} \right).
\end{equation}
In these equations, $\delta_{\rm gd}$ is the effective gas-to-dust ratio within the flow, and $\bar{\rho}$ is the mean mass density of a dust grain (in this paper $1$\,g$/$cm$^3$). The minimum grain size of the distribution has been set to $s_{\rm min}=5\times10^{-7}$\,cm in the whole paper.

\begin{figure}
\center
\includegraphics[width=\columnwidth]{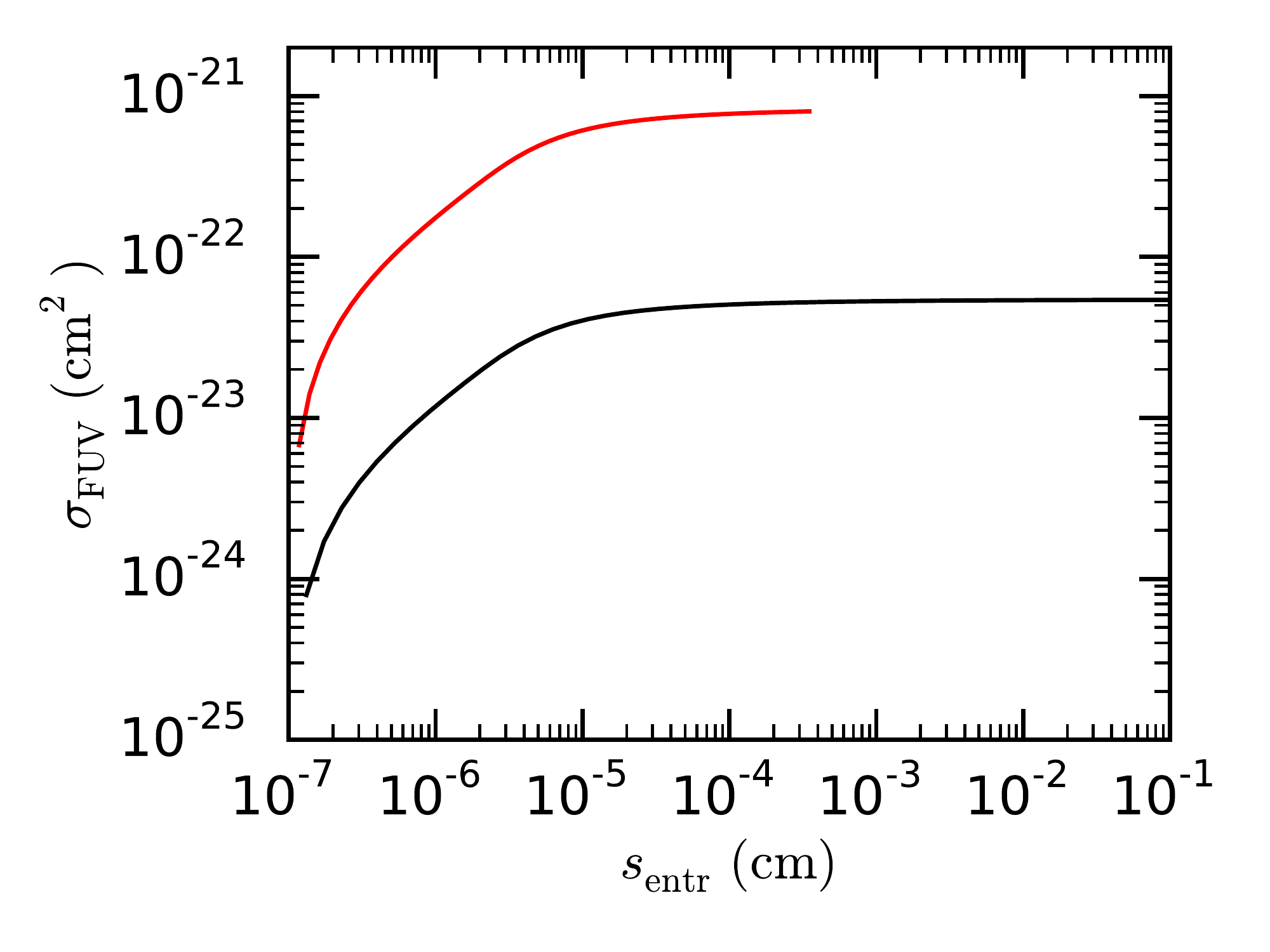}
\caption{Cross section at $\lambda=0.1\,\mu$m as a function of the maximum grain size entrained in the flow, for $s_{\rm max}=1\,$mm (black line) and $s_{\rm max}=3.5\,\mu$m (red line). The initial distribution assumes a dust-to-gas ratio of $0.01$. The cross section does not vary by much when the distribution is truncated to $s_{\rm entr}\gtrsim\lambda$, since the largest contribution to the cross section in the geometric limit comes from the smallest grains when $q>3$.
}
\label{fig:cross_section_input}
\end{figure}

The cross sections are reported in Fig. \ref{fig:cross_section_input},  as a function of the maximum grain size entrained in the flow, for $s_{\rm max}=3.5\,\mu$m (red line) and $s_{\rm max}=1\,$mm (black line). The cross section does not vary by much when the distribution is truncated to $s_{\rm entr}\gtrsim\lambda$, because the largest contribution to the cross section in the geometric limit comes from the smallest grains when $q>3$. We do not recompute the temperature $T(n)$ with the PDR code for the new truncated distributions, since it depends weakly on the maximum grain size. As discussed in Section \ref{sec:pdr}, this is due to the fact that heating is mostly generated by the photoelectric effect on PAHs and on small grains, since the total surface area per unit mass is dominated by the small grains of the distribution when $q>3$. Cooling is not significantly affected by the absence of the largest grains, since O\,\textsc{i} and C\,\textsc{ii} emission lines dominate the radiative cooling.

\subsection{Fluid equations for dust particles}
\label{subsec:hydro_dust}

In order to obtain the maximum grain size entrained in a gaseous solution, we need to solve the hydrodynamical equations of dust particles. We discretise the grain size distribution in $N_{\rm bin}=50$ bins, where the grain size is sampled logarithmically between $s_{\rm min}$ and $s_{\rm max}$. The dust mass density can be written as:

\begin{equation}
\rho_j \simeq \ \tilde{n}_j\frac{4}{3}\pi \bar{\rho} s_{{\rm max},j}^3,
\end{equation}
where the subscript $j$ refers to the $j$-th bin of dust particles. Here $\rho_j$ is the mass density of the $j$-th bin, and $s_{{\rm max},j}$ is the maximum grain size of the $j$-th bin.

The steady version of the momentum equation for dust particles is:

\begin{equation}
\label{eq:dust}
v_j \frac{dv_j}{dR} + \frac{GM_*}{R^2}-\frac{j^2}{R^3} + \frac{F_{\rm D}}{m_j}=0,
\end{equation}
where we have assumed that dust is pressureless, and that the buoyancy term related to the gas pressure gradient is negligible \citep[which is typically the case for astrophysical systems, see e.g.][]{2006MNRAS.373.1619R,2012MNRAS.420.2345L}. The fluid quantities that do not show any subscript refer to gas properties, as in Section \ref{sec:hydro}. The quantity $m_j$ is the mass of the $j$-th dust particle. The last term $F_{\rm D}$ represents the aerodynamic force experienced by the dust particle. This force can assume different expressions, depending on the physical regime. If the size of the dust particle $s\lesssim\lambda_{\rm mfp}$, the mean-free path of gas molecules within the flow, the drag force reduces to the so-called \citet{1924PhRv...23..710E} drag. If the size of the particle is larger than the mean-free path, drag is a classical fluid \citet{1851TCaPS...9....8S} drag. Within the winds addressed in this paper, $\lambda_{\rm mfp} > 10^5$\,cm (using $\lambda_{\rm mfp}\approx(\sigma_{\rm gas}n)^{-1}$, where $\sigma_{\rm gas}\sim10^{-16}$\,cm$^{2}$ is the geometrical cross-section of the gas molecules and $n<10^{11}$\,cm$^{-3}$). For the dust particles considered here, $s\ll\lambda_{\rm mfp}$, and we can therefore use the Epstein limit to estimate the drag term \citep[e.g.][]{armitage_book_10}:

\begin{equation}
F_{\rm D}=\frac{4\pi}{3}\rho s^2 v_{\rm th} (v_j-v),
\end{equation}
where $\rho=\mu m_{\rm} n$ is the gas mass density, and

\begin{equation}
v_{\rm th} = \sqrt{\frac{8k_{\rm B}T}{\pi\mu m_{\rm H}}}
\end{equation}
is the mean thermal speed of the gas molecules.

If we rewrite equation \ref{eq:dust} in the usual dimensionless form, and in the Epstein regime, we obtain:
\begin{equation}
\label{eq:dlnu_dust}
u_j\frac{du_j}{d\xi}= -\Big[ \beta\frac{\xi-1}{\xi^3} + \tilde{J}_jg\sqrt{f} (u_j-u) \Big],
\end{equation}
where:

\begin{equation}
\tilde{J}_j = \sqrt{\frac{8}{\pi}} \frac{\mu m_{\rm H} n_{\rm c}}{\bar{\rho}} \frac{R_{\rm d}}{s_{j}},
\end{equation}
and $u_j=v_j/c_{\rm s,c}$. For a given gaseous solution, we can obtain the velocity profile of the $j$-th bin of dust particles by integrating equation \ref{eq:dlnu_dust}. We assume that dust particles are launched from the disc at the same speed of gas molecules, i.e. $u_{j,{\rm d}}=u_{\rm d}$. Moreover, we require that the gas-to-dust ratio at the disc outer radius is $100$, and this automatically sets the normalisation of the grain size distribution. 

\begin{figure*}
\begin{center}
\includegraphics[width=\columnwidth]{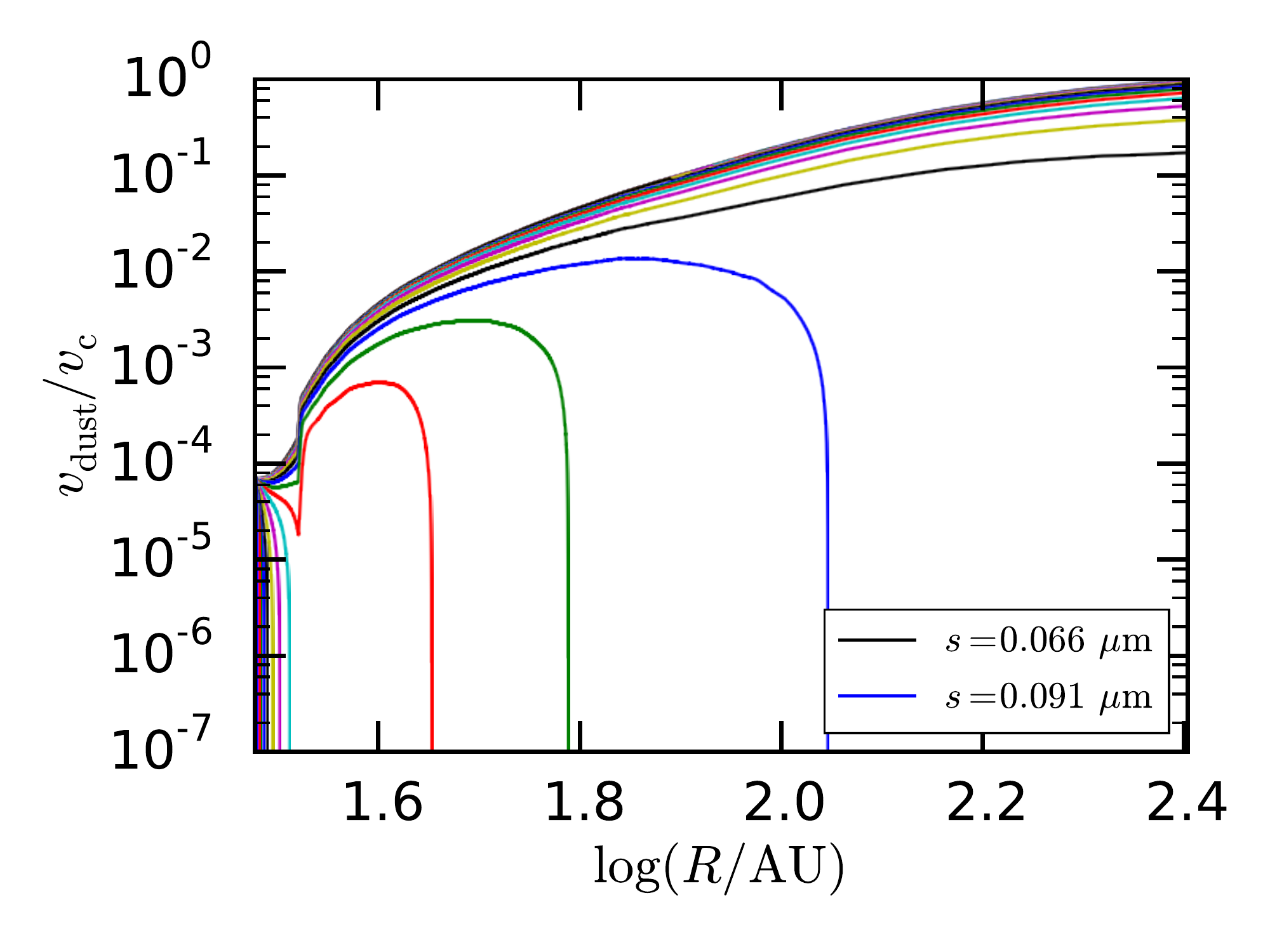}
\includegraphics[width=\columnwidth]{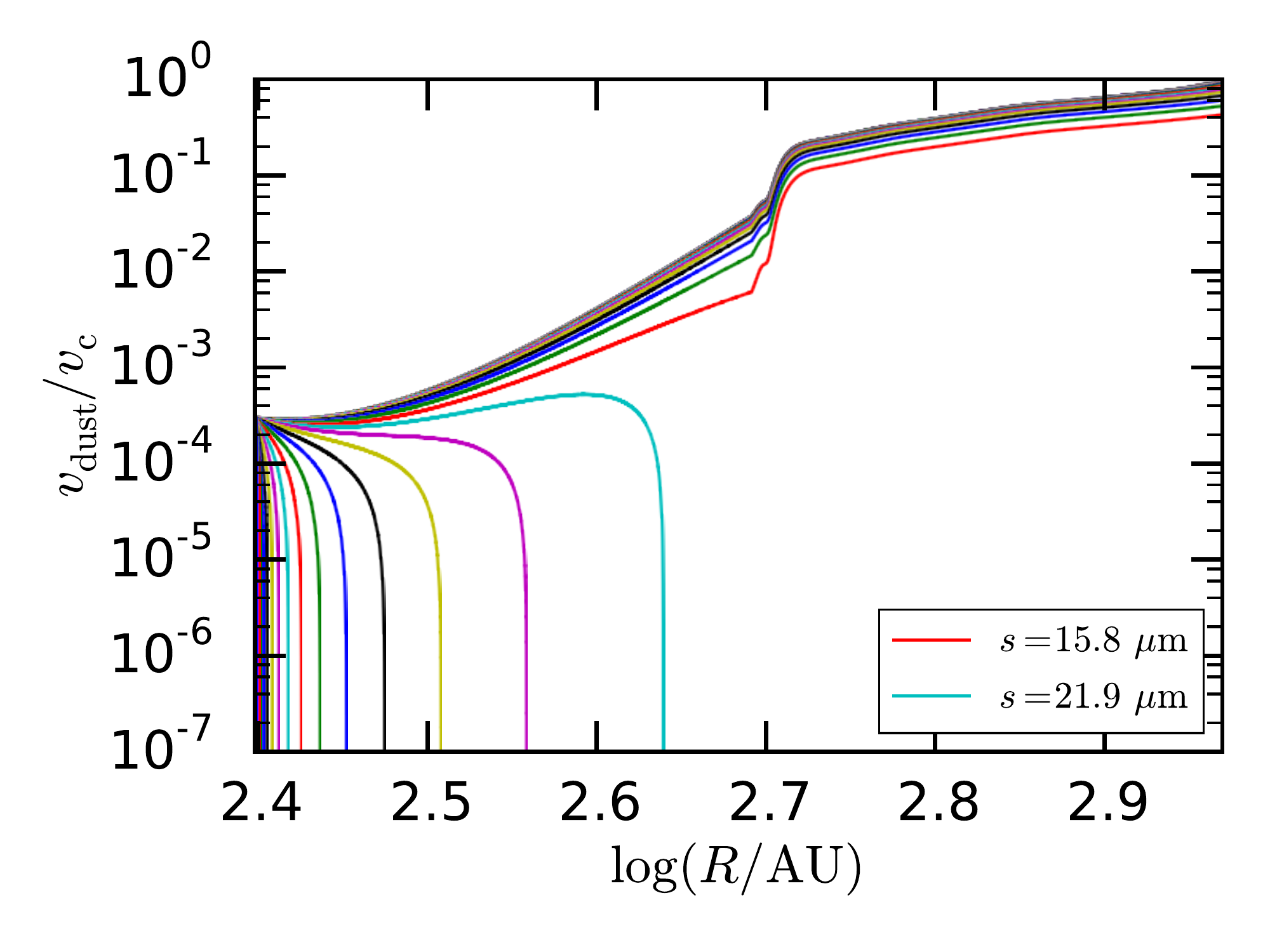}\\
\includegraphics[width=\columnwidth]{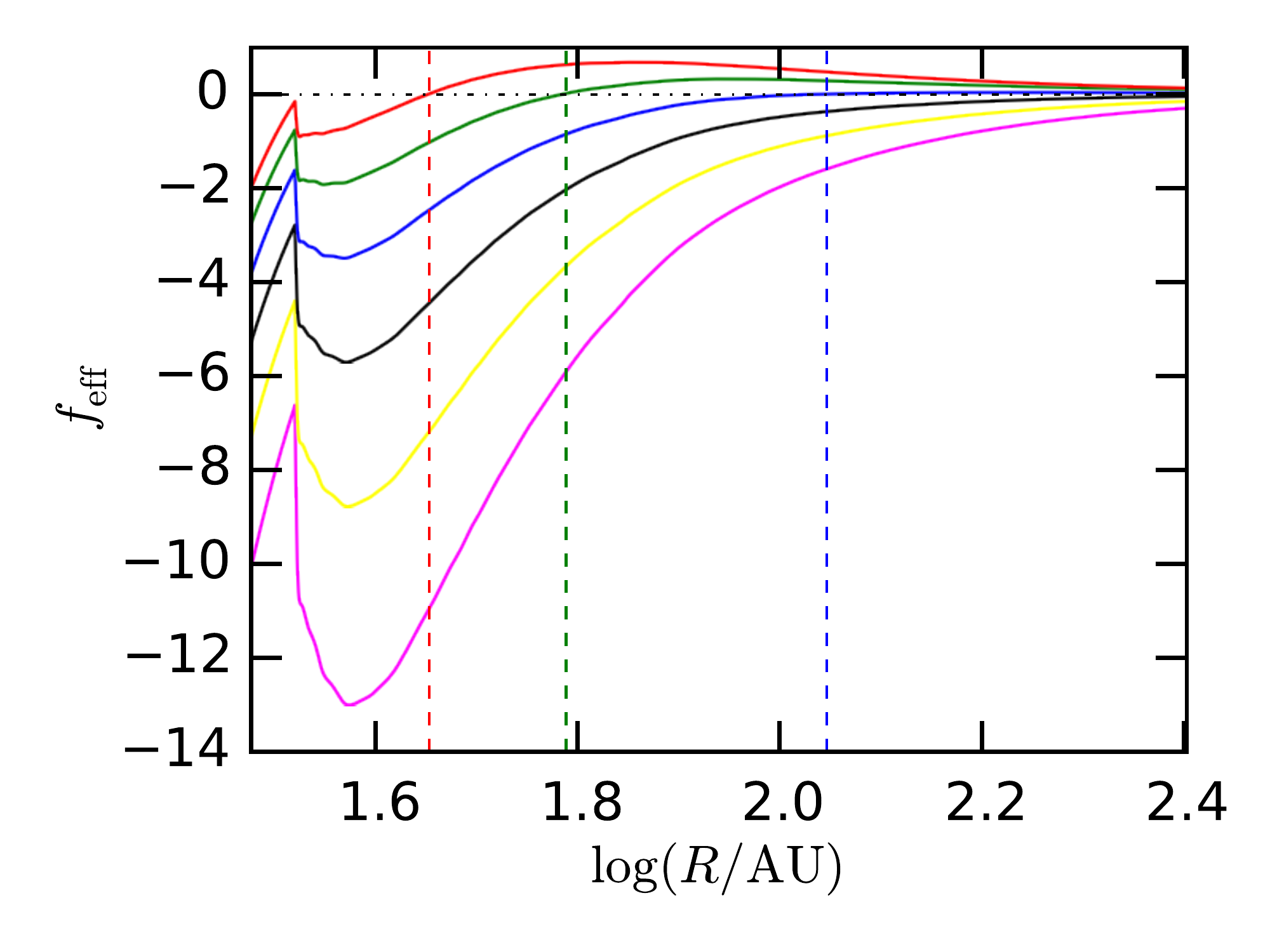}
\includegraphics[width=\columnwidth]{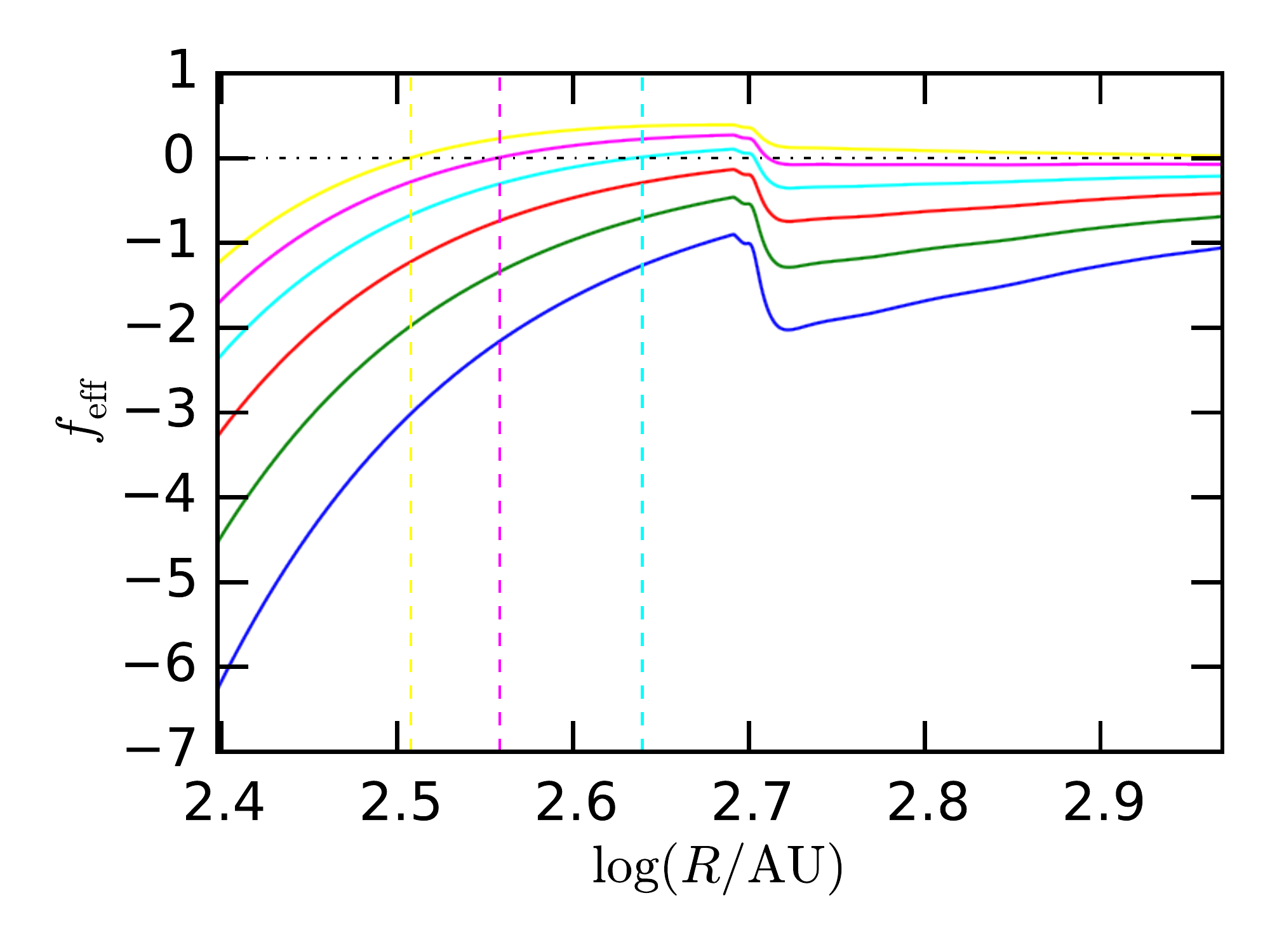}
\end{center}
\caption{Top panels: velocity profiles of the 50 dust bins between $R_{\rm d}$ and $R_{\rm c}$, for $s_{\rm max}=1\,$mm, $G_{\rm FUV}=3000\,G_0$, $P_{\rm d}\sim10^{-5}$ in cgs units, and $R_{\rm d}=30$ (left panel) and $200$\,AU (right panel). The largest grain dragged out to the critical radius and the smallest grain stalling at a finite location within the flow are highlighted in the legend. Bottom panels: $f_{\rm eff}$ computed via equation \ref{eq:feff} for six grain sizes. The six bins are selected in order to be the three largest sizes that are dragged all the way through the flow, and the three smallest sizes that stall at a finite radius, close to the disc edge. Note the perfect coincidence between the location where the grain velocity drops to $0$, and the stalling radius (i.e. where $f_{\rm eff}=0$). The stalling radii are indicated by the vertical dashed line.}
\label{fig:profiles_dust}
\end{figure*}

In order to obtain the maximum grain size entrained in the flow, we do not need to integrate equation \ref{eq:dlnu_dust} for every gaseous solution and every dust bin. If we consider the highly subsonic regime ($u_j\sim0$) equation \ref{eq:dlnu_dust} becomes:
\begin{equation}
\label{eq:feff}
\frac{du_j}{dt} = -\beta\frac{\xi - 1}{\xi^3} + \tilde{J}_jg\sqrt{f}u \equiv f_{\rm eff},
\end{equation}
where $f_{\rm eff}$ is the effective acceleration that a dust particle would feel if it were stationary. We define the stalling radius as the point at which a stationary dust grain would be in equilibrium, i.e. $f_{\rm eff}=0$, which is different for each grain size. We have demonstrated by integration of equation \ref{eq:dlnu_dust} that dust particles within the flow indeed collect at their respective stalling radii (i.e. where $f_{\rm eff}=0$) where these exist (see Fig. \ref{fig:profiles_dust}). This can be explained by the fact that equation \ref{eq:dlnu_dust} resembles the equation of motion of an overdamped harmonic oscillator, where the system returns to equilibrium without oscillating and in an exponentially decaying fashion. As expected, the results show that the smaller grains are tightly coupled to the gas flow, and are easily dragged out to the critical radius. The largest grains are dragged to the stalling radius as defined above. Note that the stalling radius does not depend on the initial velocity, i.e. this result does not depend on the initial condition we have chosen for the velocity of the dust grains at $R_{\rm d}$. 

This result confirms that in order to obtain the maximum grain size entrained in the flow $s_{\rm entr}$ we just need to solve equation $f_{\rm eff}=0$, without obtaining the exact velocity profiles of every dust bin.

\section{Iteration procedure and final solutions}
\label{sec:phys_sol}

In this section we summarise the iteration procedure we use to obtain the wind solutions by using all the ingredients reported in Sections \ref{sec:hydro}-\ref{sec:dust}. All the results depend on the temperature estimates obtained with the PDR code in Section \ref{sec:pdr}.

We initially set the main parameters of the system: the external FUV field ($G_{\rm FUV}$), the disc's outer radius ($R_{\rm d}$), and the stellar mass ($M_*$). All the results shown in this paper have $M_*=M_\odot$. We then set a critical temperature $T_{\rm c}$, which will be uniquely related to the pressure at the outer edge of the disc $P_{\rm d}$ of the final solutions. We then start the iteration procedure by assuming an initial $\sigma_{\rm FUV}=8\times10^{-22}$\,cm$^2$ for the $s_{\rm max}=3.5\,\mu$m case and $\sigma_{\rm FUV}=6\times10^{-23}$\,cm$^2$ when $s_{\rm max}=1\,$mm. We now list the following steps:

\begin{enumerate}
\item we calculate the critical radius by using equations \ref{eq:tau_c}-\ref{eq:t_crit};
\item we obtain all the boundary conditions of the gaseous flow at the critical radius, as explained in Section \ref{subsec:crit_rad};
\item we integrate the hydrodynamic equations from $R_{\rm c}$ inwards to the disc's outer radius, with the method detailed in Section \ref{subsec:num};
\item we compute the maximum grain size $s_{\rm entr}$ entrained in the flow, by using equation \ref{eq:feff};
\item we obtain the new cross section $\sigma_{\rm FUV}$;
\item we go back to step (i), and we iterate the same procedure until we reach convergence on the cross section $\sigma_{\rm FUV}$.
\end{enumerate}

As mentioned above, every solution selected via the input parameter $T_{\rm c}$ can be uniquely indicated by the gas pressure at the outer edge of the disc, which is a final output of our procedure. The boundary between the actual disc and the slow wind needs to be in pressure equilibrium (in principle there could be a contact discontinuity in both temperature and gas density). The pressure at the outer edge of the disc can be modelled by assuming simple prescriptions for the surface density and temperature profiles. For the latter, we use expression \ref{eq:t_disc}. For the former, we use the simple relation:

\begin{equation}
\label{eq:surf_dens}
\Sigma(R)=\frac{M_{\rm d}(2-p)}{2\pi R_{\rm d}^2}\left( \frac{R}{R_{\rm d}}\right)^{-p},
\end{equation}
where $M_{\rm d}$ is the mass of the disc, and we set $p=1$ for the whole paper. A final pressure-balanced solution is thus selected by the mass of the disc.

The parameter space we explore is the following. We range over the disc outer radius between $20$ and $250$\,AU, sampled every $10$\,AU. We explore three different values of external FUV field: $30$, $300$ and $3000\,G_0$, and for each of these we explore a set of initial conditions specified by the parameter $T_{\rm c}$. The critical temperature is sampled every $2$\,K, and the explored range is shown in Fig. \ref{fig:temp_init_cond} for the different field values. Not all these initial conditions lead to a final solution, in particular there are cases where the equation defining the critical radius does not have a real solution. As already explained in the paper, we focus on two grain size distributions with $q=3.5$, where the maximum grain sizes are respectively $3.5\,\mu$m and $1\,$mm. The former leads to the same cross section used in the models by \citetalias{adams_04} when $s_{\rm entr}>0.1\,\mu$m, and this will allow us to compare our results with theirs in more detail.

\section{Results}
\label{sec:results}

\begin{figure*}
\begin{center}
\includegraphics[width=\columnwidth]{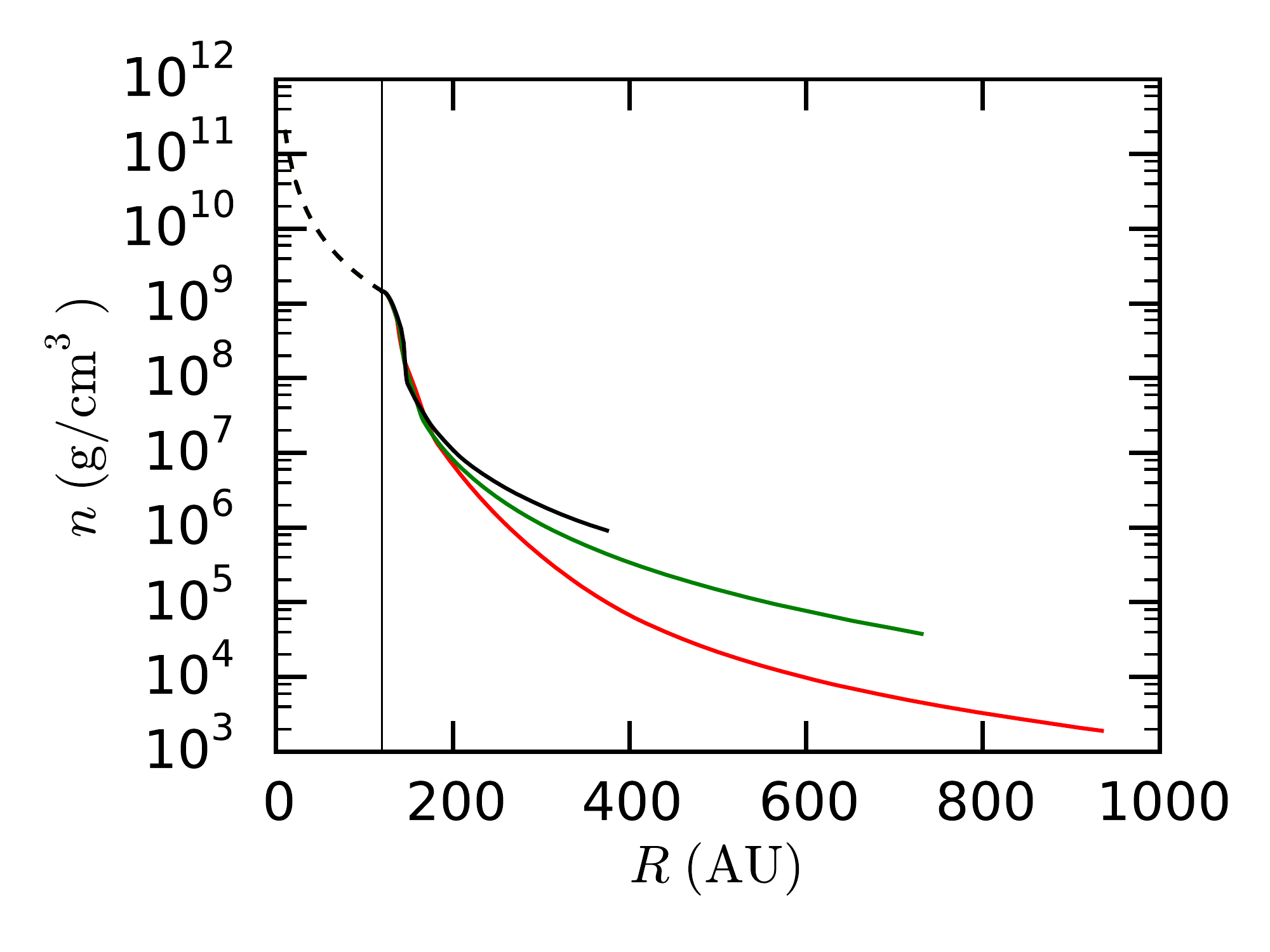}
\includegraphics[width=\columnwidth]{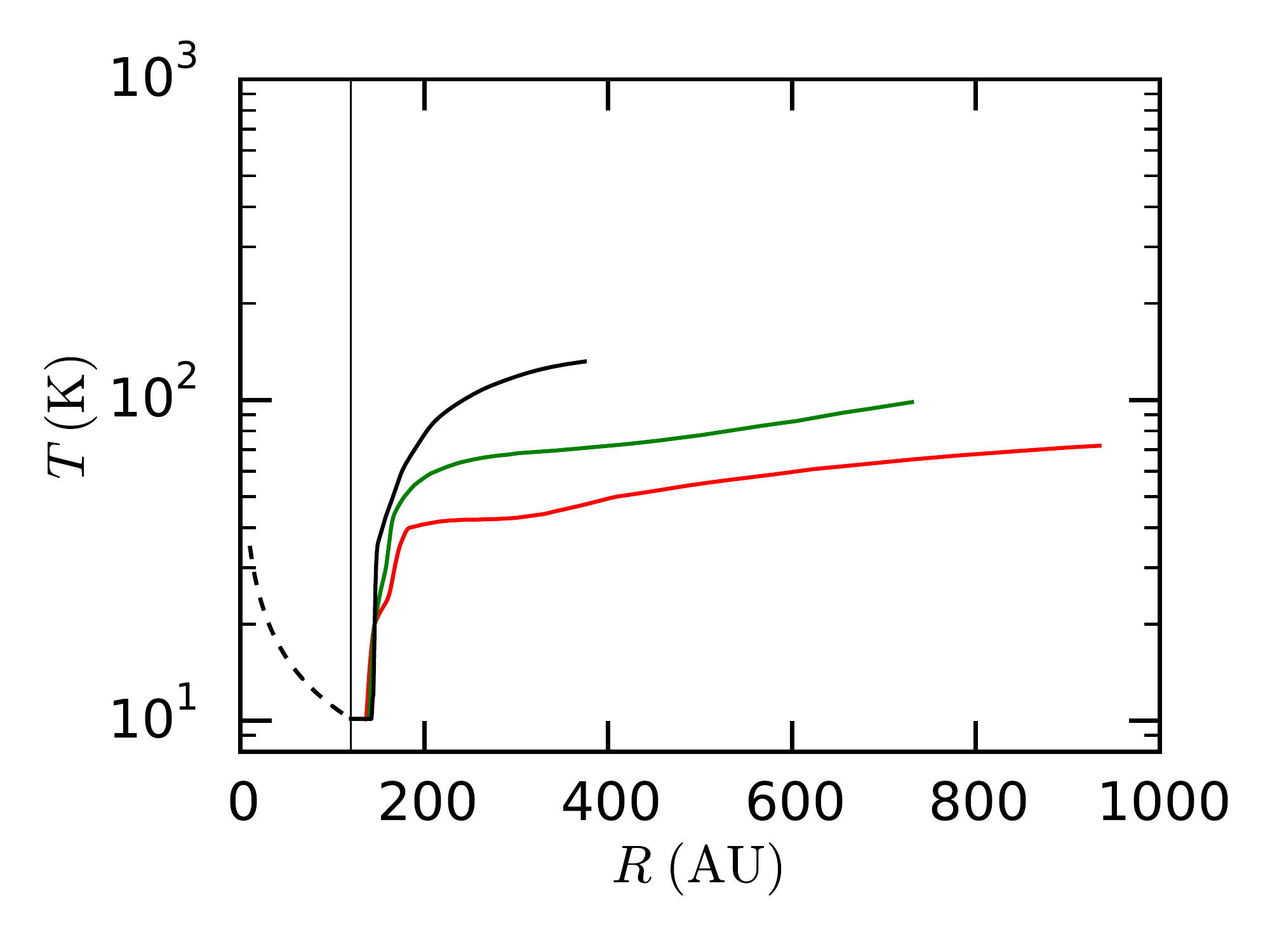}\\
\includegraphics[width=\columnwidth]{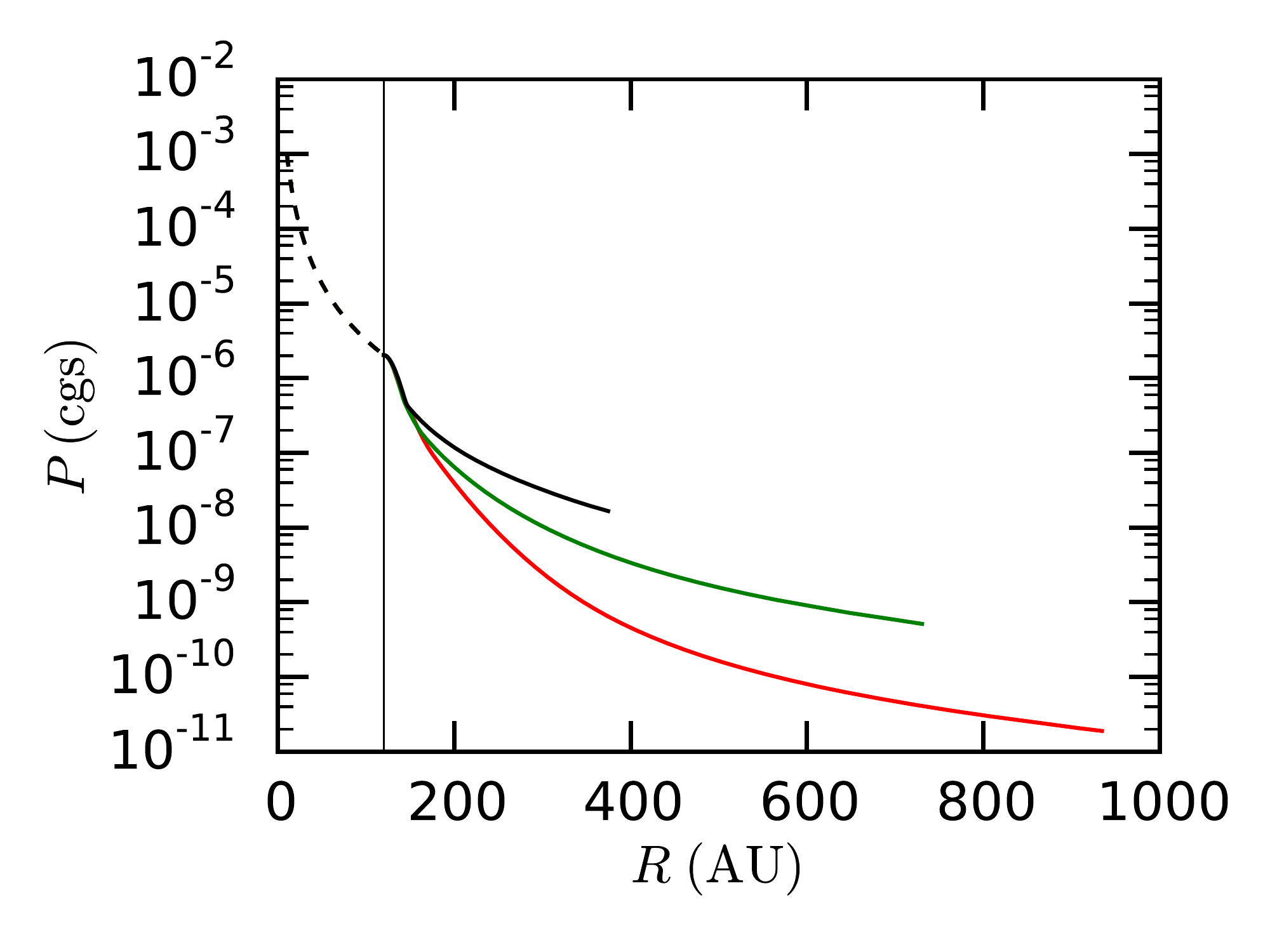}
\includegraphics[width=\columnwidth]{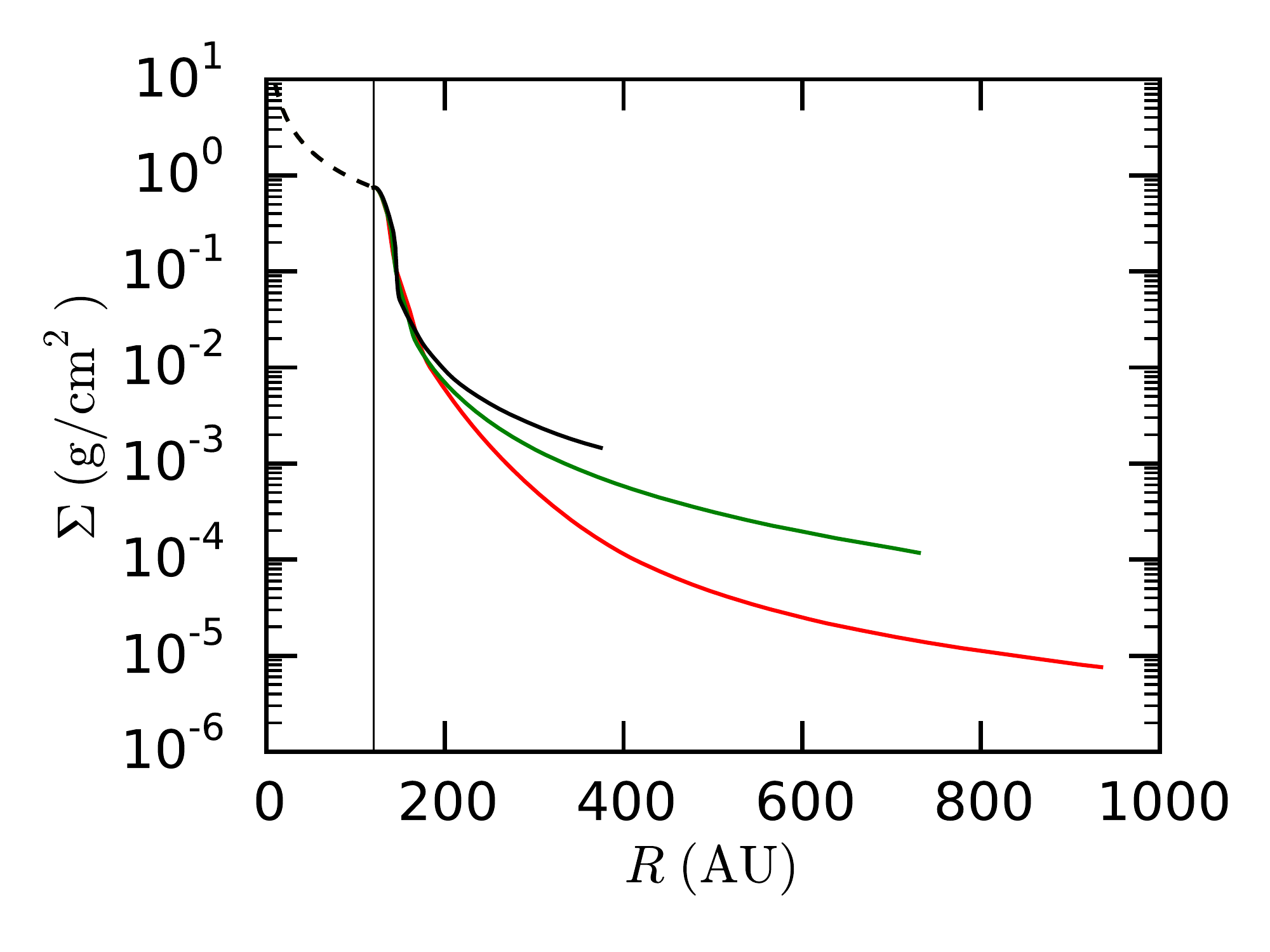}\\
\includegraphics[width=\columnwidth]{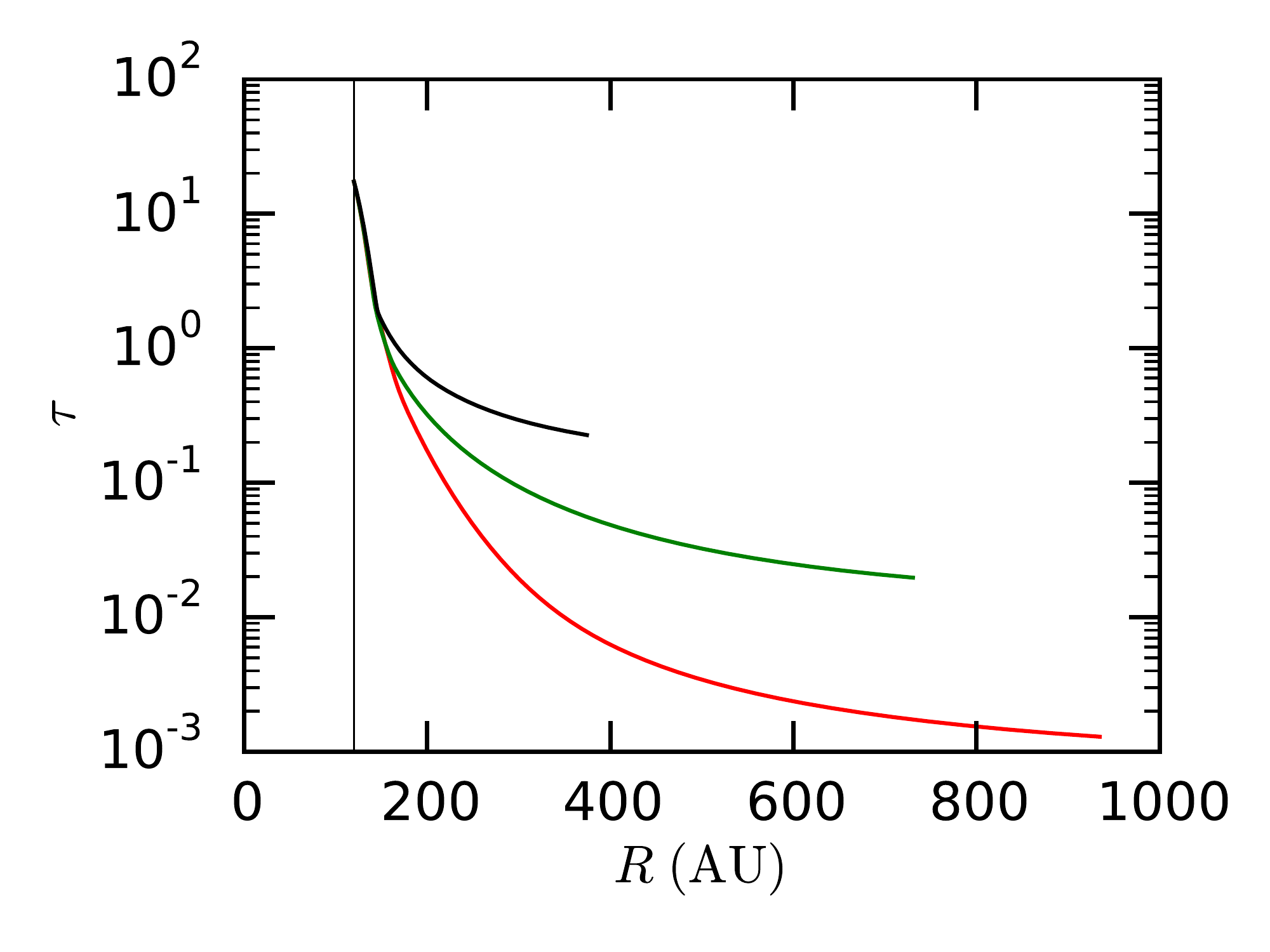}
\includegraphics[width=\columnwidth]{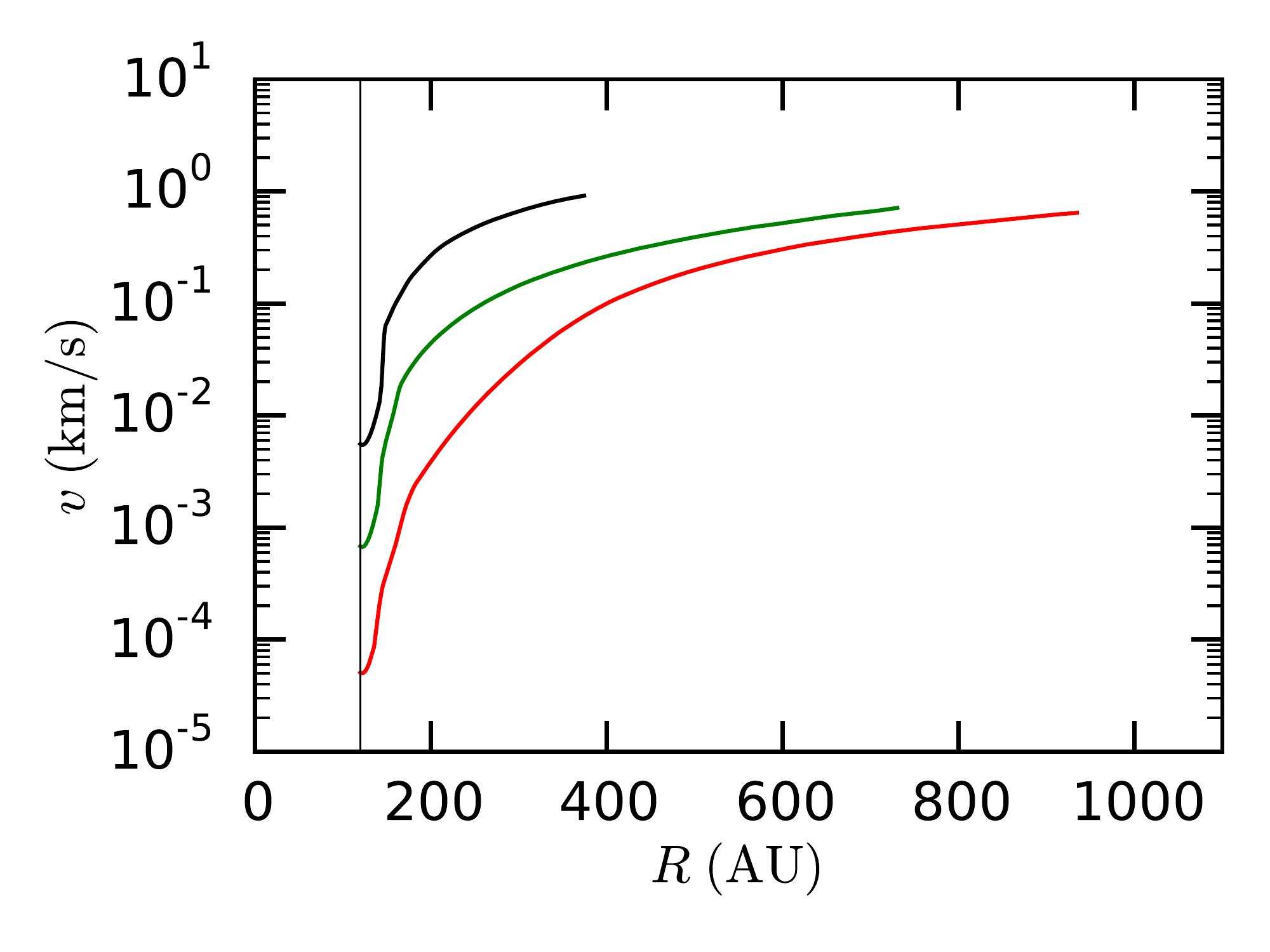}
\end{center}
\caption{Profiles of the main thermo-hydrodynamical gaseous quantities for wind solutions between $R_{\rm d}=120$ (highlighted by the black vertical line) and $R_{\rm c}$, with $s_{\rm max}=1\,$mm. Colours indicate the three ambient field intensities analysed in the paper (red: $30\,G_0$; green: $300\,G_0$; black: $3000\,G_0$). The solutions were chosen as to have $P_{\rm d}\sim10^{-6}$ in cgs units. The dashed lines represent the fluid quantities in the discs.
}
\label{fig:profiles}
\end{figure*}

We report here the global properties of the final solutions we have obtained. By construction, we obtain a class of {\it transcritical} solutions with a critical point in the heated flow. This implies that the velocity at the outer edge of the disc is subsonic, and thus all the solutions we obtain lie in the subcritical regime.

\subsection{Profiles}
\label{subsec:profiles}

The radial profiles of the gaseous quantities have characteristic features. Three examples are reported in Fig. \ref{fig:profiles}, where we show the steady-state solutions for discs with $R_{\rm d}=120$\,AU, $s_{\rm max}=1\,$mm, and $G_{\rm FUV}=30$, $300$ and $3000\,G_0$ (red, green and black lines). The solutions have been chosen such to have $P_{\rm d}\sim10^{-6}$ in cgs units. The dashed lines indicate the fluid quantities in the disc, parametrised by the temperature and surface density profiles given by relations \ref{eq:t_disc} and \ref{eq:surf_dens}. The wind solutions are shown out to the critical radius, which varies with the intensity of the external FUV field. 

In the outer regions of the photoevaporative wind, the temperature increases with radius, as the flow becomes optically thinner to the external FUV radiation. Approaching the disc outer rim, the temperature drops drastically, as the gas becomes optically thick. In this radially narrow region, the gas is close to an isothermal hydrostatic solution, where the gas density $n$ grows exponentially, and the velocity drops accordingly. As we go inwards in the radial direction, very close to the disc outer edge, the velocity has a null gradient, which is expected for an isothermal Parker wind with the addition of the centrifugal term. Note that in all
the solutions shown the optical depth at the critical point is less
than unity (though only marginally so in the case $G_{\rm FUV} = 3000\,G_0$). This justifies a posteriori our assumption that the flow is roughly isothermal and at constant velocity outward of the critical point.

\begin{figure}
\center
\includegraphics[width=\columnwidth]{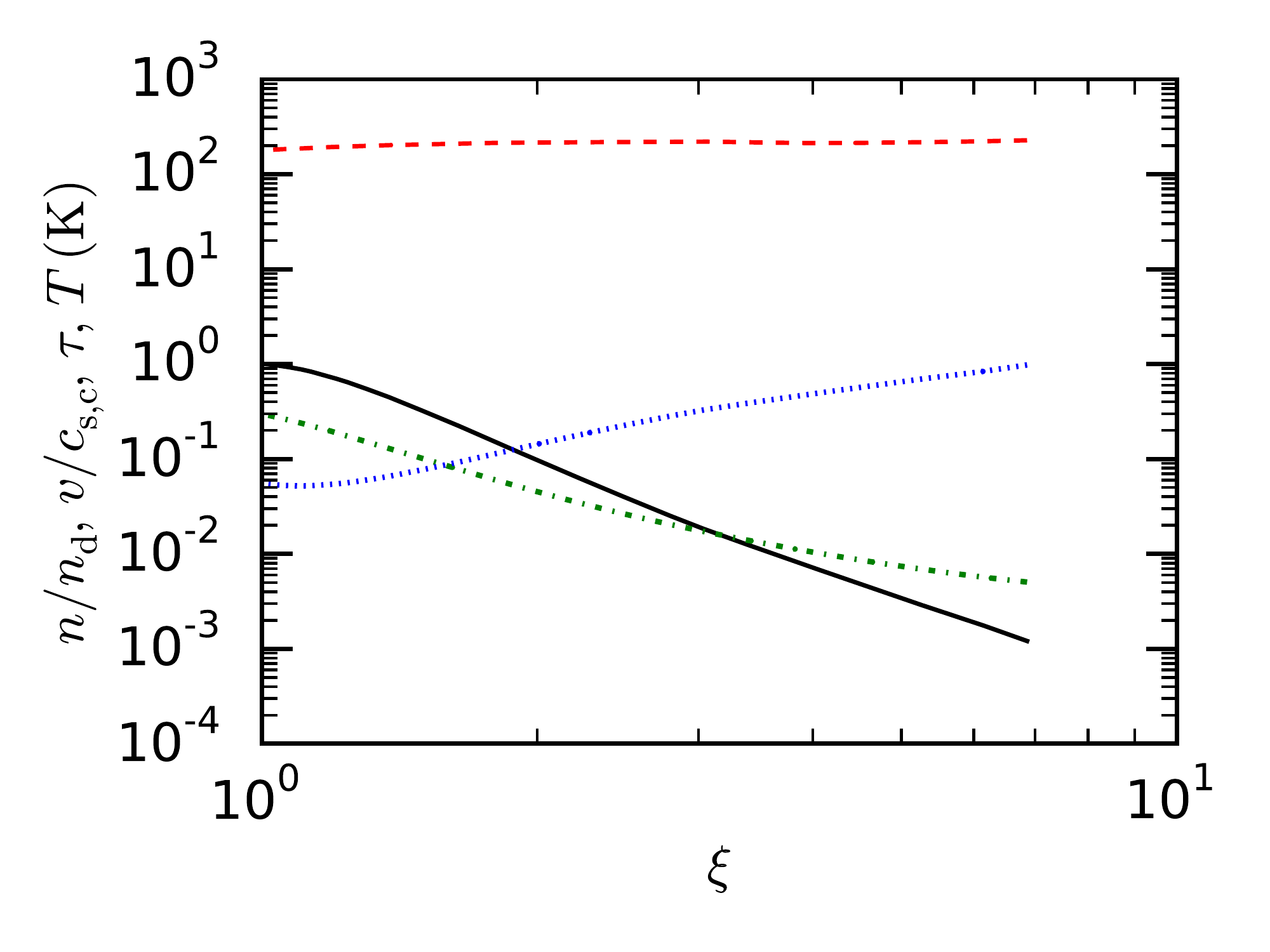}
\caption{Profiles between $R_{\rm d}$ and $R_{\rm c}$ of some relevant quantities for a disc with size $R_{\rm d}=40$\,AU, $G_{\rm FUV}=3000\,G_0$, $s_{\rm max}=1\,$mm and $T_{\rm c}=228$\,K. This solution is associated with $\dot{M}=4.52\times10^{-9}\,M_\odot/$yr, $P_{\rm d}=4.93\times10^{-7}$ in cgs units, $n_{\rm d}=1.98\times10^{-7}$ particles per cc, and $s_{\rm entr}\approx1.7\,\mu$m. Different lines indicate: black solid line:  $n/n_{\rm d}$; red dashed line: $T$\,(K); blue dotted line: $v/c_{\rm s,c}$; green dashed-dotted line: $\tau$.
}
\label{fig:profile_thin}
\end{figure}

Other solutions have profiles that are optically thin to the FUV radiation along the whole flow. These solutions are characterised by a contact discontinuity in both temperature and density at the base of the flow (but the wind at $R_{\rm d}$ is still in pressure equilibrium with the disc). An example is shown in Fig. \ref{fig:profile_thin}. The density structure is close to being a power law profile \citepalias[as in fig. 3 by][]{adams_04}, apart from the regions close to the disc edge, where the centrifugal term in the momentum equation becomes dominant.  As we see in the next section, these solutions are characterised by a low disc mass and a tenuous mass loss.

\subsection{Mass loss rates}

For any given set of initial parameters $G_{\rm FUV}$, $R_{\rm d}$ and $T_{\rm c}$ for which we obtain a solution, we can easily determine the mass loss rate from equation \ref{eq:cont}: this quantity is the most significant outcome of our model, since it determines whether the external photoevaporation mechanism is relevant for the global evolution of a protoplanetary disc, even when the environmental conditions are mild (e.g. $G_{\rm FUV}=30\,G_0$). Note that all the mass loss rate estimates presented in this paper scale linearly with the assumed value of $\mathcal{F}$ (see equation \ref{eq:cont}), i.e. depend on the scale height of the disc at its outer boundary $H_{\rm d}$.

\begin{figure*}
\begin{center}
\includegraphics[width=\columnwidth]{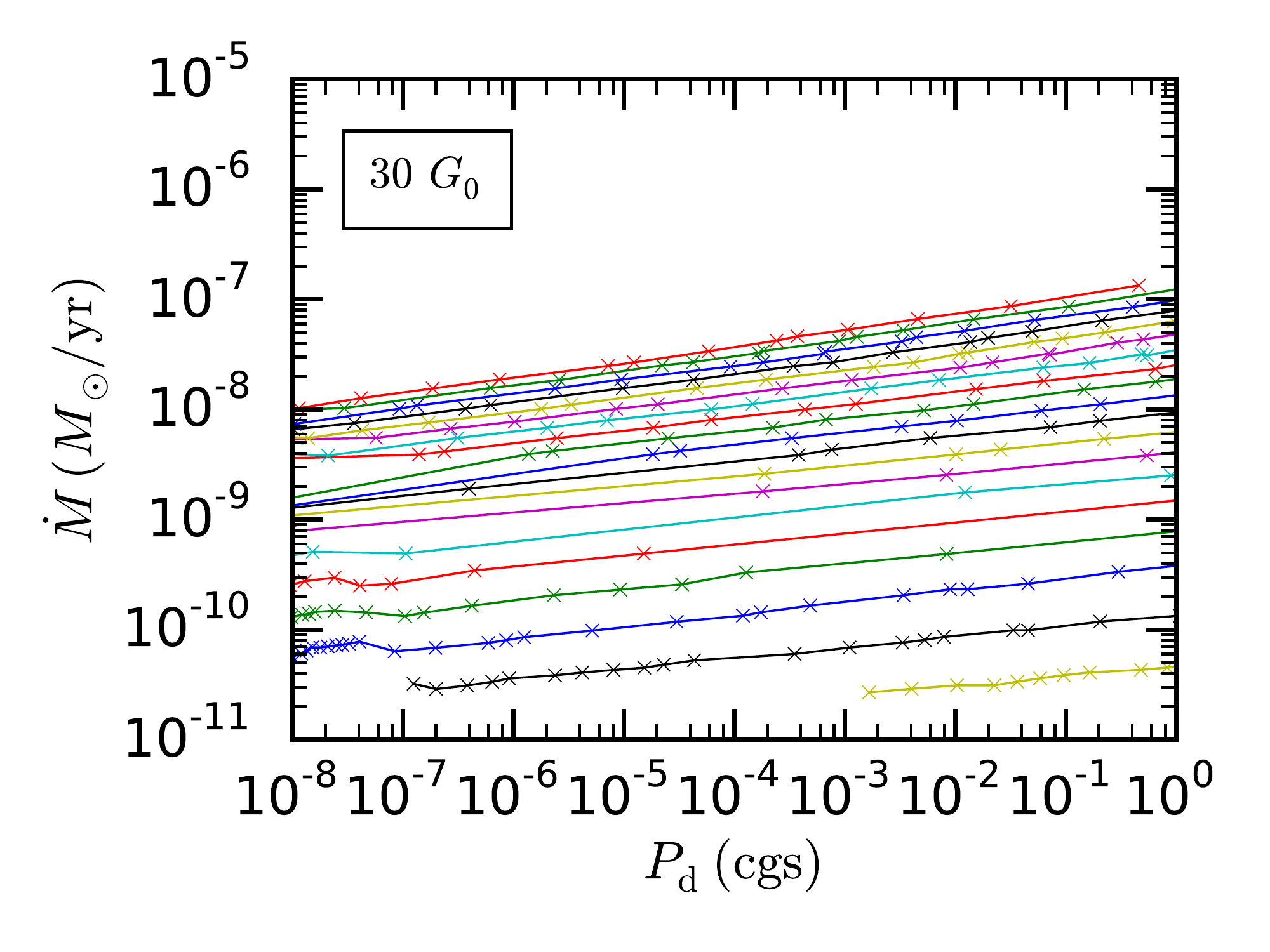}
\includegraphics[width=\columnwidth]{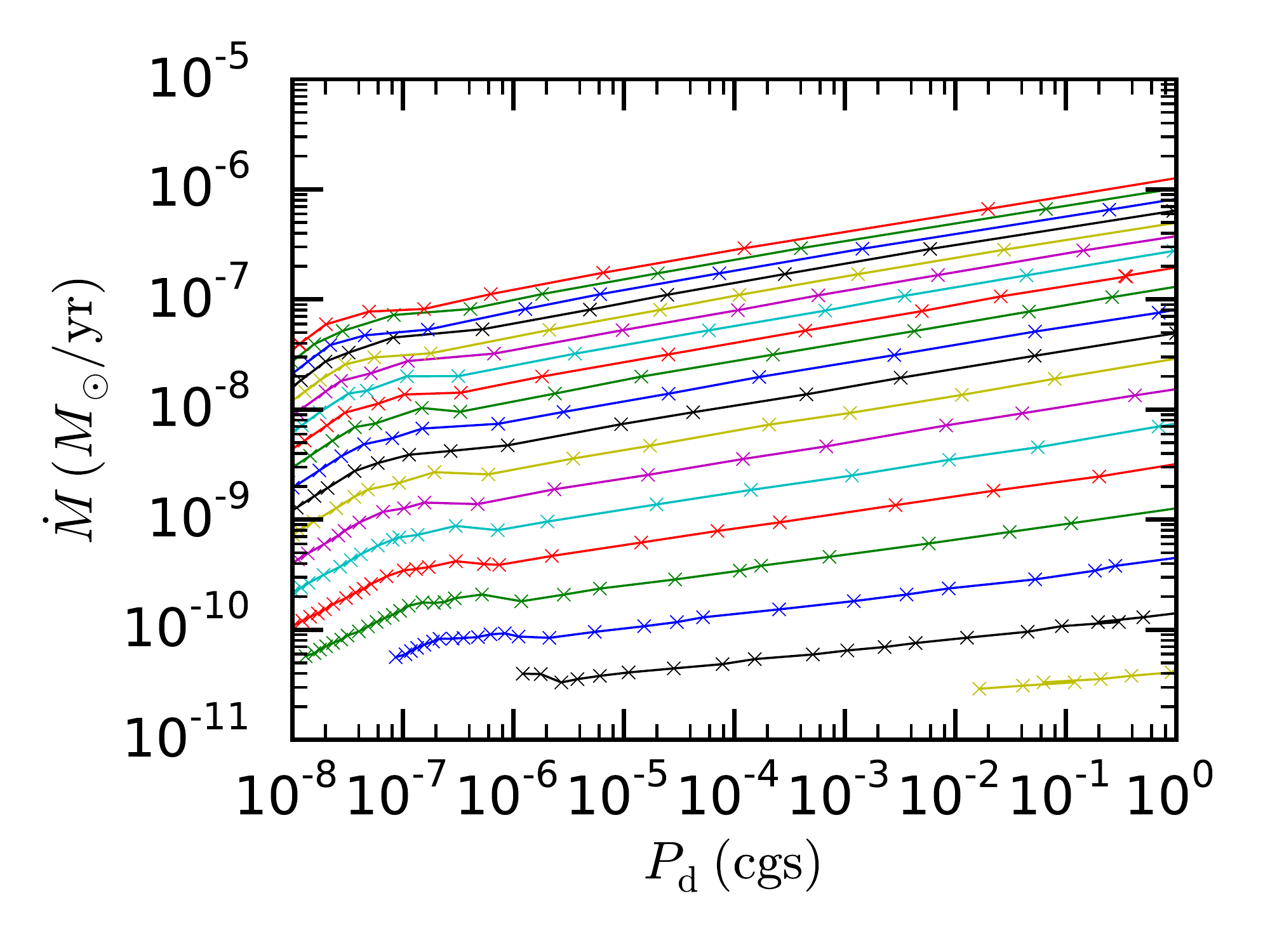}\\
\end{center}
\caption{Mass loss rates vs pressure at the edge of the disc for models with $s_{\rm max}=3.5\,\mu$m (left panel) and $s_{\rm max}=1\,$mm (right panel), with an ambient FUV intensity of $30\,G_0$. Every line represents a different outer radius $R_{\rm d}$, sampled every $10$\,AU, from $R_{\rm d}=70$\,AU (bottom yellow line) to $R_{\rm d}=250$\,AU (top red line). The markers on top of the lines indicate the output of the actual integrations. Typical values of pressure at the outer edge of protoplanetary discs range between $10^{-8}-10^{-3}$, in cgs units.
}
\label{fig:massloss_pres}
\end{figure*}

As an example, we condense most of the mass loss rates we obtain for both grain size distributions and $G_{\rm FUV}=30\,G_0$ in Fig. \ref{fig:massloss_pres}. We show them with their dependence on the pressure at the outer disc edge, as determined from the integration of the hydrodynamic equations. In the plots we focus on the range of pressure values that typify protoplanetary discs (which generally have midplane values of between $10^{-8}-10^{-3}$ in cgs units depending on the disc mass, for $R>20$\,AU). The $\times$-symbols indicate the actual outputs of the integration. Every line is associated with a disc outer radius, ranging from $70$ to $250$\,AU. For the smallest disc sizes ($R_{\rm d} < 70$\,AU), the class of solutions is truncated at high $P_{\rm d}$ values, and they do not appear on this plot. Finally, we struggle to obtain solutions in a physical range for the $s_{\rm max}=3.5\,\mu$m case, when $G_{\rm FUV}=3000\,G_0$ and the discs are larger than $\sim60\,$AU. This limit is caused by our method of solution, since in these cases equation \ref{eq:crit_rad} for the critical radius does not have a real root.

Mass loss rates increase with disc outer radius, as one would intuitively expect, since the material at the outer rim is less embedded in the gravitational potential well of the central star. Moreover, the mass loss rates also increase with external FUV intensity, since higher fluxes heat the gas up to higher temperatures. In most of the parameter space shown in Fig. \ref{fig:massloss_pres}, there is a very weak dependence of the mass loss rates on the pressure in the outer disc (which is related, for canonical parameters, to the total disc mass). The only exception is in the case of low values of $P_{\rm d}$. The reason is the following: a lower $\dot{M}$ corresponds to a roughly optically thin (and thus isothermal) flow (see e.g. Fig. \ref{fig:profile_thin}). The flow solution is then just a centrifugally modified Parker wind solution. In this case the velocity structure is independent of the density normalisation. For a disc lying in this regime, with fixed $R_{\rm d}$, the density normalisation at the base of the wind (and hence the mass loss rate) just scales linearly with the density in the disc. Hence (for a given disc temperature at $R_{\rm d}$), $\dot{M}$ is linear with pressure.

\begin{figure*}
\center
\includegraphics[width=\columnwidth]{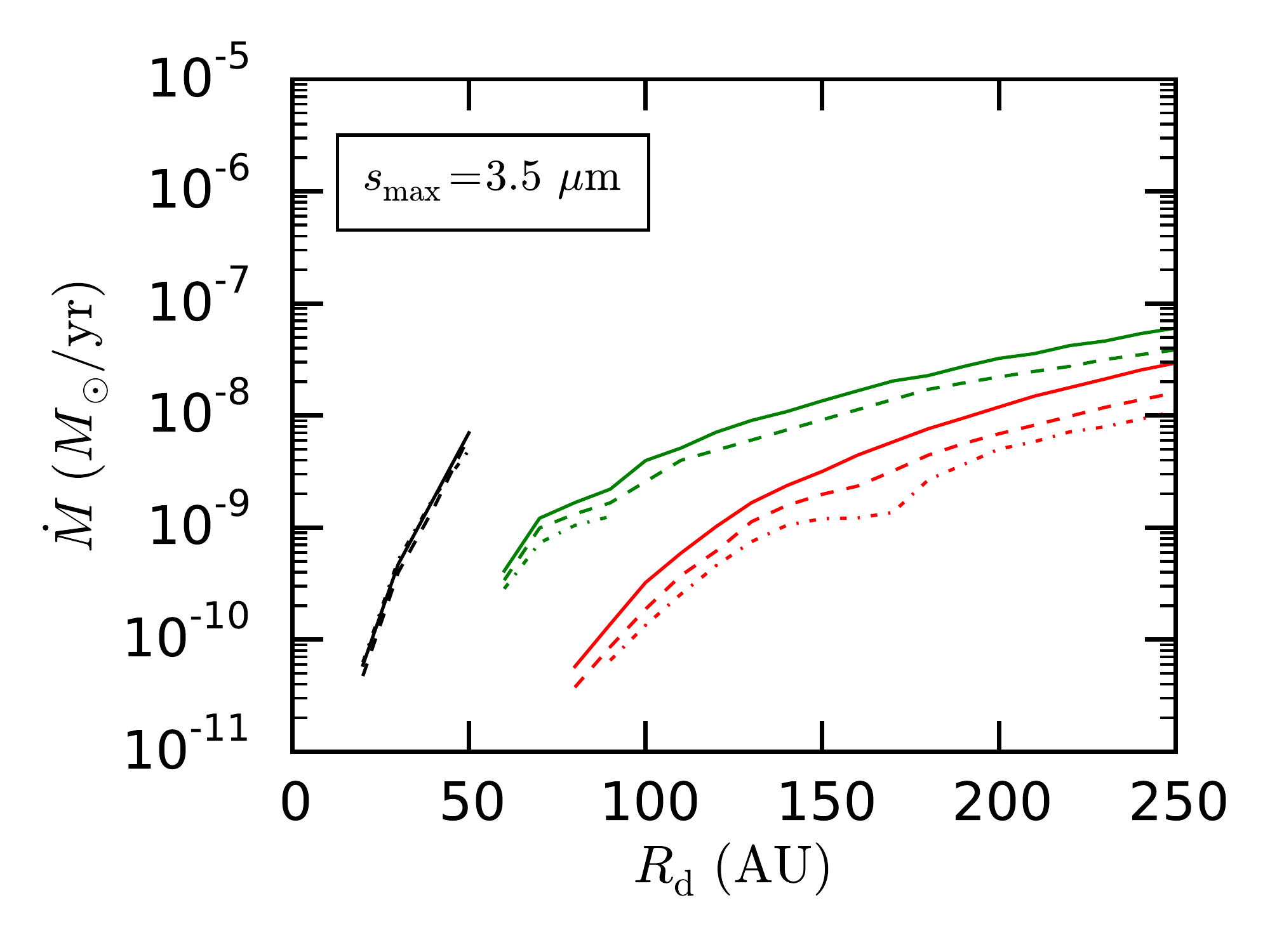}
\includegraphics[width=\columnwidth]{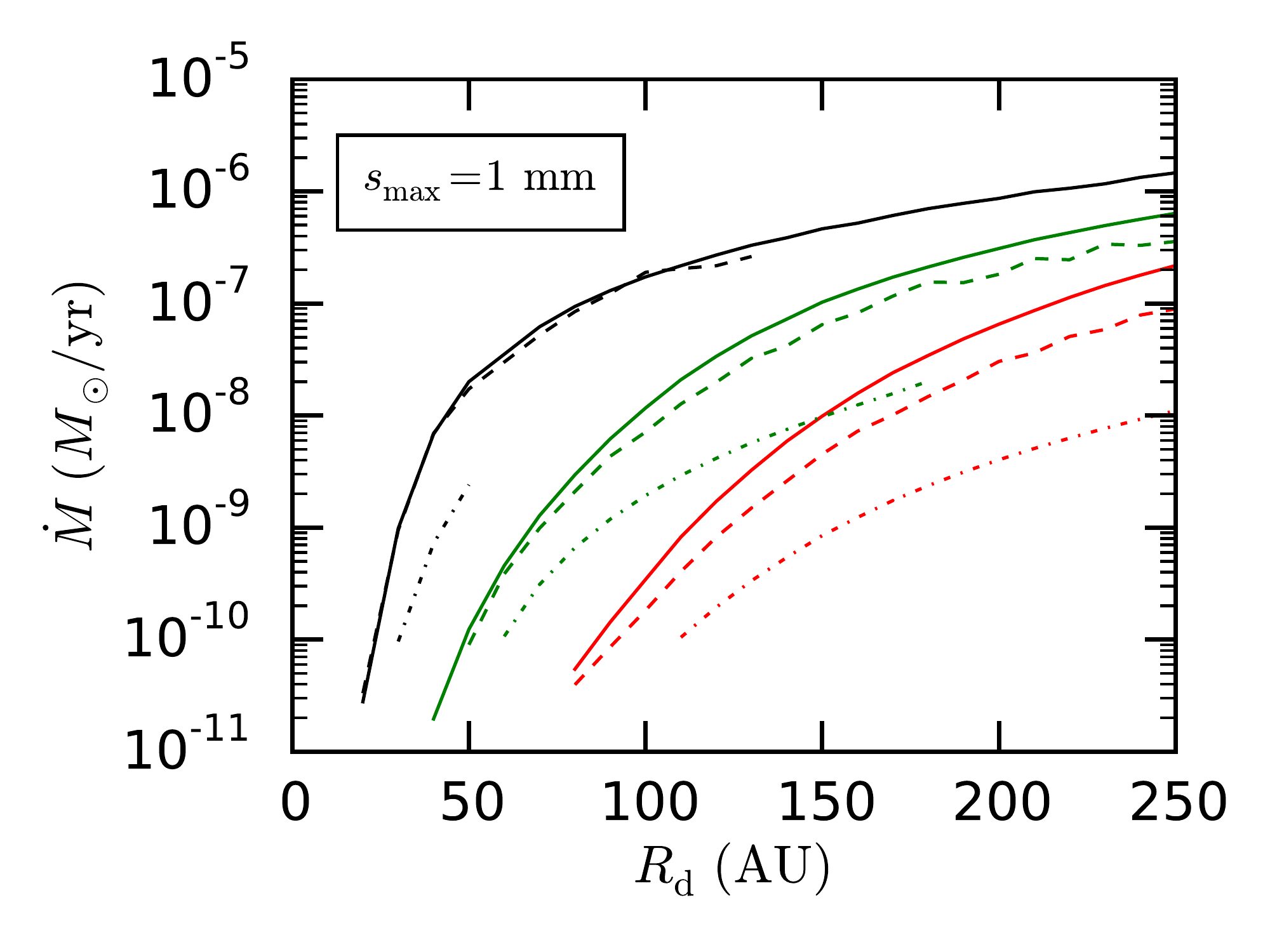}
\caption{The dependence of mass loss rate on disc outer radius, for the two different grain size distributions with $s_{\rm max}=3.5\,\mu$m (left panel) and $1\,$mm (right panel). Different colours (red, green and black) indicate different external FUV field intensities, respectively $G_{\rm FUV}=30$, $300$ and $3000\,G_0$. We show mass loss rates for disc masses $M_{\rm d}=M_{{\rm d},0}(R_{\rm d}/250\,{\rm AU})^2$, where $M_{{\rm d},0}/M_\odot=1$, $10^{-2}$ and $10^{-4}$ (solid, dashed and dashed-dotted lines, respectively).}
\label{fig:mass_loss_vs_rad}
\end{figure*}

\begin{figure*}
\center
\includegraphics[width=\columnwidth]{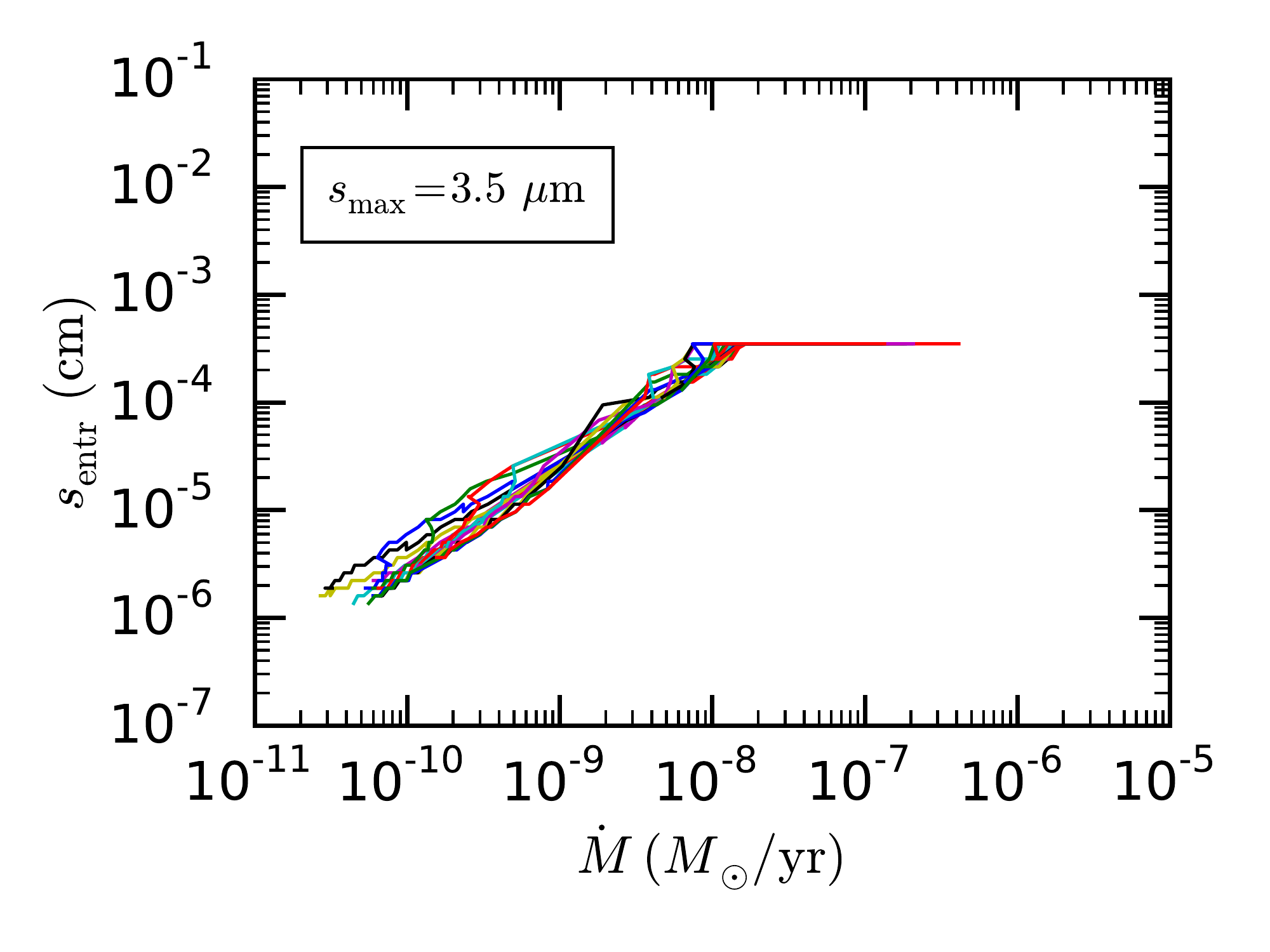}
\includegraphics[width=\columnwidth]{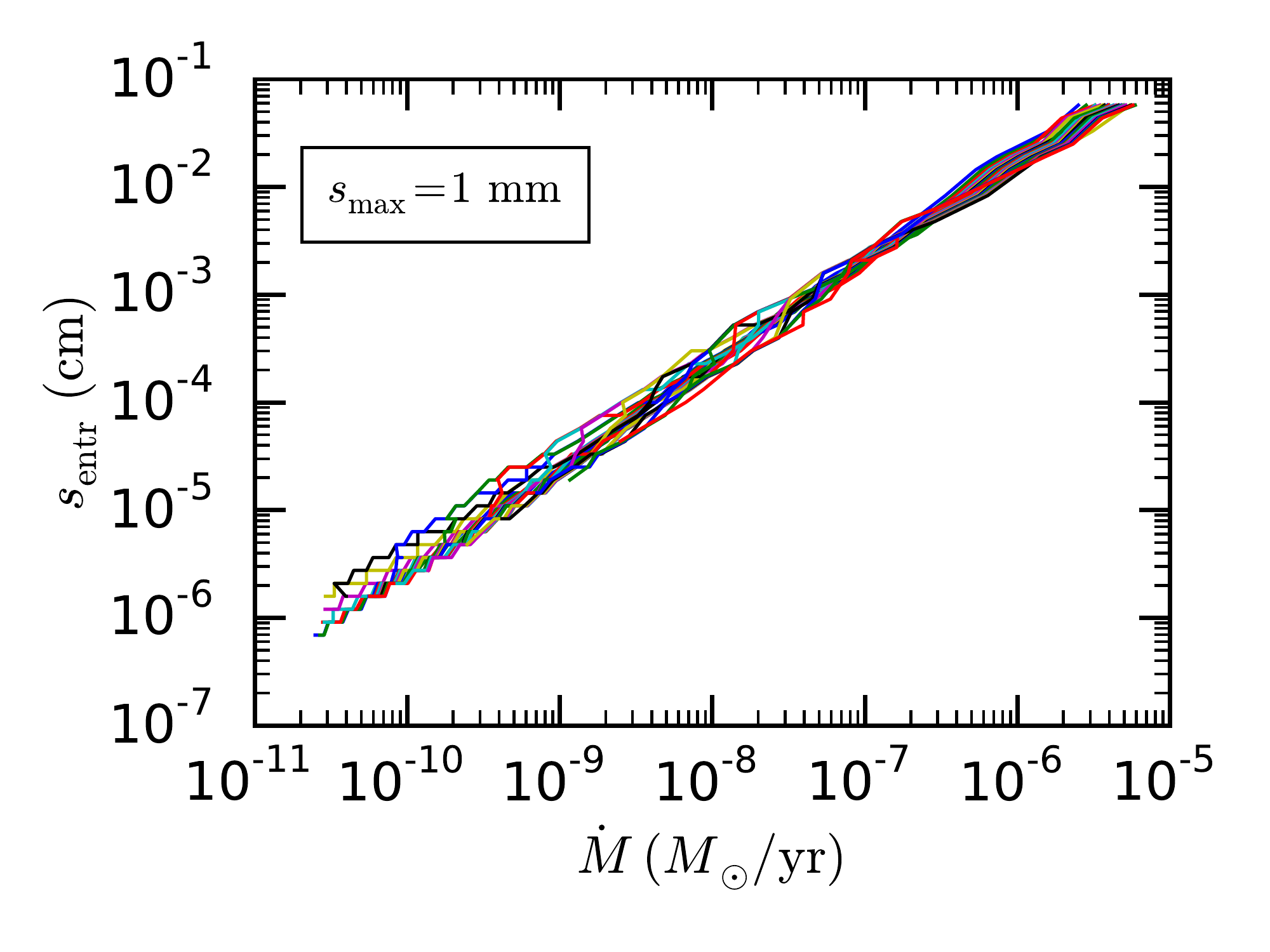}\\
\includegraphics[width=\columnwidth]{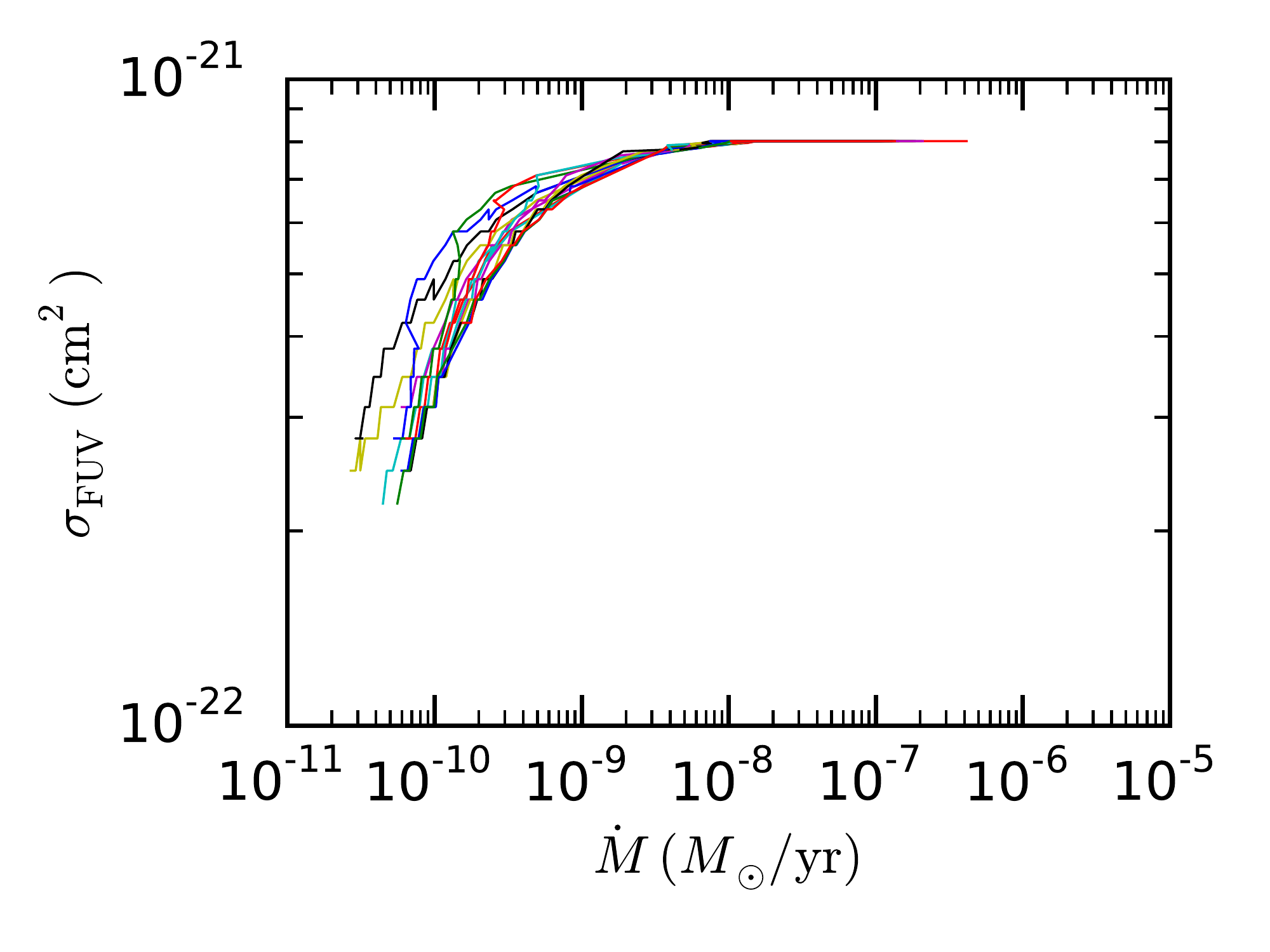}
\includegraphics[width=\columnwidth]{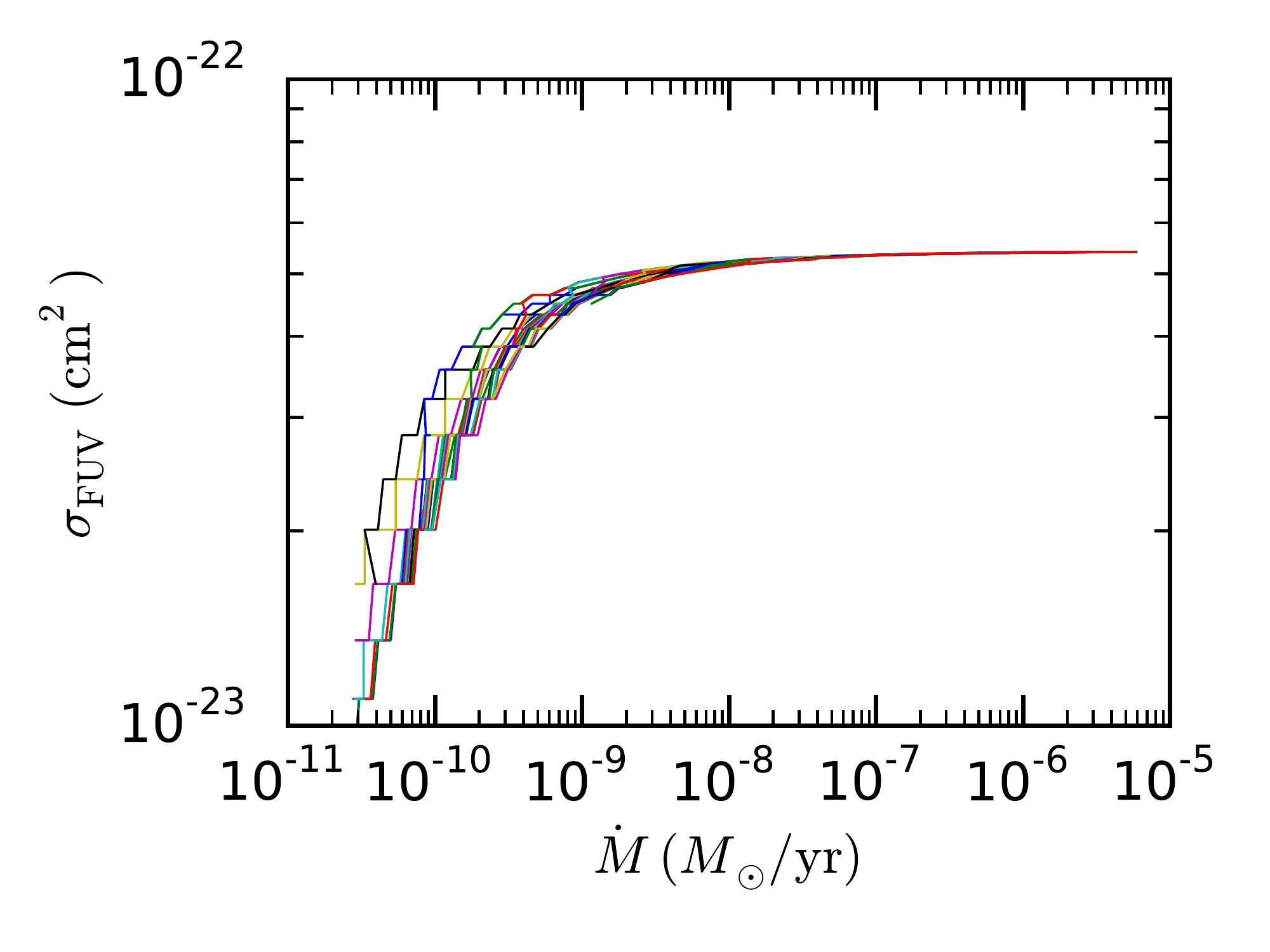}
\caption{Top panels: maximum grain size entrained in the photoevaporative wind, in the case of $G_{\rm FUV}=30\,G_0$. As usual, every line represents a different disc size. There is a strong correlation between $s_{\rm entr}$ and the mass loss rates: the more vigorous is the flow, the larger the entrained maximum grains are. Bottom panels: cross sections of the same solutions reported in the top panels. Note the different range on the $y$-axis. As expected, the cross sections become very sensitive to $s_{\rm entr}$ when $s_{\rm entr}\lesssim \lambda = 0.1\,\mu$m, as shown in Fig. \ref{fig:cross_section_input}.}
\label{fig:max_grain_size}
\end{figure*}

Fig. \ref{fig:mass_loss_vs_rad} shows the same results reported in Fig. \ref{fig:massloss_pres} for the three values of $G_{\rm FUV}=30,\ 300$ and $3000\,G_0$, where the mass loss rates are shown as a function of the disc outer radius, for different disc masses. The masses have been chosen such that $M_{\rm d}=M_{{\rm d},0}(R_{\rm d}/250\,{\rm AU})^2$ in order to maintain the same surface density normalisation in the disc, where $M_{{\rm d},0}/M_\odot=1$, $10^{-2}$ and $10^{-4}$ (solid, dashed and dashed-dotted lines, respectively). It is again apparent that the mass loss rates increase both with disc outer radius, and with the intensity of the impinging FUV radiation. As before we see a weak dependence on disc mass apart from where this is low, where the mass loss rate is nearly linear with disc mass at given $R_{\rm d}$.

Another important feature is that the mass loss rates are very sensitive to the grain size distribution for large discs. When there is substantial grain growth, the cross section $\sigma_{\rm FUV}$ decreases significantly (see Fig. \ref{fig:cross_section_input}), allowing a larger penetrating depth in the photoevaporative wind, and therefore a more significant mass loss rate. This is clearly shown in Fig. \ref{fig:mass_loss_vs_rad}, where the mass loss rates from large discs are higher for the grain size distribution with larger maximum grain size, by roughly the ratio of the cross sections, as expected. However, the mass loss rates associated with the two grain size distributions are comparable when the discs are small ($R_{\rm d}\lesssim100$ when $G_{\rm FUV}=30\,G_0$, $R_{\rm d}\lesssim70$ when $G_{\rm FUV}=300\,G_0$).

The fact that the gaseous flow is depleted of the larger grains has a relevant impact for the lowest mass loss rates, where $s_{\rm entr}\lesssim \lambda = 0.1\,\mu$m. There is in fact a strong correlation between the maximum grain size entrained in the flow, and the mass loss rate due to the photoevaporative wind. The top panels of Fig. \ref{fig:max_grain_size} report such correlation for all the obtained solutions. As usual, every line represents a different disc size. The empirical understanding is straightforward: the more vigorous is the wind, the larger are the grains that are dragged out with the gaseous flow. Obviously the maximum grain size entrained in the flow cannot be larger the the maximum grain size of the whole distribution, as well represented by the upper limit of $3.5\,\mu$m in the left panel of Fig. \ref{fig:max_grain_size}. If we neglect the centrifugal term in equation \ref{eq:dust} (or in equation \ref{eq:feff}), and we set $f_{\rm eff}=0$ to determine $s_{\rm entr}$, we obtain:

\begin{equation}
s_{\rm entr} \approx \frac{v_{\rm th}}{GM_*}\frac{1}{4\pi\mathcal{F}\bar{\rho}}\dot{M},
\end{equation}
where we have used the continuity equation \ref{eq:cont} to extract $\dot{M}$. Since $v_{\rm th}$ weakly depends on temperature only, which does not vary substantially among different solutions, the relation between $s_{\rm entr}$ and $\dot{M}$ is roughly linear, as apparent in Fig. \ref{fig:max_grain_size}. The cross sections are significantly affected only at the low end of the mass loss rates (see bottom panel of Fig. \ref{fig:max_grain_size}), where the wind is so tenuous that $s_{\rm entr}\lesssim \lambda = 0.1\,\mu$m, and the cross section drops significantly.

For the $s_{\rm max}=3.5\,\mu$m case, the mass loss rates we obtain are generally lower (by a factor of $\sim10$) than the ones derived by \citetalias{adams_04} (see their fig. 3), in the regions of parameter space that overlap, i.e. for small discs ($R_{\rm d}\lesssim60$\,AU). Since we have used a very similar cross section in this case, the difference cannot be due to a different penetrating depth. The factors that could be at the origin of such difference are at least two. The first one is the lower temperature in our models (see Section \ref{sec:pdr}). In order to check whether the temperature has a significant effect in determining the mass loss rates, we have obtained other solutions with higher temperatures, of the same order as the ones presented in fig. 3 by \citetalias{adams_04} by using the prescription by \citet{1994ApJ...427..822B} for the PAH photoelectric heating. The mass loss rates are not affected significantly, thus we can rule out that the lower mass loss rates in our study are due to the lower temperatures. The second factor, which is the most important one, is that in our non-isothermal treatment, the critical radius is always larger than the sonic radius by a factor of a few. Since the density scales very steeply with radius (more steeply than $R^{-2}$), a larger outer radius as boundary condition implies a much lower density $n_{\rm c}$, in order to have a final pressure-balanced solution at $R_{\rm d}$. This implies a lower mass loss rate (see equation \ref{eq:cont}).

\section{Discussion}
\label{sec:discussion}

The mass loss rates shown in Fig. \ref{fig:mass_loss_vs_rad} indicate that external photoevaporation can have a significant impact in the global evolution of protoplanetary discs. In particular, large discs ($R_{\rm d}>150$\,AU) show vigorous mass losses even for very low ambient fields ($G_{\rm FUV}=30\,G_0$). Depending on the grain size distribution of the dusty material entrained in the flow, these discs can have mass loss rates as high as $\sim10^{-7}M_\odot/$yr, if grain growth has been effective even at these outer regions of the disc. Thus, this effect needs to be considered when we model the evolution of protoplanetary discs.

\citet{2007MNRAS.376.1350C}, and later \citet{2013ApJ...774....9A} \citep[see also][]{2010ApJ...722.1115M}, have shown that these photoevaporative winds affect disc evolution not on account of the mass lost in the flow, but by limiting the viscous spreading of the disc, and thus accelerating disc clearing. When the discs are large, the photoevaporative wind removes mass from the outer edge on a timescale shorter than the timescale with which the disc is able to replenish the mass in these regions. Thus the disc shrinks, until it reaches a radius where these two timescales are equal \citepalias[see also][for a similar discussion]{adams_04}. Another way of looking at the same mechanism is that when the mass flux at the base of the flow is higher than the one within the disc due to viscous spreading, the disc shrinks, whereas it expands as a standard viscous disc when the opposite happens. \citet{2013ApJ...774....9A} already noticed that FUV external photoevaporation can have a significant impact in reducing the lifetime of protoplanetary discs for moderate external field intensities ($G_{\rm FUV}=300\,G_0$). Both \citet{2007MNRAS.376.1350C} and \citet{2013ApJ...774....9A} used the mass loss rates derived by \citetalias{adams_04}. It will be worth investigating how the disc evolution will be affected by the lower mass loss rates we derived for compact discs, and the much higher ones when grain growth has occurred. The high mass loss rates we obtain in the latter case suggest that external photoevaporation might be a dominant mechanism in determining the outer radius of protoplanetary discs even in very mild environments, once large dust grains have formed in the discs' outer regions.

In principle, in order to test this model in the low ambient field regime, we could follow two paths. The first one is statistical. For example, we could sample the outer radii of protoplanetary discs in different star forming regions, where the external FUV field can be considered quite uniform within the whole star forming region. Since \citet{2007MNRAS.376.1350C} showed that present disc sizes are very insensitive to the initial disc sizes when external photoevaporation is effective, in principle there should be a correlation between the present average disc size, and the external FUV flux. Note however it is very challenging to obtain good estimates of disc sizes (although ALMA is now overcoming the problem), and even more it is very difficult to measure the FUV field in these regions. A prediction for the FUV intensity within the cluster can be estimated by integrating the contribution from all the known radiation sources, even though the 3D structure of these regions is still not known. Another theoretical uncertainty is that the expected outer radii strongly depend on the assumed viscosity in the disc, and on its initial mass. Note that \citet{2013ApJ...774....9A} have already conducted a similar exploratory study for the ONC, where the FUV radiation within the cluster can be estimated well, since it is dominated by the radiation of the massive $\theta^1$C star.

A second approach is to observe this photoevaporative flow directly in discs where we know that the external radiation is mild, since there are no close energetic sources. Examination of Fig. \ref{fig:profiles} suggests that the surface densities in the flow can be high enough ($\sim10^{-3}-10^{-2}$\,g/cm$^2$) to be detectable in molecular line emission (e.g. $^{12}$CO) with reasonable sensitivities. The outer regions are typified by a steep cliff in surface density surrounded by a gently sloping plateau of low density gas. The same regions would also be highly depleted in large grains ($s_{\rm entr}\lesssim1\,\mu$m), and thus the disc outer radius estimated from dust continuum emission will be smaller than the one traced by molecular line emission. A good example of this feature is the huge 114-426 protoplanetary disc observed in silhouette by HST in Orion \citep{2008AJ....136.2136R}, where the outer regions associated with a photoevaporative flow \citep[e.g.][]{2012ApJ...757...78M} are not observed in dust continuum emission at submm wavelengths \citep[][]{2014ApJ...784...82M,2015ApJ...808...69B}, and present hints of radial gradient in the maximum grain size entrained in the flow \citep{2012ApJ...757...78M}. Note however the in this case the disc is so large ($R_{\rm d}\sim1000$\,AU) that it is expected to lie in the supercritical regime of FUV photoevaporation.

The presence of these dust depleted structures in the outskirts of protoplanetary discs can be interpreted as the outcome of other physical processes, such as the radial drift of dust particles in the absence of a wind \citep{birnstiel_14}. However, discs affected by external photoevaporation will present other two features that are not compatible with the pure radial drift scenario. The first one is a positive temperature gradient in the regions associated with the flow. The second signature is kinematic: since the specific angular momentum in the flow equals the specific angular momentum at $R_{\rm d}$, the angular velocity of the flow will scale as $R^{-2}$, instead of the typical Keplerian $R^{-3/2}$. The observability of this feature favours small discs, where the absolute deviation from a Keplerian velocity in the flow will be more significant. The radial motions of the photoevaporative flow will however be hardly detectable, since the wind reaches velocities of $\sim1$\,km/s at the critical radius where molecular gas is likely to be highly photodissociated.

\section{Conclusions}
\label{sec:concl}

In this paper we present new mass loss rates of externally photoevaporated discs covering a large parameter space. In particular, we are able to self-consistently solve the hydrodynamical equations of the photoevaporative winds  for large discs (up to $R_{\rm d}=250$\,AU) and low FUV fluxes (down to $G_{\rm FUV}=30\,G_0$), where previous studies have been unable to obtain numerical solutions. We have shown that the method of solution can significantly affect the final estimates of the mass loss rates. More specifically, by properly treating the non-isothermal nature of the flow, we obtain lower mass loss rates than had been previously determined for an ISM-like dust component of the disc.

We have shown that the grain size distribution of the dust component of the disc has a significant effect on the mass loss rate. When grain growth has occurred in the midplane of the disc, out to its outer radius, the mass loss is remarkably more vigorous, mostly because of the consequent reduction of the cross section of dust particles at FUV wavelengths. Note that we have assumed that grain growth does not affect the PAH abundance, where PAHs are the most significant heating source at almost all gas densities and optical depths considered here. Thus our assumption implies that the total gas heating is not significantly affected by grain growth (see Sections \ref{subsection:pdr_dust} - \ref{subsec:pdr_results}). For a moderate maximum size of the dust grain size distribution ($s_{\rm max}=1\,$mm), and a standard dust to gas ratio of $0.01$, we obtain a mass loss rate of $\dot{M}>10^{-8}M_\odot/$yr when $R_{\rm d}>150$\,AU and $G_{\rm FUV}=30\,G_0$, and $M_{\rm d}\gtrsim10^{-3}M_\odot$. For the same parameters the mass loss rate can be as high as $10^{-7}M_\odot/$yr when $R_{\rm d}\sim250$\,AU. These high mass loss rates are expected to significantly affect the global evolution of protoplanetary discs. Our results indicate that even very mild environments can lead to significant truncation of the disc outer radius, especially when grain growth is effective. Such truncation is also expected to yield an acceleration of disc clearing. {\it We thus predict} that external photoevaporation is a more significant mechanism leading to disc clearing than previously considered. For such low values of the external field, future disc evolution models should include photoevaporation caused by both the external FUV field and the FUV flux from the central star \citep{2009ApJ...690.1539G,2015ApJ...804...29G}, as suggested by \citet{2013ApJ...774....9A}.

We have suggested observable characteristic features that can probe a currently photoevaporating disc. In particular, our results show that photoevaporative winds are highly dust depleted, and that their surface densities can be observed in molecular line emission with present facilities (ALMA in particular). Strong evidence of ongoing external photoevaporation in this mild regime is a positive thermal radial gradient in the flow region, which can be probed by spatially resolved line ratios of different rovibrational transitions, and a non-Keplerian rotation curve that can be obtained via line kinematics.

Finally we emphasise that the present work is restricted to
solar mass stars. As noted by \citetalias{adams_04}, we expect the effect of external photoevaporation to be yet more significant in the case of lower mass stars since the gravitational well is shallower.

\section*{Acknowledgements}
We thank the anonymous referee whose comments helped improving the quality of the paper and the description of the PDR modelling. We are grateful to Mark Wyatt for providing us the code to compute the dust opacities. We thank James Owen, Serena Viti, Richard Alexander, Grant Kennedy and Ewine van Dishoeck for stimulating discussion. We also thank Benedetta Caggioni for helping with Fig. \ref{fig:cartoon}. SF thanks the Science and Technology Facility Council and the Isaac Newton Trust for the award of a studentship. This work has been supported by the DISCSIM project, grant agreement 341137 funded by the European Research Council under ERC-2013-ADG. The work of TGB was funded by STFC grant ST/J001511. All the figures were generated with the \textsc{python}-based package \textsc{matplotlib} \citep{2007CSE.....9...90H}.

\bibliography{photo_bib}

\begin{thebibliography}{102}
\expandafter\ifx\csname natexlab\endcsname\relax\def\natexlab#1{#1}\fi

\bibitem[{{Adams}(2010)}]{2010ARA&A..48...47A}
{Adams} F.~C., 2010, \araa, 48, 47

\bibitem[{{Adams} {et~al}\mbox{.}(2004){Adams}, {Hollenbach}, {Laughlin}, \&
  {Gorti}}]{adams_04}
{Adams} F.~C., {Hollenbach} D., {Laughlin} G., {Gorti} U., 2004, \apj, 611, 360

\bibitem[{{Adams} {et~al}\mbox{.}(2006){Adams}, {Proszkow}, {Fatuzzo}, \&
  {Myers}}]{2006ApJ...641..504A}
{Adams} F.~C., {Proszkow} E.~M., {Fatuzzo} M., {Myers} P.~C., 2006, \apj, 641,
  504

\bibitem[{{Allen} {et~al}\mbox{.}(2007){Allen}, {Megeath}, {Gutermuth},
  {Myers}, {Wolk}, {Adams}, {Muzerolle}, {Young}, \&
  {Pipher}}]{2007prpl.conf..361A}
{Allen} L. {et~al.}, 2007, Protostars and Planets V, 361

\bibitem[{{Anderson}, {Adams} \& {Calvet}(2013){Anderson}, {Adams}, \&
  {Calvet}}]{2013ApJ...774....9A}
{Anderson} K.~R., {Adams} F.~C., {Calvet} N., 2013, \apj, 774, 9

\bibitem[{{Andrews} \& {Williams}(2005)}]{2005ApJ...631.1134A}
{Andrews} S.~M., {Williams} J.~P., 2005, \apj, 631, 1134

\bibitem[{{Andrews} \& {Williams}(2007)}]{2007ApJ...671.1800A}
{Andrews} S.~M., {Williams} J.~P., 2007, \apj, 671, 1800

\bibitem[{{Andrews} {et~al}\mbox{.}(2012){Andrews}, {Wilner}, {Hughes}, {Qi},
  {Rosenfeld}, {{\"O}berg}, {Birnstiel}, {Espaillat}, {Cieza}, {Williams},
  {Lin}, \& {Ho}}]{2012ApJ...744..162A}
{Andrews} S.~M. {et~al.}, 2012, \apj, 744, 162

\bibitem[{{Armitage}(2010)}]{armitage_book_10}
{Armitage} P.~J., 2010, {Astrophysics of Planet Formation}

\bibitem[{{Asplund} {et~al}\mbox{.}(2009){Asplund}, {Grevesse}, {Sauval}, \&
  {Scott}}]{2009ARA&A..47..481A}
{Asplund} M., {Grevesse} N., {Sauval} A.~J., {Scott} P., 2009, \araa, 47, 481

\bibitem[{{Bakes} \& {Tielens}(1994)}]{1994ApJ...427..822B}
{Bakes} E.~L.~O., {Tielens} A.~G.~G.~M., 1994, \apj, 427, 822

\bibitem[{{Bally} {et~al}\mbox{.}(2015){Bally}, {Mann}, {Eisner}, {Andrews},
  {Di Francesco}, {Hughes}, {Johnstone}, {Matthews}, {Ricci}, \&
  {Williams}}]{2015ApJ...808...69B}
{Bally} J. {et~al.}, 2015, \apj, 808, 69

\bibitem[{{Bally}, {O'Dell} \& {McCaughrean}(2000){Bally}, {O'Dell}, \&
  {McCaughrean}}]{2000AJ....119.2919B}
{Bally} J., {O'Dell} C.~R., {McCaughrean} M.~J., 2000, \aj, 119, 2919

\bibitem[{{Balog} {et~al}\mbox{.}(2007){Balog}, {Muzerolle}, {Rieke}, {Su},
  {Young}, \& {Megeath}}]{2007ApJ...660.1532B}
{Balog} Z., {Muzerolle} J., {Rieke} G.~H., {Su} K.~Y.~L., {Young} E.~T.,
  {Megeath} S.~T., 2007, \apj, 660, 1532

\bibitem[{{Bergin} {et~al}\mbox{.}(2003){Bergin}, {Calvet}, {D'Alessio}, \&
  {Herczeg}}]{2003ApJ...591L.159B}
{Bergin} E., {Calvet} N., {D'Alessio} P., {Herczeg} G.~J., 2003, \apjl, 591,
  L159

\bibitem[{{Birnstiel} \& {Andrews}(2014)}]{birnstiel_14}
{Birnstiel} T., {Andrews} S.~M., 2014, \apj, 780, 153

\bibitem[{{Birnstiel}, {Ormel} \& {Dullemond}(2011){Birnstiel}, {Ormel}, \&
  {Dullemond}}]{2011A&A...525A..11B}
{Birnstiel} T., {Ormel} C.~W., {Dullemond} C.~P., 2011, \aap, 525, A11

\bibitem[{{Bisbas} {et~al}\mbox{.}(2014){Bisbas}, {Bell}, {Viti}, {Barlow},
  {Yates}, \& {Vasta}}]{2014MNRAS.443..111B}
{Bisbas} T.~G., {Bell} T.~A., {Viti} S., {Barlow} M.~J., {Yates} J., {Vasta}
  M., 2014, \mnras, 443, 111

\bibitem[{{Bisbas} {et~al}\mbox{.}(2012){Bisbas}, {Bell}, {Viti}, {Yates}, \&
  {Barlow}}]{bisbas_12}
{Bisbas} T.~G., {Bell} T.~A., {Viti} S., {Yates} J., {Barlow} M.~J., 2012,
  \mnras, 427, 2100

\bibitem[{{Bisbas}, {Papadopoulos} \& {Viti}(2015){Bisbas}, {Papadopoulos}, \&
  {Viti}}]{2015ApJ...803...37B}
{Bisbas} T.~G., {Papadopoulos} P.~P., {Viti} S., 2015, \apj, 803, 37

\bibitem[{{Bohren} \& {Huffman}(1983)}]{1983asls.book.....B}
{Bohren} C.~F., {Huffman} D.~R., 1983, {Absorption and scattering of light by
  small particles}

\bibitem[{{Breslau} {et~al}\mbox{.}(2014){Breslau}, {Steinhausen}, {Vincke}, \&
  {Pfalzner}}]{2014A&A...565A.130B}
{Breslau} A., {Steinhausen} M., {Vincke} K., {Pfalzner} S., 2014, \aap, 565,
  A130

\bibitem[{{Bruderer} {et~al}\mbox{.}(2012){Bruderer}, {van Dishoeck}, {Doty},
  \& {Herczeg}}]{2012A&A...541A..91B}
{Bruderer} S., {van Dishoeck} E.~F., {Doty} S.~D., {Herczeg} G.~J., 2012, \aap,
  541, A91

\bibitem[{{Cabrit} {et~al}\mbox{.}(2006){Cabrit}, {Pety}, {Pesenti}, \&
  {Dougados}}]{2006A&A...452..897C}
{Cabrit} S., {Pety} J., {Pesenti} N., {Dougados} C., 2006, \aap, 452, 897

\bibitem[{{Clarke} \& {Carswell}(2007)}]{clarke_book_07}
{Clarke} C., {Carswell} B., 2007, {Principles of Astrophysical Fluid Dynamics}

\bibitem[{{Clarke}(2007)}]{2007MNRAS.376.1350C}
{Clarke} C.~J., 2007, \mnras, 376, 1350

\bibitem[{{Croxall} {et~al}\mbox{.}(2012){Croxall}, {Smith}, {Wolfire},
  {Roussel}, {Sandstrom}, {Draine}, {Aniano}, {Dale}, {Armus}, {Beir{\~a}o},
  {Helou}, {Bolatto}, {Appleton}, {Brandl}, {Calzetti}, {Crocker}, {Galametz},
  {Groves}, {Hao}, {Hunt}, {Johnson}, {Kennicutt}, {Koda}, {Krause}, {Li},
  {Meidt}, {Murphy}, {Rahman}, {Rix}, {Sauvage}, {Schinnerer}, {Walter}, \&
  {Wilson}}]{2012ApJ...747...81C}
{Croxall} K.~V. {et~al.}, 2012, \apj, 747, 81

\bibitem[{{Dai} {et~al}\mbox{.}(2015){Dai}, {Facchini}, {Clarke}, \&
  {Haworth}}]{2015MNRAS.449.1996D}
{Dai} F., {Facchini} S., {Clarke} C.~J., {Haworth} T.~J., 2015, \mnras, 449,
  1996

\bibitem[{{de Gregorio-Monsalvo} {et~al}\mbox{.}(2013){de Gregorio-Monsalvo},
  {M{\'e}nard}, {Dent}, {Pinte}, {L{\'o}pez}, {Klaassen}, {Hales},
  {Cort{\'e}s}, {Rawlings}, {Tachihara}, {Testi}, {Takahashi}, {Chapillon},
  {Mathews}, {Juhasz}, {Akiyama}, {Higuchi}, {Saito}, {Nyman}, {Phillips},
  {Rod{\'o}n}, {Corder}, \& {Van Kempen}}]{2013A&A...557A.133D}
{de Gregorio-Monsalvo} I. {et~al.}, 2013, \aap, 557, A133

\bibitem[{{de Juan Ovelar} {et~al}\mbox{.}(2012){de Juan Ovelar}, {Kruijssen},
  {Bressert}, {Testi}, {Bastian}, \& {C{\'a}novas}}]{2012A&A...546L...1D}
{de Juan Ovelar} M., {Kruijssen} J.~M.~D., {Bressert} E., {Testi} L., {Bastian}
  N., {C{\'a}novas} H., 2012, \aap, 546, L1

\bibitem[{{Draine}(2011)}]{2011piim.book.....D}
{Draine} B.~T., 2011, {Physics of the Interstellar and Intergalactic Medium}

\bibitem[{{Draine} \& {Li}(2007)}]{2007ApJ...657..810D}
{Draine} B.~T., {Li} A., 2007, \apj, 657, 810

\bibitem[{{Epstein}(1924)}]{1924PhRv...23..710E}
{Epstein} P.~S., 1924, Physical Review, 23, 710

\bibitem[{{Fang} {et~al}\mbox{.}(2012){Fang}, {van Boekel}, {King}, {Henning},
  {Bouwman}, {Doi}, {Okamoto}, {Roccatagliata}, \&
  {Sicilia-Aguilar}}]{2012A&A...539A.119F}
{Fang} M. {et~al.}, 2012, \aap, 539, A119

\bibitem[{{Fatuzzo} \& {Adams}(2008)}]{2008ApJ...675.1361F}
{Fatuzzo} M., {Adams} F.~C., 2008, \apj, 675, 1361

\bibitem[{{Fedele} {et~al}\mbox{.}(2010){Fedele}, {van den Ancker}, {Henning},
  {Jayawardhana}, \& {Oliveira}}]{2010A&A...510A..72F}
{Fedele} D., {van den Ancker} M.~E., {Henning} T., {Jayawardhana} R.,
  {Oliveira} J.~M., 2010, \aap, 510, A72

\bibitem[{{France} {et~al}\mbox{.}(2014){France}, {Schindhelm}, {Bergin},
  {Roueff}, \& {Abgrall}}]{2014ApJ...784..127F}
{France} K., {Schindhelm} E., {Bergin} E.~A., {Roueff} E., {Abgrall} H., 2014,
  \apj, 784, 127

\bibitem[{{Gaches} {et~al}\mbox{.}(2015){Gaches}, {Offner}, {Rosolowsky}, \&
  {Bisbas}}]{2015ApJ...799..235G}
{Gaches} B.~A.~L., {Offner} S.~S.~R., {Rosolowsky} E.~W., {Bisbas} T.~G., 2015,
  \apj, 799, 235

\bibitem[{{Geers} {et~al}\mbox{.}(2006){Geers}, {Augereau}, {Pontoppidan},
  {Dullemond}, {Visser}, {Kessler-Silacci}, {Evans}, {van Dishoeck}, {Blake},
  {Boogert}, {Brown}, {Lahuis}, \& {Mer{\'{\i}}n}}]{2006A&A...459..545G}
{Geers} V.~C. {et~al.}, 2006, \aap, 459, 545

\bibitem[{{Geers} {et~al}\mbox{.}(2007){Geers}, {van Dishoeck}, {Visser},
  {Pontoppidan}, {Augereau}, {Habart}, \& {Lagrange}}]{2007A&A...476..279G}
{Geers} V.~C., {van Dishoeck} E.~F., {Visser} R., {Pontoppidan} K.~M.,
  {Augereau} J.-C., {Habart} E., {Lagrange} A.~M., 2007, \aap, 476, 279

\bibitem[{{Gorti} \& {Hollenbach}(2009)}]{2009ApJ...690.1539G}
{Gorti} U., {Hollenbach} D., 2009, \apj, 690, 1539

\bibitem[{{Gorti}, {Hollenbach} \& {Dullemond}(2015){Gorti}, {Hollenbach}, \&
  {Dullemond}}]{2015ApJ...804...29G}
{Gorti} U., {Hollenbach} D., {Dullemond} C.~P., 2015, \apj, 804, 29

\bibitem[{{Guarcello} {et~al}\mbox{.}(2009){Guarcello}, {Micela}, {Damiani},
  {Peres}, {Prisinzano}, \& {Sciortino}}]{2009A&A...496..453G}
{Guarcello} M.~G., {Micela} G., {Damiani} F., {Peres} G., {Prisinzano} L.,
  {Sciortino} S., 2009, \aap, 496, 453

\bibitem[{{Guarcello} {et~al}\mbox{.}(2007){Guarcello}, {Prisinzano}, {Micela},
  {Damiani}, {Peres}, \& {Sciortino}}]{2007A&A...462..245G}
{Guarcello} M.~G., {Prisinzano} L., {Micela} G., {Damiani} F., {Peres} G.,
  {Sciortino} S., 2007, \aap, 462, 245

\bibitem[{{Habing}(1968)}]{1968BAN....19..421H}
{Habing} H.~J., 1968, \bain, 19, 421

\bibitem[{{Haisch}, {Lada} \& {Lada}(2001){Haisch}, {Lada}, \&
  {Lada}}]{2001ApJ...553L.153H}
{Haisch}, Jr. K.~E., {Lada} E.~A., {Lada} C.~J., 2001, \apjl, 553, L153

\bibitem[{{Hall}(1997)}]{1997MNRAS.287..148H}
{Hall} S.~M., 1997, \mnras, 287, 148

\bibitem[{{Hollenbach} {et~al}\mbox{.}(1994){Hollenbach}, {Johnstone},
  {Lizano}, \& {Shu}}]{hollenbach_94}
{Hollenbach} D., {Johnstone} D., {Lizano} S., {Shu} F., 1994, \apj, 428, 654

\bibitem[{{Hollenbach} \& {Tielens}(1999)}]{1999RvMP...71..173H}
{Hollenbach} D.~J., {Tielens} A.~G.~G.~M., 1999, Reviews of Modern Physics, 71,
  173

\bibitem[{{Hunter}(2007)}]{2007CSE.....9...90H}
{Hunter} J.~D., 2007, Computing in Science and Engineering, 9, 90

\bibitem[{{Johnstone}, {Hollenbach} \& {Bally}(1998){Johnstone}, {Hollenbach},
  \& {Bally}}]{johnstone_98}
{Johnstone} D., {Hollenbach} D., {Bally} J., 1998, \apj, 499, 758

\bibitem[{{Kaufman} {et~al}\mbox{.}(1999){Kaufman}, {Wolfire}, {Hollenbach}, \&
  {Luhman}}]{1999ApJ...527..795K}
{Kaufman} M.~J., {Wolfire} M.~G., {Hollenbach} D.~J., {Luhman} M.~L., 1999,
  \apj, 527, 795

\bibitem[{{Lada} \& {Lada}(2003)}]{2003ARA&A..41...57L}
{Lada} C.~J., {Lada} E.~A., 2003, \araa, 41, 57

\bibitem[{{Laibe} \& {Price}(2012)}]{2012MNRAS.420.2345L}
{Laibe} G., {Price} D.~J., 2012, \mnras, 420, 2345

\bibitem[{{Laor} \& {Draine}(1993)}]{1993ApJ...402..441L}
{Laor} A., {Draine} B.~T., 1993, \apj, 402, 441

\bibitem[{{Li} \& {Greenberg}(1997)}]{1997A&A...323..566L}
{Li} A., {Greenberg} J.~M., 1997, \aap, 323, 566

\bibitem[{{Maaskant} {et~al}\mbox{.}(2014){Maaskant}, {Min}, {Waters}, \&
  {Tielens}}]{2014A&A...563A..78M}
{Maaskant} K.~M., {Min} M., {Waters} L.~B.~F.~M., {Tielens} A.~G.~G.~M., 2014,
  \aap, 563, A78

\bibitem[{{Mamajek}(2009)}]{2009AIPC.1158....3M}
{Mamajek} E.~E., 2009, in American Institute of Physics Conference Series, Vol.
  1158, American Institute of Physics Conference Series, {Usuda} T., {Tamura}
  M., {Ishii} M., eds., pp. 3--10

\bibitem[{{Mann} {et~al}\mbox{.}(2015){Mann}, {Andrews}, {Eisner}, {Williams},
  {Meyer}, {Di Francesco}, {Carpenter}, \& {Johnstone}}]{2015ApJ...802...77M}
{Mann} R.~K., {Andrews} S.~M., {Eisner} J.~A., {Williams} J.~P., {Meyer} M.~R.,
  {Di Francesco} J., {Carpenter} J.~M., {Johnstone} D., 2015, \apj, 802, 77

\bibitem[{{Mann} {et~al}\mbox{.}(2014){Mann}, {Di Francesco}, {Johnstone},
  {Andrews}, {Williams}, {Bally}, {Ricci}, {Hughes}, \&
  {Matthews}}]{2014ApJ...784...82M}
{Mann} R.~K. {et~al.}, 2014, \apj, 784, 82

\bibitem[{{Mann} \& {Williams}(2010)}]{2010ApJ...725..430M}
{Mann} R.~K., {Williams} J.~P., 2010, \apj, 725, 430

\bibitem[{{Mathis}, {Rumpl} \& {Nordsieck}(1977){Mathis}, {Rumpl}, \&
  {Nordsieck}}]{1977ApJ...217..425M}
{Mathis} J.~S., {Rumpl} W., {Nordsieck} K.~H., 1977, \apj, 217, 425

\bibitem[{{McElroy} {et~al}\mbox{.}(2013){McElroy}, {Walsh}, {Markwick},
  {Cordiner}, {Smith}, \& {Millar}}]{2013A&A...550A..36M}
{McElroy} D., {Walsh} C., {Markwick} A.~J., {Cordiner} M.~A., {Smith} K.,
  {Millar} T.~J., 2013, \aap, 550, A36

\bibitem[{{Miotello} {et~al}\mbox{.}(2012){Miotello}, {Robberto}, {Potenza}, \&
  {Ricci}}]{2012ApJ...757...78M}
{Miotello} A., {Robberto} M., {Potenza} M.~A.~C., {Ricci} L., 2012, \apj, 757,
  78

\bibitem[{{Mitchell} \& {Stewart}(2010)}]{2010ApJ...722.1115M}
{Mitchell} T.~R., {Stewart} G.~R., 2010, \apj, 722, 1115

\bibitem[{{Murray-Clay}, {Chiang} \& {Murray}(2009){Murray-Clay}, {Chiang}, \&
  {Murray}}]{murray-clay_09}
{Murray-Clay} R.~A., {Chiang} E.~I., {Murray} N., 2009, \apj, 693, 23

\bibitem[{{{\"O}berg} {et~al}\mbox{.}(2015){{\"O}berg}, {Furuya}, {Loomis},
  {Aikawa}, {Andrews}, {Qi}, {van Dishoeck}, \& {Wilner}}]{2015ApJ...810..112O}
{{\"O}berg} K.~I., {Furuya} K., {Loomis} R., {Aikawa} Y., {Andrews} S.~M., {Qi}
  C., {van Dishoeck} E.~F., {Wilner} D.~J., 2015, \apj, 810, 112

\bibitem[{{O'Dell}, {Wen} \& {Hu}(1993){O'Dell}, {Wen}, \&
  {Hu}}]{1993ApJ...410..696O}
{O'Dell} C.~R., {Wen} Z., {Hu} X., 1993, \apj, 410, 696

\bibitem[{{Offner} {et~al}\mbox{.}(2014){Offner}, {Bisbas}, {Bell}, \&
  {Viti}}]{2014MNRAS.440L..81O}
{Offner} S.~S.~R., {Bisbas} T.~G., {Bell} T.~A., {Viti} S., 2014, \mnras, 440,
  L81

\bibitem[{{Offner} {et~al}\mbox{.}(2013){Offner}, {Bisbas}, {Viti}, \&
  {Bell}}]{2013ApJ...770...49O}
{Offner} S.~S.~R., {Bisbas} T.~G., {Viti} S., {Bell} T.~A., 2013, \apj, 770, 49

\bibitem[{{Oliveira} {et~al}\mbox{.}(2010){Oliveira}, {Pontoppidan},
  {Mer{\'{\i}}n}, {van Dishoeck}, {Lahuis}, {Geers}, {J{\o}rgensen},
  {Olofsson}, {Augereau}, \& {Brown}}]{2010ApJ...714..778O}
{Oliveira} I. {et~al.}, 2010, \apj, 714, 778

\bibitem[{{Pani{\'c}} {et~al}\mbox{.}(2009){Pani{\'c}}, {Hogerheijde},
  {Wilner}, \& {Qi}}]{2009A&A...501..269P}
{Pani{\'c}} O., {Hogerheijde} M.~R., {Wilner} D., {Qi} C., 2009, \aap, 501, 269

\bibitem[{{Parker}(1958)}]{1958ApJ...128..664P}
{Parker} E.~N., 1958, \apj, 128, 664

\bibitem[{{Pi{\'e}tu}, {Dutrey} \& {Guilloteau}(2007){Pi{\'e}tu}, {Dutrey}, \&
  {Guilloteau}}]{2007A&A...467..163P}
{Pi{\'e}tu} V., {Dutrey} A., {Guilloteau} S., 2007, \aap, 467, 163

\bibitem[{{Pinte} {et~al}\mbox{.}(2008){Pinte}, {Padgett}, {M{\'e}nard},
  {Stapelfeldt}, {Schneider}, {Olofsson}, {Pani{\'c}}, {Augereau},
  {Duch{\^e}ne}, {Krist}, {Pontoppidan}, {Perrin}, {Grady}, {Kessler-Silacci},
  {van Dishoeck}, {Lommen}, {Silverstone}, {Hines}, {Wolf}, {Blake}, {Henning},
  \& {Stecklum}}]{2008A&A...489..633P}
{Pinte} C. {et~al.}, 2008, \aap, 489, 633

\bibitem[{{Porras} {et~al}\mbox{.}(2003){Porras}, {Christopher}, {Allen}, {Di
  Francesco}, {Megeath}, \& {Myers}}]{2003AJ....126.1916P}
{Porras} A., {Christopher} M., {Allen} L., {Di Francesco} J., {Megeath} S.~T.,
  {Myers} P.~C., 2003, \aj, 126, 1916

\bibitem[{{Ricci}, {Robberto} \& {Soderblom}(2008){Ricci}, {Robberto}, \&
  {Soderblom}}]{2008AJ....136.2136R}
{Ricci} L., {Robberto} M., {Soderblom} D.~R., 2008, \aj, 136, 2136

\bibitem[{{Ricci} {et~al}\mbox{.}(2010{\natexlab{a}}){Ricci}, {Testi}, {Natta},
  \& {Brooks}}]{2010A&A...521A..66R}
{Ricci} L., {Testi} L., {Natta} A., {Brooks} K.~J., 2010{\natexlab{a}}, \aap,
  521, A66

\bibitem[{{Ricci} {et~al}\mbox{.}(2010{\natexlab{b}}){Ricci}, {Testi}, {Natta},
  {Neri}, {Cabrit}, \& {Herczeg}}]{2010A&A...512A..15R}
{Ricci} L., {Testi} L., {Natta} A., {Neri} R., {Cabrit} S., {Herczeg} G.~J.,
  2010{\natexlab{b}}, \aap, 512, A15

\bibitem[{{Ricci} {et~al}\mbox{.}(2012){Ricci}, {Trotta}, {Testi}, {Natta},
  {Isella}, \& {Wilner}}]{2012A&A...540A...6R}
{Ricci} L., {Trotta} F., {Testi} L., {Natta} A., {Isella} A., {Wilner} D.~J.,
  2012, \aap, 540, A6

\bibitem[{{Rice} {et~al}\mbox{.}(2006){Rice}, {Armitage}, {Wood}, \&
  {Lodato}}]{2006MNRAS.373.1619R}
{Rice} W.~K.~M., {Armitage} P.~J., {Wood} K., {Lodato} G., 2006, \mnras, 373,
  1619

\bibitem[{{Richert} {et~al}\mbox{.}(2015){Richert}, {Feigelson}, {Getman}, \&
  {Kuhn}}]{2015ApJ...811...10R}
{Richert} A.~J.~W., {Feigelson} E.~D., {Getman} K.~V., {Kuhn} M.~A., 2015,
  \apj, 811, 10

\bibitem[{{Richling} \& {Yorke}(2000)}]{2000ApJ...539..258R}
{Richling} S., {Yorke} H.~W., 2000, \apj, 539, 258

\bibitem[{{Roccatagliata} {et~al}\mbox{.}(2011){Roccatagliata}, {Bouwman},
  {Henning}, {Gennaro}, {Feigelson}, {Kim}, {Sicilia-Aguilar}, \&
  {Lawson}}]{2011ApJ...733..113R}
{Roccatagliata} V., {Bouwman} J., {Henning} T., {Gennaro} M., {Feigelson} E.,
  {Kim} J.~S., {Sicilia-Aguilar} A., {Lawson} W.~A., 2011, \apj, 733, 113

\bibitem[{{Rollig} {et~al}\mbox{.}(2007){Rollig}, {Abel}, {Bell}, \& {et
  al.}}]{rollig_07}
{Rollig} M., {Abel} N.~P., {Bell} T., {et al.}, 2007, \aap, 467, 187

\bibitem[{{Rosenfeld} {et~al}\mbox{.}(2013){Rosenfeld}, {Andrews}, {Hughes},
  {Wilner}, \& {Qi}}]{2013ApJ...774...16R}
{Rosenfeld} K.~A., {Andrews} S.~M., {Hughes} A.~M., {Wilner} D.~J., {Qi} C.,
  2013, \apj, 774, 16

\bibitem[{{Rosotti} {et~al}\mbox{.}(2014){Rosotti}, {Dale}, {de Juan Ovelar},
  {Hubber}, {Kruijssen}, {Ercolano}, \& {Walch}}]{2014MNRAS.441.2094R}
{Rosotti} G.~P., {Dale} J.~E., {de Juan Ovelar} M., {Hubber} D.~A., {Kruijssen}
  J.~M.~D., {Ercolano} B., {Walch} S., 2014, \mnras, 441, 2094

\bibitem[{{Salyk} {et~al}\mbox{.}(2014){Salyk}, {Pontoppidan}, {Corder},
  {Mu{\~n}oz}, {Zhang}, \& {Blake}}]{2014ApJ...792...68S}
{Salyk} C., {Pontoppidan} K., {Corder} S., {Mu{\~n}oz} D., {Zhang} K., {Blake}
  G.~A., 2014, \apj, 792, 68

\bibitem[{{Smith}, {Bally} \& {Morse}(2003){Smith}, {Bally}, \&
  {Morse}}]{2003ApJ...587L.105S}
{Smith} N., {Bally} J., {Morse} J.~A., 2003, \apjl, 587, L105

\bibitem[{{Stokes}(1851)}]{1851TCaPS...9....8S}
{Stokes} G.~G., 1851, Transactions of the Cambridge Philosophical Society, 9, 8

\bibitem[{{St{\"o}rzer} \& {Hollenbach}(1999)}]{1999ApJ...515..669S}
{St{\"o}rzer} H., {Hollenbach} D., 1999, \apj, 515, 669

\bibitem[{{Testi} {et~al}\mbox{.}(2014){Testi}, {Birnstiel}, {Ricci},
  {Andrews}, {Blum}, {Carpenter}, {Dominik}, {Isella}, {Natta}, {Williams}, \&
  {Wilner}}]{2014prpl.conf..339T}
{Testi} L. {et~al.}, 2014, Protostars and Planets VI, 339

\bibitem[{{Tielens}(2008)}]{2008ARA&A..46..289T}
{Tielens} A.~G.~G.~M., 2008, \araa, 46, 289

\bibitem[{{van der Marel} {et~al}\mbox{.}(2013){van der Marel}, {van Dishoeck},
  {Bruderer}, {Birnstiel}, {Pinilla}, {Dullemond}, {van Kempen}, {Schmalzl},
  {Brown}, {Herczeg}, {Mathews}, \& {Geers}}]{2013Sci...340.1199V}
{van der Marel} N. {et~al.}, 2013, Science, 340, 1199

\bibitem[{{Vicente} \& {Alves}(2005)}]{2005A&A...441..195V}
{Vicente} S.~M., {Alves} J., 2005, \aap, 441, 195

\bibitem[{{Vincke}, {Breslau} \& {Pfalzner}(2015){Vincke}, {Breslau}, \&
  {Pfalzner}}]{2015A&A...577A.115V}
{Vincke} K., {Breslau} A., {Pfalzner} S., 2015, \aap, 577, A115

\bibitem[{{Visser} {et~al}\mbox{.}(2007){Visser}, {Geers}, {Dullemond},
  {Augereau}, {Pontoppidan}, \& {van Dishoeck}}]{2007A&A...466..229V}
{Visser} R., {Geers} V.~C., {Dullemond} C.~P., {Augereau} J.-C., {Pontoppidan}
  K.~M., {van Dishoeck} E.~F., 2007, \aap, 466, 229

\bibitem[{{Weingartner} \& {Draine}(2001)}]{2001ApJS..134..263W}
{Weingartner} J.~C., {Draine} B.~T., 2001, \apjs, 134, 263

\bibitem[{{Woitke} {et~al}\mbox{.}(2015){Woitke}, {Min}, {Pinte}, {Thi},
  {Kamp}, {Rab}, {Anthonioz}, {Antonellini}, {Baldovin-Saavedra}, {Carmona},
  {Dominik}, {Dionatos}, {Greaves}, {G{\"u}del}, {Ilee}, {Liebhart},
  {M{\'e}nard}, {Rigon}, {Waters}, {Aresu}, {Meijerink}, \&
  {Spaans}}]{2015arXiv151103431W}
{Woitke} P. {et~al.}, 2015, ArXiv e-prints

\bibitem[{{Wolfire} {et~al}\mbox{.}(2003){Wolfire}, {McKee}, {Hollenbach}, \&
  {Tielens}}]{2003ApJ...587..278W}
{Wolfire} M.~G., {McKee} C.~F., {Hollenbach} D., {Tielens} A.~G.~G.~M., 2003,
  \apj, 587, 278

\bibitem[{{Wright} {et~al}\mbox{.}(2012){Wright}, {Drake}, {Drew}, {Guarcello},
  {Gutermuth}, {Hora}, \& {Kraemer}}]{2012ApJ...746L..21W}
{Wright} N.~J., {Drake} J.~J., {Drew} J.~E., {Guarcello} M.~G., {Gutermuth}
  R.~A., {Hora} J.~L., {Kraemer} K.~E., 2012, \apjl, 746, L21

\bibitem[{{Wyatt} \& {Dent}(2002)}]{2002MNRAS.334..589W}
{Wyatt} M.~C., {Dent} W.~R.~F., 2002, \mnras, 334, 589

\end{thebibliography}

\label{lastpage}
\end{document}